\documentclass[11pt]{JHEP3}

\usepackage{amsmath,epsfig}
\usepackage{amssymb,amsfonts}
\usepackage{latexsym}
\usepackage{tocvsec2}
\usepackage{subeqnarray}
\usepackage{xcolor}

\usepackage{graphicx}
\usepackage{amsmath}
\usepackage{amssymb,amsfonts}
\usepackage{longtable}

\newcommand{\be}{\begin{equation}}
\newcommand{\ee}{\end{equation}}
\newcommand{\beq}{\begin{equation}}
\newcommand{\eq}{\begin{equation}}
\newcommand{\eeq}{\end{equation}}
\newcommand{\eqa}{\begin{eqnarray}}
\newcommand{\eeqa}{\end{eqnarray}}
\newcommand{\bea}{\begin{eqnarray}}
\newcommand{\eea}{\end{eqnarray}}
\newcommand{\bsea}{\begin{subeqnarray}}
\newcommand{\esea}{\end{subeqnarray}}

\newcommand{\lp}{\left(}
\newcommand{\rp}{\right)}

\newcommand{\mc}[1]{\mathcal{#1}}
\newcommand{\p}{{\partial}}

\newcommand{\dd}{\mathrm{d}}
\newcommand{\DD}{\mathrm{D}}
\newcommand{\Dp}{\mathrm{D}^\perp}
\newcommand{\Tr}{\textrm{Tr}}
\newcommand{\ud}{\mathrm{d}}

\newcommand{\al}{\alpha}

\newcommand{\ba}{\beta}
\newcommand{\da}{\delta}
\newcommand{\ka}{\kappa}
\newcommand{\la}{\lambda}
\newcommand{\La}{\Lambda}

\newcommand{\si}{\sigma}

\newcommand{\munu}{{\mu\nu}}

\newcommand{\half}{\frac{1}{2}}

\def\nn{\nonumber}

\newcommand{\hs}{{\hat{s}}}

\newcommand{\hT}{{\hat{T}}}

\newcommand{\hg}{{\hat{g}}}

\newcommand{\hS}{{\hat{S}}}
\newcommand{\hR}{{\hat{R}}}

\newcommand{\tR}{{\tilde{R}}}

\newcommand{\tT}{{\tilde{T}}}

\newcommand{\tu}{{\tilde{u}}}

\newcommand{\mfg}{{\mathfrak{g}}}

\def\hri#1#2{\href{http://arxiv.org/abs/#1}{[ArXiv:#1]#2}}
\def\hre#1#2{\href{http://arxiv.org/abs/#1/#2}{[ArXiv:#1/#2]}}

\allowdisplaybreaks[3]

\title{AdS/Ricci-flat correspondence}
\author{\large Marco M.~Caldarelli$^{a}$, Joan Camps$^{b}$, Blaise Gout\'eraux$^{c}$ and Kostas Skenderis$^{a}$\\
~\\
~\\
$^a$ Mathematical Sciences and STAG research centre, University of Southampton,\\
Highfield, Southampton SO17 1BJ, United Kingdom\\
$^b$ DAMTP, Cambridge University\\
Wilberforce Road, Cambridge CB3 0WA, United Kingdom\\
$^c$\href{http://www.nordita.org}{Nordita}, KTH Royal Institute of Technology and Stockholm University\\
Roslagstullsbacken 23, SE-106 91 Stockholm, Sweden
\\\\
E-mail: \email{M.M.Caldarelli@soton.ac.uk, J.Camps@damtp.cam.ac.uk, blaise@kth.se, K.Skenderis@soton.ac.uk}
}

\preprint{NORDITA-2013-118}

\abstract{We present a comprehensive analysis of the AdS/Ricci-flat correspondence, a map between a class of asymptotically locally AdS spacetimes  and a class of Ricci-flat spacetimes.
We provide a detailed derivation of the map, discuss a number of extensions and apply it to a number of important examples, such as AdS on a torus, AdS black branes and fluids/gravity metrics. In particular,  the correspondence links the hydrodynamic regime of asymptotically flat black $p$-branes or the Rindler fluid with that of AdS. It implies that this class of Ricci-flat spacetimes inherits from AdS a generalized conformal symmetry and has a holographic structure. We initiate the discussion of holography by analyzing how the map acts on boundary conditions and holographic 2-point functions.}

\keywords{Holography, Hydrodynamics}

\begin{document}
\section{Introduction and summary of results}

Over the last fifteen years a lot effort was invested in developing and applying holographic dualities in a variety of directions.
The prototype example, the AdS/CFT duality \cite{Maldacena:1997re}, still remains to date the best understood case. In the current formulation
of holography the detailed structure of AdS gravity appears to be crucial and many of the properties of Asymptotically AdS solutions are linked to
properties of the dual QFT. Holographic dualities with non-AdS asymptotics exist but they are either less developed or their holographic dictionary can be
linked to that of AdS. A prime example of such cases is that of the dualities obtained by a decoupling limit of the non-conformal branes \cite{Itzhaki:1998dd} whose holographic dictionary
(worked out in \cite{Wiseman:2008qa,Kanitscheider:2008kd}) can indeed be linked to that of AdS via generalized dimensional reduction (a consistent reduction over a compact manifold followed by a continuation in the dimension of the compactification manifold) \cite{Kanitscheider:2009as}.

In a recent work \cite{Caldarelli:2012hy} we presented a map between a class of Asymptotically AdS spacetimes and a class of Einstein solutions with zero cosmological constant, the AdS/Ricci-flat correspondence.  The correspondence was obtained via generalized dimensional reduction and like the case of non-conformal branes one may wish to use it in order to understand holography for Ricci-flat manifolds. This is an ambitious task that we plan to undertake elsewhere. Here we will mostly focus on setting up the correspondence and explaining the results announced in \cite{Caldarelli:2012hy}.

The AdS/Ricci-flat correspondence applies to a class of solutions of Einstein's equations. In order to use this map we need to know the solutions for general dimensions.
On the AdS side, we are considering solutions of Einstein's equations with a cosmological constant $\Lambda=-d(d-1)/2 \ell^2$ for the metric $G_{MN}$ in $d+1$ dimensions,
\begin{equation}
R_{MN} =- \frac{d}{\ell} G_{MN}\,, \qquad (M,N=0, \ldots, d)\,,
\end{equation}
which are of the form
\be
    \ud s_\Lambda^2=\ud \hs^2_{p+2}(r,x;d) + e^{2\hat{\phi}(r,x;d)/(d-p-1)}\ud \vec{y}^2,
    \label{KKDiagonal_intro}
\ee
where we label the coordinates as $x^M = \{r, z^\mu\}, (\mu=0,...,d-1)$ and $z^\mu=\{x^a, \vec{y}\}, (a=0,...,p)$. The conformal boundary of the spacetime
is at $r=0$ and has coordinates $z^\mu$. The coordinates $\vec{y}$ are coordinates on a $(d-p-1)$-torus.  The metric is specified by a $(p+2)$-dimensional metric $\hat{g}(r,x;d)$ and a scalar field $\hat{\phi}(r,x;d)$, where we explicitly indicate that the metric and scalar field may depend on $d$.

On the Ricci-flat side, we are considering Ricci-flat solutions,
\be
R_{AB}=0\,, \qquad (A,B=0, \ldots, n+p+2)
\ee
in $D=n+p+3$ dimensions, which are the form
\be
    \ud s_0^2= e^{2\tilde{\phi}(r,x;n)/(n+p+1)}\left(\ud \tilde{s}^2_{p+2}(r,x;n) + \ell^2 \ud \Omega_{n+1}^2\right),
    \label{KKRicciflat_intro}
\ee
where $\ud\Omega_{n+1}^2$ is the metric of the unit round $(n+1)$-sphere. This solution is also determined by a $(p+2)$-dimensional metric $\tilde{g}(r,x;n)$ and a scalar field $\tilde{\phi}(r,x;n)$, and (as in the AdS case) we explicitly indicate that this metric and scalar field may depend on $n$.

The AdS/Ricci-flat correspondence is the statement that given a solution to Einstein's equations with a negative cosmological constant of the form (\ref{KKDiagonal_intro}), known as a function of $d$,
there is a Ricci-flat solution obtained by extracting $\hat{g}$ and a scalar field $\hat{\phi}$ from (\ref{KKDiagonal_intro}), setting
\be  \label{map_intro}
\tilde{g}(r,x;n)=\hat{g}(r,x;-n)\,; \qquad \tilde{\phi}(r,x;n)=\hat{\phi}(r,x;-n)\,,
\ee
and then substituting in (\ref{KKRicciflat_intro}). Similarly, starting from a Ricci-flat solution (\ref{KKRicciflat_intro}) we may obtain a solution  to Einstein's equations with a negative cosmological constant by $n \to -d$. We emphasize that the map is local and does not depend on (\ref{KKDiagonal_intro}) being locally Asymptotically AdS.

One may generalize this correspondence in a variety of ways. Firstly, one may replace the torus by any $(d-p-1)$-dimensional compact Ricci-flat manifold $\mathbf X$
in (\ref{KKDiagonal_intro})
and the sphere by any $(n+1)$-dimensional compact Einstein manifold $\mathbf{\tilde{X}}$ of constant positive curvature in (\ref{KKRicciflat_intro}). Secondly, by changing the Einstein manifolds $\mathbf X$ and $\mathbf{\tilde{X}}$ to have positive, negative or zero constant curvature one may arrange for similar correspondences between Asymptotically AdS solutions, Asymptotically de Sitter solutions and Ricci-flat solutions. All these cases are obtained by showing that the reduction over the compact manifold keeping only the $(p+2)$ metric and the scalar parametrizing the overall size of the compact manifold is a consistent truncation, the resulting $(p+2)$-dimensional actions contain $d$ and $n$ only as a parameter and the two actions are mapped to each other by $d \leftrightarrow -n$. As a generalization to cases where (\ref{KKDiagonal_intro}) is not locally Asymptotically AdS and has a degenerate boundary metric, one could consider solvmanifolds \cite{Hervik:2007zz,Lauret} with an $\bf R^{d-p-1}$ subspace.\footnote{See \cite{BrittoPacumio:1999sn,Taylor:2000xf} for work on holography on particular solvmanifolds/manifolds with a degenerate boundary metric.} Further generalizations are possible by considering more general consistent truncations, see \cite{Gouteraux:2011qh,Smolic:2013gx} for related work, or starting from gravity coupled to matter and then consider similar reductions, see \cite{Gouteraux:2011ce,Gouteraux:2012yr} for related work.

In this paper we will focus on the simplest case described by (\ref{KKDiagonal_intro}) and (\ref{KKRicciflat_intro}) and discuss a number of examples and applications. Asymptotically locally AdS spacetimes in $(d+1)$-dimensions are characterized by a boundary conformal structure $[g_{(0)}]$, i.e.\ a boundary metric $g_{(0)}$ up to Weyl transformations, and by a conserved $d$-dimensional symmetric tensor $T_{ij}$ with its trace determined by the boundary conformal structure. In AdS/CFT correspondence, $g_{(0)}$ plays the role of the source of the stress energy tensor of the dual QFT and $T_{ij}$ is its expectation value \cite{de Haro:2000xn}. The Ansatz (\ref{KKDiagonal_intro}) implies that we are considering Asymptotically locally AdS spacetimes with the ($d$-dimensional) boundary metric having at least a $U(1)^{d-p-1}$ isometry. Modulo this restriction, $g_{(0)}$ can be arbitrary which is essential for extracting the implication of this correspondence for Ricci-flat holography. In this paper however we will restrict our attention to $g_{(0)}$ being flat leaving a more general analysis for subsequent work. With this restriction the holographic dual of the bulk AdS solutions we consider is a QFT on $Mink_{p+1} \times T^{d-p-1}$ on non-trivial states characterized by the vacuum expectation value of the energy momentum tensor.

We will now discuss in some detail the two simplest cases, namely  AdS and AdS with a small excitation, as this illustrates many of the salient issues of the correspondence while keeping the technicalities to a minimum.
The simplest case is to have a CFT on $Mink_{p+1} \times T^{d-p-1}$ on its vacuum. The corresponding bulk solution is
Anti-de Sitter with $(d-p-1)$ boundary coordinates compactified on a torus,
\be \label{AdS_intro}
\ud s_\Lambda^2 = \frac{\ell^2}{r^2} \left(\ud r^2 + \eta_{ab} \ud x^a \ud x^b + \ud \vec{y}^2\right) .
\ee
Matching with (\ref{KKDiagonal_intro}) we find
\be
\ud \hat{s}_{p+2}^2(r,x;d) = \frac{\ell^2}{r^2} \left(\ud r^2 + \eta_{ab} \ud x^a \ud x^b \right); \qquad
\hat{\phi}(r,x;d)=(p+1-d) \log\frac r\ell\,,
\ee
and applying (\ref{map_intro}) gives
\be
\ud \tilde{s}_{p+2}^2(r,x;n) = \frac{\ell^2}{r^2} \left(\ud r^2 + \eta_{ab} \ud x^a \ud x^b \right); \qquad
\tilde{\phi}(r,x;n)=(p+1+n) \log\frac r\ell\,.
\ee
Substituting in (\ref{KKRicciflat_intro}), we obtain
\be \label{Mink_intro}
\ud s^2_0 = (\ud r^2 + r^2 \ud\Omega^2_{n+1}) +  \eta_{ab} \ud x^a \ud x^b\,,
\ee
which is simply $D$-dimensional Minkowski spacetime. On the AdS side, we fixed $\eta_{ab}$ as the boundary condition for the metric on (the noncompact part of) the boundary.
This maps on the Ricci-flat side to a metric on a $p$-brane located at $r=0$,  the origin in transverse space. The radial direction in AdS becomes the transverse distance from the $p$-brane.

Let us briefly comment on symmetries. Prior to compactification on $T^{d-p-1}$, the isometry group of (\ref{AdS_intro}) is the conformal group in $d$ dimensions. In particular this involves dilatations and special conformal transformations, whose infinitesimal action on $x^M$ is given respectively by
\be  \label{sym_intro}
\delta_\lambda x^M = \lambda x^M\qquad\text{and}\qquad \delta_b z^\mu = b^\mu z^2 - 2 z^\mu (z \cdot b) + r^2 b^\mu, \quad \delta_b r = - 2 (z \cdot b) r\,,
\ee
where $z^2 = z^\mu z^\nu \eta_{\mu \nu}$ and $z \cdot b = z^\mu b^\nu \eta_{\mu \nu}$. Compactifying the $y$ coordinates breaks both of these symmetries: the theory obtained by reduction over the torus is not conformal anymore. It still has a generalized conformal stucture,
which controls some of its properties, see \cite{Kanitscheider:2008kd}. In the current context, after we reduce over the torus, the relevant fields are the metric $\hat{g}$ and the scalar field $\hat{\phi}$.
Then the transformations
\be  \label{sym2_intro}
\delta_\lambda x^a = \lambda x^a, \quad \delta_\lambda r = \lambda r \qquad
\text{and}\qquad \delta_b x^a = b^a x^2 - 2 x^a (x \cdot b) + r^2 b^a, \quad \delta_b r = - 2 (x \cdot b) r\,,
\ee
where $x^2 = x^a x^b \eta_{ab}$ and $x \cdot b = x^a b^b \eta_{ab}$, are isometries of the metric $\hat{g}$, but the scalar field $\hat{\phi}$ transforms non-trivially,
\be
\delta_\lambda \hat{\phi} =  (p+1-d) \lambda\,; \qquad \delta_b \hat{\phi} = -2  (p+1-d) (x \cdot b)\,.
\ee
While (\ref{sym2_intro}) is not a symmetry, it is a solution-generating transformation: one can explicitly check that the field equations are solved by $(\hat{g}, \hat{\phi} + \delta \hat{\phi})$.

Let us now see how these transformations act on the Ricci-flat side. Under (\ref{sym2_intro}) the metric transforms as
\be
\delta g_{0 AB} = 2 \sigma(x) g_{0 AB}
\ee
where $\sigma = \lambda$ for dilatations and $\sigma = - 2  (x \cdot b)$ for special conformal transformations. In other words, this is a specific conformal transformation of the $D$-dimensional Ricci-flat metric. In general a Ricci-flat metric does not remain Ricci-flat after a conformal transformation, but for the transformation discussed here this turns out to be the case. Indeed, recall that under a Weyl transformation the Ricci tensor transforms as
\be
\delta R_{AB} = - (D-2) \nabla_A \nabla_B \sigma - g_{AB} \Box \sigma
\ee
and since $\sigma$ is either constant (dilatations) or linear in $x$  (special conformal transformations) the resulting metric is still Ricci-flat. In other words, while (\ref{sym2_intro}) is not an isometry of (\ref{Mink_intro}), it is a conformal isometry that preserves the Ricci-flat condition, i.e.\ it is a solution-generating transformation.

The next simplest case to consider is to add an excitation on AdS. This is achieved by adding a normalizable perturbation and, in Fefferman-Graham (FG) gauge with a flat boundary metric, the bulk metric reads
\cite{de Haro:2000xn}
\be \label{FG}
\ud s_\Lambda^2 = \frac{\ud \rho^2}{4\rho^2}+\frac{1}{\rho} \left(\eta_{\mu \nu}
+ \rho^{\frac{d}2} g_{(d)\mu\nu} + \cdots \right)
\ud z^\mu\ud z^\nu,
\ee
where we set the AdS radius $\ell=1$. The coefficient $g_{(d)\mu\nu}$ is  related to the expectation value of the dual stress energy tensor \cite{de Haro:2000xn},
\be \label{tmn_intro}
T_{\mu \nu} = \frac{d}{ 16 \pi G_{d+1}} g_{(d)\mu \nu}
\ee
($G_{d+1}$ is Newton's constant),
which satisfies
 \be \label{WI}
\partial^\mu T_{\mu \nu}=0\,, \qquad  T_\mu^{\phantom{1}\mu}=0\,,
\ee
as a consequence of the gravitational field equations \cite{Henningson:1998gx, de Haro:2000xn}.
These reflect the fact that the dual theory is conformal.

Compactifying on a $(d-p-1)$ torus,
the metric $\hat{g}$ and the scalar field $\hat{\phi}$ can be extracted from equation (\ref{KKDiagonal_intro}):
\bea
\ud \hs^2&{=}&\frac{\ud \rho^2}{4\rho^2}{+}\frac{1}{\rho^2}\left(\eta_{ab}{+}\rho^{\frac d2} (\hat{g}_{(d)ab}{+}\rho \hat{g}_{(d+2)ab}{+} \ldots) \right)\ud x^a \ud x^b, \nonumber\\
\hat{\phi} &=&\frac12(p+1-d) \ln\rho +\rho^{\frac d2} \hat{\phi}_{(d)} + \rho^{\frac d2+1} \hat{\phi}_{(d+2)} +  \ldots\label{pFG_intro}
\eea
Note that now both the scalar field and the $(p+2)$-metric depend explicitly on $d$.
The displayed coefficients are related to the expectation values of the stress energy tensor
$\hat{T}_{ab}$ and of the scalar operator $ \hat{\cal O}_\phi$ of the $(p+2)$-dimensional theory
\cite{Kanitscheider:2009as}:
\be \label{non-conf_intro}
\hT_{ab}=\frac{d}{16\pi G_{p+2}}\hg_{(d)ab}\,, \quad \hat{\mathcal O}_\phi=-\frac{d (d-p-1)}{32 \pi G_{p+2}}\hat{\phi}_{(d)}\,,
\ee
while
\be \label{HO_intro}
\hat{g}_{(d+2)ab} = - \frac{1}{2 d (d+2)}\Box \hat{g}_{(d)ab}\,, \qquad \hat{\phi}_{(d+2)}= - \frac{1}{2 d (d+2)}\Box \hat{\phi}_{(d)}\,,
\ee
and $G_{p+2}=G_{d+1}/vol(T)$ is the $(p+2)$-dimensional Newton's constant and $vol(T)$ is the volume of the torus.
The stress energy tensor satisfies the expected trace and diffeomorphism Ward identities
\be \label{WI_non-conf_intro}
\partial^a \hT_{ab}=0\,, \qquad \hT_a^{\phantom{1}a}=(d-p-1)\hat{\mathcal O}_{\phi}\,.
\ee
Note that the stress energy tensor is not traceless anymore, reflecting the fact that the compactification
breaks dilatations and special conformal transformations.

We can now apply the AdS/Ricci-flat correspondence and, to further simplify the presentation, we will consider a linearized perturbation only, i.e.\ we will linearize in the strength of the perturbation.
The corresponding Ricci-flat solution is then (setting $\rho=r^2$):
\bea
\ud s_0^2&&=(\eta_{AB}+ h_{AB}+\dots) \ud x^A \ud x^B \nonumber \\
 &&= \left(1-\frac{16\pi G_{p+2}\, \,}{n\,r^{n}}\left(1+
\frac{r^2}{2 (n-2)} \Box_x\right) \hat{\mathcal O}_\phi(x;n) \right)
\left(\ud r^2+\eta_{ab} \ud x^a \ud x^b + r^2\ud \Omega_{n+1}^2\right) \nonumber \\
&&\hphantom{=}
\  -\frac{16\pi G_{p+2}\, \,}{n\,r^{n}}\left(1+\frac{r^2}{2 (n-2)} \Box_x\right) \hT_{ab}(x;n) \ud x^a \ud x^b +\dots,
\label{asAdSmaptoFlat_intro}
\eea
where $x^A$ are $D$-dimensional coordinates, $\Box_x$ is the Laplacian along the brane, and $\hat{\mathcal O}_\phi(x;n)$, $\hT_{ab}(x;n)$ indicate $\hat{\mathcal O}_\phi(x)$, $\hT_{ab}(x)$ with $d=-n$.

In AdS the holographic stress energy tensor appears at order $r^d$ in the asymptotic expansion near infinity.
Higher order coefficients contain additional derivatives relative to the stress energy tensor (see \eqref{HO_intro}) and vanish faster as we approach the conformal boundary
(see \eqref{pFG_intro}). Moving to the Ricci-flat side, the map $d \to -n$ implies that the holographic stress energy tensor appears at order $1/r^n$, and now this term vanishes as $r \to \infty$, i.e.\ near infinity in Minkowski spacetime. This is precisely the order where one expects to find the ADM energy due to a $p$-brane located at $r=0$. Indeed, defining
${\bar{h}}_{AB}=h_{AB}-\frac12{}h^C{}_{\!C}\,\eta_{AB}$, where $h_{AB}$ is given in \eqref{asAdSmaptoFlat_intro},
we find that ${\bar{h}}_{AB}$ satisfies
\be
\Box{\bar h}_{AB}=16\pi G_{p+2} \Omega_{n+1}\delta_A^{\phantom{1}a}\delta_B^{\phantom{1}b} \hT_{ab}(x;n) \delta^{n+2}(r)
\ee
through second order terms in (boundary) derivatives.
Comparing with the linearized Einstein's equations we conclude that the
holographic stress energy tensor takes a new meaning: it is (proportional to) minus the stress energy tensor due to a $p$-brane located at $r=0$ that sources the linearized gravitational field $h_{AB}$.

Note however that while in AdS the higher order terms vanish faster than the leading order term as we
approach the conformal boundary,  the corresponding terms in (\ref{asAdSmaptoFlat_intro}) vanish slower as $r \to \infty$. This means that if we start from an AdS solution containing a sufficiently high number of subleading terms the corresponding Ricci-flat solution would now contain terms that blow up as $r \to \infty$. Higher order terms, however, contain additional derivatives (along boundary directions) relative to lower order terms and are subleading in a derivative expansion.

In AdS/CFT the standard observables are correlation functions of local operators of the dual QFT. These are computed gravitationally by finding regular solutions of the bulk field equations satisfying appropriate Dirichlet boundary conditions. While a complete analysis of the map of these observables is left for future work, we do analyze a specific case of the computation of a 2-point function. The map in this case is canonical in the sense that the regular solution that yields the 2-point correlator is mapped on the Ricci-flat side to the solution with the expected fall-off in the radial coordinate, see section \ref{section:correlation}.

Another interesting case to consider is to have a thermal state on the boundary, which corresponds to a black hole in AdS. The planar AdS black brane reads,
\be \label{AdSBB}
\ud s_\Lambda^2 = \frac1{r^2} (-f(r)d\tau^2+d\vec{x}^2+d\vec{y}^2)+\frac{\ud r^2}{r^2f(r)}\,,
\ee
where $x^a=\{\tau, \vec{x}\}$, the $\vec{y}$ coordinates parametrize a torus, as before, and
$ f(r)=1-r^d/b^{d}$.
Applying the map, we obtain
\be \label{BlackpBrane}
	\ud s_0^2=-f(r)\ud \tau^2 + \frac{\ud r^2}{f(r)}+r^2\ud \Omega_{n+1}^2+\ud \vec{x}^2,
\ee
where $ f( r)=1-(b/{r})^{n}$, i.e.\ the Schwarzschild black $p$-brane. A special case is when there is no sphere, i.e.\ $n=-1$. In this case, a further change of coordinates to $(t= \tau/b + \log f(r),\, \mathfrak{r}=b^2 f(r))$, shows that the metric describes Minkowski spacetime in ingoing Rindler coordinates,
\be \label{Rindler}
\ud s_0^2= -\mathfrak{r} \ud t^2 + 2 \ud \mathfrak{r} \ud t + \ud \vec{x}^2.
\ee

One can now use the AdS/Ricci-flat map and the AdS/CFT duality to extract some of the basic properties of the Ricci-flat solutions. Recall that the AdS black branes are dual to a conformal ideal fluid, with equation of state that follows from \eqref{WI}
\be
 \varepsilon=(d-1)P
\ee
where $\varepsilon$ and $P$ are the energy and pressure densities. Applying the AdS/Ricci-flat map we find,
 \be \label{eqS}
 \tilde{\varepsilon}= -(n+1)\tilde{P}\,,\qquad c_s^2=\frac{\partial \tilde{P}}{\partial \tilde{\varepsilon}}=-\frac1{n+1}\,,
\ee
where $\tilde{\varepsilon}$ and $\tilde{P}$ are the ADM energy and pressure densities.
When $n=-1$, we find that $\tilde{\varepsilon}=0$, which is indeed the energy density of the Rindler fluid \cite{Compere:2011dx}, while when $n\geq 1$ the speed of sound $c_s$ is imaginary and there is an instability for the sound modes. This is precisely \cite{Emparan:2009at} the Gregory-Laflamme (GL) instability \cite{Gregory:1993vy, Gregory:1994bj}.

In QFT, the long-wavelength behavior close to thermal equilibrium is described by hydrodynamics. In AdS/CFT
there are bulk solutions describing holographically this hydrodynamic regime in a derivative expansion \cite{Bhattacharyya:2008jc,Bhattacharyya:2008mz}. Applying the AdS/Ricci-flat to these solutions when $n \geq 1$ leads to solutions describing the nonlinear evolution of the GL instability to second order in gradients, which we will present in detail here. In \cite{Caldarelli:2012hy} we linearized these solutions (in the amplitude of perturbations)
and obtained the dispersion relation of the GL unstable modes to cubic order in the wavenumber of the perturbation, generalizing the quadratic approximation in \cite{Camps:2010br}.
This dispersion relation agrees remarkably well with numerical data \cite{Figueras} all the way up to the threshold mode, when $n$ is sufficiently large.

When $n=-1$ the AdS hydrodynamic solutions \cite{Bhattacharyya:2008jc,Bhattacharyya:2008mz} map exactly to the hydrodynamic solutions of the Rindler fluid \cite{Bredberg:2011jq, Compere:2011dx, Compere:2012mt, Eling:2012ni}. In particular, the AdS boundary maps to the position of the accelerated observer, which is precisely where boundary conditions were imposed in \cite{Bredberg:2011jq, Compere:2011dx, Compere:2012mt, Eling:2012ni}. The first and second order transport coefficients are also linked by the map, as well as the position of the horizons and the entropy currents.

It is well known that on the AdS side the structure of the hydrodynamic expansion is organized by the Weyl invariance of the dual CFT \cite{Baier:2007ix}. The AdS/Ricci-flat map implies that the hydrodynamic solutions on the Ricci-flat side are also organized similarly, due to the generalized conformal structure mentioned earlier.

This paper is organized as follows. In section \ref{section:map}, we go through the derivation of the map, relegating technical details to Appendix~\ref{appendix:KKred}. We comment on the implications for holography in section \ref{section:HoloDicRicciFlat} and in particular on 2-point correlation functions in \ref{section:correlation}. Some generalizations of the map are presented in \ref{section:generalizations}. Throughout section \ref{secFluidsGravity}, we work in the hydrodynamic limit, starting from the fluids/gravity metrics in \ref{section:FluidsGravityMetrics} and mapping them to asymptotically flat, second-order in viscous corrections metrics in \ref{section:RicciHydro}. In \ref{section:EntropyCurrent}, we examine the entropy current, and \ref{section:GL} derives the Gregory-Laflamme dispersion relation to cubic order in momentum. Section \ref{subsection:n=1,2} deals with the regularization of the $n=1,2$ cases. Many technical details are relegated in the appendices. Section \ref{section:Rindler} studies the limit of vanishing sphere and Rindler hydrodynamics, and finally we conclude in \ref{section:ccl}.

\section{AdS/Ricci-flat correspondence}

\subsection{Derivation of the map \label{section:map}}

In this section we derive the AdS/Ricci-flat map.
Let us start from the AdS-Einstein action in $d+1$ dimensions:
\be
    S_\Lambda=\frac1{16\pi G_{d+1}}\int_{\mathcal M} \ud^{d+1}x\,\sqrt{-g}\left [R-2 \La \right],
    \label{EinsteinAdSAction}
\ee
where $\La=-d(d-1)/2 \ell^2$. This theory can be consistently reduced to a $(p+2)$-dimensional theory
using the reduction Ansatz (\ref{KKDiagonal_intro}). The proof of this consists of showing that the field equations of (\ref{EinsteinAdSAction}) evaluated on the reduction Ansatz (\ref{KKDiagonal_intro}) lead to equations for $\hat{g}$ and $\hat{\phi}$ which can be obtained as the
field equations of the $(p+2)$-dimensional theory
with action
\be
    \hat{S}=\frac1{16\pi G_{p+2}}\int_{\mathcal M} \ud^{p+2}x\,\sqrt{-\hg}
e^{\phi}\left [\hR+\frac{d-p-2}{d-p-1}\left(\partial\hat{\phi}\right)^2-2\La \right],
    \label{ED2LiouvilleAction}
\ee
where $G_{p+2}=G_{d+1}/vol(T^{p-d-1})$.
In particular, this implies that all solutions of (\ref{ED2LiouvilleAction}) uplift to solutions
(\ref{EinsteinAdSAction}). We relegate the details (which follow closely the discussion in \cite{Gouteraux:2011qh})
to appendix~\ref{appendix:KKred}.

Alternatively, we can start from Einstein gravity in $D=p+n+3$ dimensions,
\be
S_0 = \frac{1}{16 \pi G_{p+n+3}} \int_{\mathcal M} \ud^{p+n+3}x\,\sqrt{-g} R
    \label{EinsteinAction}
\ee
and consider the reduction Ansatz (\ref{KKRicciflat_intro}). As we show in the appendix~\ref{appendix:KKred},
this reduction is also consistent, leading to the following $(p+2)$-dimensional theory,
 \be
    \tilde{S}=\frac1{16\pi \tilde{G}_{p+2}}\int_{\mathcal M} \ud^{p+2}x\,\sqrt{-\tilde{g}}
e^{\tilde{\phi}}\left [\tilde{R}+\frac{n+p+2}{n+p+1}\left(\partial\tilde{\phi}\right)^2+\tR_{n+1} \right ],
    \label{ED2LiouvilleAction2}
\ee
where $\tR_{n+1}=n(n+1)/\ell^2$ is the curvature scalar of a $(n+1)$-dimensional sphere of radius $\ell$ (the compactification manifold in (\ref{KKRicciflat_intro})) and $\tilde{G}_{p+2} = G_{n+p+3}/(\ell^{n+1} \Omega_{n+1})$ ($\Omega_{n+1}$ is the volume of the unit radius $(n+1)$ sphere).

In both actions, \eqref{ED2LiouvilleAction} and \eqref{ED2LiouvilleAction2},
the parameters $d$ and $n$ (related to the number of reduced dimensions) appear only in the  coefficients of the various terms. Thus, both $d$ and $n$ may be analytically continued to any value provided the effective actions remain well-defined, i.e.\ the coefficients of all terms remain bounded and satisfy appropriate conditions (for example, the coefficient of the kinetic term is positive etc.). Such a reduction, which is both consistent and analytically continuable in the number of reduced dimensions, has been called \emph{generalized}, \cite{Kanitscheider:2009as,Gouteraux:2011ce,Gouteraux:2011qh}. Note also that we end up in the so-called dual frame,\footnote{Recent developments in cosmology call it the Galileon frame \cite{VanAcoleyen:2011mj,Charmousis:2012dw} and it may be useful to extend the AdS/Ricci-flat map in presence of higher-derivative terms.} which is very useful for holographic applications as the reduced holographic dictionary is very easily derived, \cite{Kanitscheider:2009as}.

Inspection of \eqref{ED2LiouvilleAction}, \eqref{ED2LiouvilleAction2} shows that they are related by
\be \label{Map1}
	d=-n\,; \qquad \tilde{g}(r,x;n)=\hat{g}(r,x;d=-n)\,; \qquad \tilde{\phi}(r,x;n)=\hat{\phi}(r,x;d=-n)\,.
\ee
This then provides a derivation of the AdS/Ricci-flat correspondence.
Note that without loss of generality we can set $G_{p+2}=\tilde{G}_{p+2}$ (this just a choice of units)
and we will do so throughout this paper.

\subsection{Towards holography  \label{section:HoloDicRicciFlat}}

In this subsection we make a few comments about setting up holography using the correspondence.

Let us first briefly recall the holographic setup for Asymptotically AdS spacetimes.
It is well-known that one may always bring Asymptotically AdS solutions to a Fefferman-Graham form, at least in a small enough neighbourhood of the conformal boundary,
\bea
	\frac{\ud s^2_\Lambda}{\ell^2} &=& \frac{\ud \rho^2}{4\rho^2}+\frac{1}{\rho} \mfg_{\mu\nu}(\rho,z^\la)\ud z^\mu\ud z^\nu\,, \nonumber \\
	 \mfg(\rho,z)&=& g_{(0)}(z)+\rho g_{(2)}(z)+\ldots+\rho^{d/2} \left(g_{(d)}(z)+h_{(d)}(z)\log \rho\right)+\ldots, \label{gExpansion}
\eea
where $\rho$ is a radial bulk coordinate parametrizing geodesics orthogonal to the boundary, while $g_{(0)}(z)$ is the boundary metric (more properly a representative of the boundary conformal structure). $g_{(0)}$ is also the source for the stress energy tensor $T_{\mu \nu}$ of the dual theory, and as such it must be unconstrained if we are to extract holographically the QFT data, namely the correlation functions of $T_{\mu \nu}$.
The subleading metric coefficients $g_{(2)}$, etc., are locally related to $g_{(0)}$, and this holds until order $d$, where only the divergence and the trace of $g_{(d)}$ are fixed. $g_{(d)}$ itself is related to the holographic stress-tensor \cite{de Haro:2000xn}. Subsequent orders are determined from the field equations and involve both $g_{(0)}$ and $g_{(d)}$ (we give some of these terms in Appendix~\ref{app:SubleadingFG}). The logarithmic contribution $h_{(d)}$ appears only for even $d$, and it is the metric variation of the conformal anomaly \cite{de Haro:2000xn,Henningson:1998gx}.

Correlation functions of the stress energy tensor can be obtained by functionally differentiating $g_{(d)}$ with respect to $g_{(0)}$. In practice, the relation between $g_{(d)}$ and $g_{(0)}$ is obtained perturbatively by expanding around a background solution, with linearized fluctuations giving 2-point functions, quadratic fluctuations 3-point functions, etc.\ and requiring regularity in the interior.

In this paper we restrict ourselves to the case of a flat boundary metric $g_{(0)\mu\nu}=\eta_{\mu\nu}$ (with notable exception the next subsection). This means we can only discuss 1-point functions (apart from the next subsection, where we discuss 2-point functions).
This also implies that all coefficients $g_{(k)}$ with $0<k<d$ are identically zero, and so is the logarithmic term $h_{(d)}$.
In that case, extracting the holographic stress-tensor is particularly easy -- the formula is given in (\ref{tmn_intro}).

Let us now review the holographic dictionary for the dilatonic theory \eqref{ED2LiouvilleAction}.
From the perspective of the $(p+2)$-dimensional theory we have a metric $\hat{g}$ which is dual to the stress energy tensor $\hat{T}_{ab}$ and a scalar field $\hat{\phi}$ which is dual to a scalar operator $\hat{{\cal O}}_\phi$. The scalar operator is essentially the Lagrangian of the dual theory \cite{Kanitscheider:2008kd}. The holographic dictionary can be worked out from first principles using the standard procedure (analyzing the structure of the asymptotic solutions of the field equations of \eqref{ED2LiouvilleAction} etc.) and it has indeed been obtained in this way in \cite{Wiseman:2008qa, Kanitscheider:2008kd}.
However, thanks to the generalized dimensional reduction, all this information can be simply obtained by reducing the corresponding  AdS expressions \cite{Kanitscheider:2009as}. This leads to the results we reviewed in the introduction, equations (\ref{pFG_intro})-(\ref{WI_non-conf_intro}).

The AdS/Ricci-flat correspondence may be used now to obtain a Ricci-flat solution corresponding to (\ref{gExpansion}). After linearizing in $g_{(d)}$ and keeping terms with up to two boundary derivatives, the explicit form of the solution is given in (\ref{asAdSmaptoFlat_intro}). We would now like to understand where the AdS holographic data sit in this solution. As discussed in the introduction the choice of a boundary metric translates on the Ricci-flat side to the choice of a metric on the $p$-brane.  Indeed, asymptotically the metric (\ref{asAdSmaptoFlat_intro})
is a Ricci-flat $p$-brane.  The second piece of holographic data, the holographic stress energy tensor, appears now at the order where one expects to find the ADM energy due to a $p$-brane located at $r=0$. To make this more precise, let us
consider the combination\footnote{To check this one may use \eqref{WI_non-conf_intro} with $d=-n$.}
\be
	\bar{h}_{AB}=h_{AB}-\frac{h^C{}_C}{2}\eta_{AB}=
	-
\delta^a_A\delta^b_B  \frac{16\pi G_{p+2}\,\ell^{n+1}}{n\,r^{n}} \left(\hT_{ab}(x;n)+
\frac{r^2}{2(n-2)}\Box_x\hT_{ab}(x;n) \right),
\ee
where $h^C{}_C=\eta^{CD} h_{AB}$. Recall that AdS compactified on a torus was mapped to Minkowski spacetime in $p$-brane coordinates, \eqref{Mink_intro}; $x^a$ are now the coordinates along the $p$-brane and $r$ is the distance transverse to it. Correspondingly, the Laplacian splits into a Laplacian along the brane, $\Box_x$, and a Laplacian along the transverse space $\Box_{\perp}$, $\Box = \Box_x + \Box_{\perp}$.
It is now straightforward to show that $\bar{h}_{AB}$ satisfies
\be
	\Box\bar{h}_{AB}=\left[\Box_x + \Box_{\perp}\right]\bar{ h}_{AB}
	=16\pi G_{p+n+3}\,\delta^a_A\delta^b_B\hT_{ab}(x;n)\delta^{n+2}(r)
\label{LinEqT}
\ee
to second order in derivatives along the brane (that is, up to terms like $\Box_x\Box_x\hT_{ab}(x)$), and zero for the other components. Here we used $G_{p+n+3}=G_{p+2} \Omega_{n+1}\ell^{n+1}$ (see the discussion below (\ref{ED2LiouvilleAction2})) and $\Box_{\perp} (1/r^{n}) = - n \Omega_{n+1} \delta^{n+2}(r)$. Note that for this result to hold the precise coefficient of the subleading term $g_{(d+2)}$ in the Fefferman-Graham expansion plays a crucial role, otherwise the terms with $\Box_x \hT_{ab}(x)$ would not cancel.

Now recall that the linearized Einstein equations in flat space read
\be
\Box \bar{h}_{AB}=-16\pi G_{p+n+3} \tT_{AB},
\ee
which implies the identification,
\beq
 \tT_{ab}=\delta_a^A\delta_b^B\tT_{AB}=-\hT_{ab}(x;n)\,\delta^{n+2}( r)\,.
\label{stresstensorrelation}
\eeq
In other words, the linearized perturbations are sourced by the stress tensor $\tT_{ ab }$ sitting at $r=0$ and localized along the brane directions. As it will become apparent when we discuss examples in later sections, the minus sign in the definition of $\tT_{ab}$, combined with additional signs coming from $d \to -n$,
ensures that the energy density is positive and is also consistent with the results in \cite{Camps:2010br}.
Note also that $\tT_{ab}$ is conserved, thanks to \eqref{WI_non-conf_intro}.

Terms proportional to $\hat{\mathcal O}_\phi$ still appear in the metric \eqref{asAdSmaptoFlat_intro}, but there is no longer any bulk scalar. During the oxydation from $p+2$ to $n+p+3$ dimensions, we pulled it back in the metric as an overall scale factor -- $\hat{\mathcal O}_\phi$ is the coefficient of the $1/r^n$ fall-off of the scale factor. A non-zero value of $\hat{\mathcal O}_\phi$ now implies (via \eqref{WI_non-conf_intro}) that the matter that sources $h_{AB}$ is not conformal.

There are two further remarks about \eqref{asAdSmaptoFlat_intro}. First, it is clear that if one includes subleading terms containing more derivatives, this will eventually destroy the asymptotics at finite $n$. Each extra derivative comes accompanied by a unit increase of its scaling with $r$, such that for enough extra derivatives, this power becomes positive and diverges when $r\to\infty$. When that happens, the metric is still Ricci-flat, but it no longer asymptotes to the $p$-brane. Second, we have restricted to flat boundaries: it is straightforward to see that the same kind of issues will arise for curved boundaries, unless the parent Fefferman-Graham expansion truncates.\footnote{Such a truncation occurs if the bulk metric is conformally flat or the boundary metric is conformally Einstein \cite{Skenderis:1999nb,de Haro:2000xn}.} To address this issue one needs exact solutions
and we will discuss this in the next subsection. Another way to proceed is to truncate the derivative expansion. This is precisely what happens in the hydrodynamic regime and we will discuss this in detail in section \ref{secFluidsGravity}.

\subsubsection{Correlation functions\label{section:correlation}}

In the previous subsection we discussed how the AdS/Ricci-flat correspondence acts on the asymptotic part of an asymptotically AdS solution and on the corresponding holographic data. In particular, we saw that (within a linearized and boundary derivative expansion) the 1-point function for the holographic stress energy tensor $T_{ab}$ maps to the stress energy tensor due to a $p$-brane localized in the interior of an asymptotically flat spacetime (with $p$-brane boundary conditions). In this subsection we would like to discuss the map of the 2-point function of $T_{ab}$.

Let us briefly recall how we compute correlation functions in gauge/gravity duality.  We start from a background solution with holographic data $(g_{(0)\mu \nu}^B, g_{(d)\mu \nu}^B)$ and then turn on a source $h_{(0)\mu \nu}$ perturbatively, i.e.\ we consider solutions satisfying the boundary condition
\be
g_{(0)\mu \nu}= g_{(0)\mu \nu}^B +  h_{(0)\mu \nu}\,.
\ee
Solving Einstein's equations perturbatively in $h_{(0)}$ and imposing regularity in the interior leads to a unique solution from which one can extract $g_{(d)\mu \nu}$,
\be
g_{(d)\mu \nu} = g_{(d)\mu \nu}^B + {\cal T}_{\mu \nu \mu_1 \nu_1} h_{(0)}^{\mu_1 \nu_1} + \frac{1}{2}
{\cal T}_{\mu \nu \mu_1 \nu_1 \mu_2 \nu_2} h_{(0)}^{\mu_1 \nu_1} h_{(0)}^{\mu_2 \nu_2} + \cdots
\ee
where the dots indicate higher order terms in $h_{(0)}$. It then follows from (\ref{tmn_intro}) that up to numerical factors, ${\cal T}_{\mu \nu \mu_1 \nu_1}$ is the 2-point function of $T_{\mu\nu}$, ${\cal T}_{\mu \nu \mu_1 \nu_1 \mu_2 \nu_2}$ is the 3-point function, etc. It also follows from this analysis that if we want to compute 2-point functions it suffices to linearize around the background solution. Then the 2-point function is extracted from the asymptotics of the (regular) linearized solution $h_{\mu \nu}$:
\be
h_{\mu \nu}= \frac{1}{r^2}h_{(0)}^{\mu_1 \nu_1}\left( g_{(0)\mu \mu_1}^B g_{(0)\nu \nu_1}^B + \cdots + r^d  {\cal T}_{\mu \nu \mu_1 \nu_1} + \cdots \right).
\ee

We will now discuss this in the simplest possible setup relevant for us, namely we will consider as a background solution AdS on $T^{d-p-1}$, (\ref{AdS_intro}). In this case the fluctuations should preserve the symmetries of the torus. Furthermore, we will focus on the fluctuations that do not depend on the torus, are in the radial axial gauge $h_{rM}=0$ and are transverse-traceless in the boundary $(p+1)$ directions,
\beq
h_{ab}(k,r) = h_{(0)ab}(k) \varphi_\Lambda(k,r)\,, \qquad  \eta^{ab} h_{(0) ab}= k^a h_{(0)ab}=0\,,
\eeq
where we Fourier transformed over the non-compact boundary directions $x^a$, with $k_a$ being the corresponding momentum. This corresponds to computing the transverse-traceless part of the
2-point function of $T_{ab}$ of the QFT dual to the dilatonic theory \eqref{ED2LiouvilleAction}.
This computation was done in section 7 of \cite{Kanitscheider:2008kd} (where the entire set of 2-point functions was computed) by considering fluctuations in the reduced $(p+2)$-dimensional theory and we borrow from there the relevant results. The fluctuation equation reduces to the equation (see (7.23) of \cite{Kanitscheider:2008kd} and use $\rho=r^2$):
\be \label{fl_AdS}
r^2 \varphi_\Lambda'' + (1-d) r \varphi_\Lambda' - r^2 k^2 \varphi_\Lambda =0\,,
\ee
where the prime indicates the derivative with respect to $r$.
This is actually the same equation as that satisfied by a massless scalar in $AdS_{d+1}\,$:
\beq \label{massless_eq_AdS}
0=\Box_{AdS} \varphi_\Lambda-r^2 k^2\varphi_\Lambda = r^{d+1} \partial_r (r^{1-d} \partial_r \varphi_\Lambda) - r^2 k^2 \varphi_\Lambda\,.
\eeq
Indeed, the generalized dimensional reduction implies that the transverse traceless fluctuation of the $p+2$ theory are inherited from those of the parent $d+1$ AdS gravity (and as is well known the transverse traceless fluctuations of that theory are governed by a massless scalar).

The general solution of (\ref{fl_AdS}) that is regular in the interior and approaches 1 as $r \to 0$ is
\be \label{reg_AdS}
\varphi_\Lambda(k,r) = \frac{1}{2^{d/2-1}\Gamma(d/2)} (kr)^{d/2} K_{d/2}(kr)\,.
\ee
The asymptotic expansion of linearized solution when $d$ is not even is then
\begin{align}
h_{ab}(r,k)=& h_{(0)ab}(k) \left[\vphantom{ \left(\frac{k}{2}\right)^d}\left(1-\frac{1}{2(d-2)}k^2r^2+O(k^4 r^4)\right) \right. \nonumber \\
& \qquad \qquad \left.+\frac{\Gamma(-d/2)}{\Gamma(d/2)} \left(\frac{k}{2}\right)^d\left(r^d+\frac{1}{2(d+2)}k^2r^{d+2}+O(k^4 r^4)\right)\right].
\label{Lin_exp}
\end{align}
The coefficient of the $r^d$ term, $k^d$, is essentially the 2-point function.\footnote{When $d$ is an even integer, the
asymptotic expansion now contains a logarithmic term, $(r k)^d \log (r k)$, and the 2-point function is now proportional to $k^d \log k$.}

We now want to apply the AdS/Ricci-flat correspondence. Letting $d \to -n$ in (\ref{fl_AdS}) we find that
the equation now becomes
\be \label{fl_flat}
0=r^2 \varphi_0'' + (1+n) r \varphi_0' - r^2 k^2 \varphi_0
= r^2 \left( \frac{1}{r^{n+1}} \partial_r (r^{n+1} \partial_r )\varphi_0 - k^2 \varphi_0 \right) = r^2 (\Box_{n+2} -k^2) \varphi_0\,.
\ee
This is precisely the equation for a massless field in flat $(n+2)$ dimensions with metric
\be
ds^2_{n+2}=dr^2 + r^2 d \Omega_{n+1}^2\,.
\ee
A short computation shows that a perturbation of (\ref{Mink_intro}) that is
a transverse traceless perturbation along the $p$-brane and an S-mode in the transverse directions is given by
\be
h_{ab}(k,r)=h_{(0)ab}(k) \varphi_0(k,r)\,, \qquad \eta^{ab} h_{(0)ab}=0,\ k^a h_{(0)ab} =0\,.
\ee
So indeed the equation for transverse traceless mode of AdS on a torus is mapped to a corresponding equation in Minkowski.

Let us now act with the map on the regular solution (\ref{reg_AdS}).
The corresponding solution is
\be \label{AF_fl}
\varphi_0(k,r) =
\varphi_{(0)0}(k) \frac{K_{n/2}(kr)}{(k r)^{n/2}}\,,
\ee
where we used $K_{-\nu}=K_{\nu}$ and we kept free the overall normalization $\varphi_{(0)0}(k)$ (this factor can be absorbed in $h_{(0)ab}(k)$).
This solution falls off exponentially fast as $kr \to \infty$, so the perturbation is asymptotically flat in the directions transverse to the $p$-brane. The second solution of (\ref{fl_flat}) which is linearly independent is $I_{n/2}(k r)/(k r)^{n/2}$. This solution blows up at infinity.
So we see that the regularity of the solution in AdS
(which is what selected the $K$ Bessel function) is linked with the Ricci-flat perturbation being asymptotically flat.

Let us now consider the behavior for $k r \to 0$. The solution (\ref{reg_AdS}) diverges as $1/r^n$ in this limit, but this is precisely what is expected. Recall that $\Box_{n+2} 1/r^n \sim \delta^{n+2}(r)$ so the singularity at $r=0$ is just a reflection of the fact that there is a source term localized at the position of the brane at $r=0$, confirming what we found earlier in (\ref{stresstensorrelation}).

Finally, let us now address an issue we raised in the previous subsection. There we noticed that subleading terms in the Fefferman-Graham expansion may become leading after the AdS/Ricci-flat map. Indeed, when $n=2$, corresponding to $d=-2$, the subleading term of order $r^{d+2}$ in (\ref{Lin_exp}) becomes of the same order as the source term and
more subleading terms would appear to lead to a solution that blows up as $r \to \infty$ on the Ricci-flat side.
In the case at hand, we know the exact solution so the $d=-2$ case maps to
\begin{equation}
\varphi_0(k,r;n=2) = \varphi_0(k) \frac{K_{1}(kr)}{k r}\,.
\end{equation}
This falls off exponentially as $k r \to \infty$ so the subleading terms in (\ref{Lin_exp}) do not cause any problem: (at least in this example) one may safely truncate the series at order $r^d$ and apply the map.
In later sections we will indeed truncate the series in a similar fashion using a low energy expansion and one may expect that the corresponding hydrodynamic solutions will correctly capture the long distance and long wavelength behavior of exact solutions.

\subsection{Generalizations \label{section:generalizations}}

There is a natural generalization of the AdS/Ricci-flat correspondence that we discuss in this section.
The new correspondence connects pairs of Einstein metrics with cosmological constant either positive, negative or zero. All possible combinations are
allowed, i.e.\ AdS/AdS, AdS/dS, AdS/Ricci-flat, Ricci-flat/Ricci-flat etc., although in some cases the map may be trivial.

Recall that an Einstein metric $g$ of cosmological constant $\Lambda$ in $d+1$ dimensions satisfies the equation
\be\label{EinsteinSpaceCondition}
R_{MN}=-\sigma \frac{d}{\ell^2} g_{AB}\,;
\qquad \Lambda = \sigma \frac{d(d_1)}{\ell^2} \equiv \sigma |\Lambda|\,,
\ee
where $\sigma=\{+,-,0\}$ specifies whether we have positive, negative or zero cosmological constant
and $\ell$ is the (A)dS radius.

On the one side of the correspondence we have a solution
of $(d+1)$-dimensional Einstein gravity with cosmological constant $\Lambda$
of the form:
\be
    \ud s_\Lambda^2=\ud \hs^2_{p+2}(r,x;d) + e^{2\hat{\phi}(r,x;d)/(d-p-1)}\ud {\bf X}_{d-p-1}^2\,,
    \label{KK1}
\ee
where $\ud {\bf{X}}_{d-p-1}^2$ is the metric of the compact Euclidean Einstein manifold $\mathbf X^{d-p-1}$ of curvature $R_{d-p-1}\,$.

On the other side, we have a solution
of $(n+p+3)$-dimensional Einstein gravity with cosmological constant $\tilde{\Lambda}$
of the form:
\be
    \ud s_{\tilde{\Lambda}}= e^{2\tilde{\phi}(r,x;n)/(n+p+1)}\left(\ud \tilde{s}^2_{p+2}(r,x;n) + \ud \tilde{\bf {X}}_{n+1}^2\right),
    \label{KK2}
\ee
where $\ud \tilde{\bf {X}}_{n+1}^2$ is the metric of the compact Euclidean Einstein manifold $\tilde{{\bf X}}^{n+1}$ of curvature $\tilde{R}_{n+1}\,$.

The two solutions are in correspondence provided (\ref{Map1}) holds and
\be \label{general_map}
	-2 \La \leftrightarrow \tR_{n+1}\,,\qquad  R_{d-p-1} \leftrightarrow - 2\tilde\La\,.
\ee
This is proven following the same steps as in section \ref{section:map}.
The Einstein-Dilaton theory obtained by reducing the $d+1$ theory over the
compact manifold $\mathbf X^{d-p-1}$ has a scalar potential comprising
two exponential terms, whose origin is respectively the
higher-dimensional cosmological constant, $\La$, and the curvature $R_{d-p-1}$ of
the internal space,
\be
    \hS=\frac1{16\pi G_{p+2}}\int_{\mathcal M} \ud^{p+2}x\,\sqrt{-\hg}
e^{\phi}\left [\hR+\frac{d-p-2}{d-p-1}\left(\partial\hat{\phi}\right)^2-2\La+R_{d-p-1} e^{-\frac{2\hat{\phi}}{d-p-1}} \right].
    \label{ED2LiouvilleAction_gen}
\ee
Alternatively, one may start from the $(d+p+3)$-dimensional theory and reduce over the compact manifold $\tilde{\mathbf{X}}^{n+1}$. This leads to the same action (\ref{ED2LiouvilleAction_gen}) after  (\ref{Map1})- (\ref{general_map}) are applied.
A more detailed discussion can be found in appendix~\ref{appendix:KKred}.

Other generalizations include considering non-diagonal reductions and addition of matter fields in the higher-dimensional theory as well as higher-derivative gravity terms. Related work appeared in \cite{Gouteraux:2011qh,Smolic:2013gx, Gouteraux:2011ce,Charmousis:2012dw,Gouteraux:2012yr} and it would be interesting to work out these generalizations.
One case examined in \cite{Gouteraux:2011ce} can however be encompassed in the present scheme, whereby a $(p+2)$-form is included in the higher-dimensional theory, with no legs along the compact space. Upon reduction to $p+2$ dimensions, it becomes space-filling, and can be dualized to a scalar potential for the KK scalar. By the same considerations as above, this scalar potential can be exchanged against either that coming from a cosmological constant or the curvature of a compactified space.This is of interest since it would make a connection to theories containing D3, M2 or M5 branes.

\section{Non-homogeneous black branes from fluids/gravity in AdS}\label{secFluidsGravity}

In this section we present an important application of the map we have derived: the construction of geometries approximating fluctuating Ricci-flat black branes from the known geometries of the fluids/gravity correspondence \cite{Bhattacharyya:2008jc, Bhattacharyya:2008mz} (for a review, see \cite{Rangamani:2009xk,Hubeny:2011hd}).

There are various lessons we learn from this construction, the most important being the explicit metric \eqref{EFcoordsRF}. Other lessons follow from this. This geometry is a solution to Einstein's equations without a cosmological constant up to corrections with third derivatives along brane directions $x^a$, but exact otherwise. This implies, in particular, that it captures nonlinear aspects of the Gregory-Laflamme instability. Much of the physics of this spacetime is encoded in its `holographic' ADM stress tensor \eqref{Tflat}, in the same spirit as in AdS/CFT. This stress tensor is precise to second order in derivatives, and we will extract from it a dispersion relation $\Omega(k)$ for the unstable linearized modes $e^{\Omega t+ikx}$ precise up $k^4$ corrections in the small $k$ limit. This agrees remarkably with numerical data, especially for a large number of dimensions, as we will comment upon.

We also learn that a hidden conformal symmetry governs the structure of this family of solutions to the vacuum Einstein equations. This conformal symmetry has an obvious origin in the AdS side, and when correctly exploited leads to important simplifications, and for instance allows to write very compactly the geometries of the fluids-gravity correspondence \cite{Bhattacharyya:2008jc, Bhattacharyya:2008mz}. Our construction of the spacetimes \eqref{EFcoordsRF}, via action of the map on the fluids/gravity spacetimes trivially pushes forward these structures in AdS to flat space. One of the consequences of this symmetry is a reduction on the number of transport coefficients at each order in the derivative expansion. For instance, in the AdS side the boundary conformal symmetry sets the bulk viscosity of the dual fluid to zero. In the Ricci-flat case the bulk viscosity is not zero, but it turns out to be determined in terms of the shear viscosity. This is a general pattern: the inherited symmetry constrains the contributions to the stress tensor in the same way it does in AdS, and this restricts the number of independent transport coefficients to be determined to the same number as in the conformal case. In some sense the ADM stress tensor of Ricci-flat black branes is non-conformal in the simplest possible way.

\subsection{Fluids/gravity in AdS \label{section:FluidsGravityMetrics}}
The fluids/gravity correspondence contains explicit families of spacetimes that solve Einstein's equations with a negative cosmological constant to certain order in derivatives in the boundary directions. To zeroth order in derivatives, these spacetimes are the planar black holes in AdS \eqref{AdSBB}, and are characterized by a homogeneous velocity $u_\mu$ and temperature $T=d/(4\pi b)$. To first order, first derivatives of these fields appear. It is convenient to group these derivatives into objects transforming naturally under conformal transformations:
\bea
{\mc A}_\mu&\equiv&u^\lambda\partial_\lambda u_\mu-\frac{\theta}{d-1}u_\nu\,,\\
\sigma_{\mu\nu}&\equiv&P_\mu{}^\alpha P_\nu^\beta\partial_{(\alpha}u_{\beta)}-\frac{\theta}{d-1}P_{\mu\nu}\,,\\
\omega_{\mu\nu}&\equiv&P_\mu{}^\alpha P_\nu^\beta\partial_{[\alpha}u_{\beta]}\,,
\eea
where we have specialized to minkowskian coordinates in the boundary, which we will take to be flat.\footnote{This has to be relaxed to fully exploit the conformal symmetry of the system, as conformal transformations $g_{\mu\nu}\rightarrow e^{2\Phi(x)} g_{\mu\nu}$ change the curvature.} As usual, the expansion of the fluid is $\theta\equiv\partial_\lambda u^\lambda$ and $P_{\mu\nu}\equiv\eta_{\mu\nu}+u_{\mu}u_{\nu}$ projects on directions transverse to the velocity. ${\mc A}_\mu$ acts as a connection for the local conformal symmetry and both the shear $\sigma_{\mu\nu}$ and vorticity $\omega_{\mu\nu}$ transform with weights $-1$ under conformal transformations, see \cite{Bhattacharyya:2008mz, Loganayagam:2008is} for a discussion.

To second order in derivatives, the fluids/gravity metric presented in \cite{Bhattacharyya:2008mz} reads
\be \label{EFcoords}
	\ud s^2_\Lambda=-2u_\mu\, \ud x^\mu\lp \ud r+{\mc V}_\nu \ud x^\nu\rp
+{\mc G}_{\mu\nu}\ud x^\mu \ud x^\nu
\ee
where
\bea
	{\mc V}_\mu&=&r{\mc A}_\mu+\frac1{d-2}\left[\mathcal D_\la\omega^\la_{\phantom{1}\mu}-\mathcal D_\la\sigma^\la_{\phantom{1}\mu}+\frac{{\mc R}u_\mu}{2(d-1)}\right]-\frac{2L(br)}{(br)^{d-2}}P_\mu{}^\nu\mathcal D_\la\sigma^\la_{\phantom{1}\nu}\nonumber\\
	&&-\frac{u_\mu}{2(br)^{d}}\left[r^2\left(1-(br)^{d}\right)-\frac12\omega_{\al\ba}\omega^{\al\ba}-(br)^2K_2(br)\frac{\sigma_{\al\ba}\sigma^{\al\ba}}{(d-1)}\right]\label{EFmetric1}
\eea
and
\bea
	{\mc G}_{\mu\nu}&=&r^2P_{\mu\nu}-\omega_{\mu\la}\omega^{\la}_{\phantom{1}\nu}+2br^2F(br)\sigma_{\mu\nu}+2(br)^2\sigma_{\mu\la}\sigma^{\la}_{\phantom{1}\nu}\left[F^2(br)-H_1(br)\right]\nonumber\\
	&&+2(br)^2\left[H_1(br)-K_1(br)\right]\frac{\sigma_{\al\ba}\sigma^{\al\ba}}{d-1}P_{\mu\nu}+\nonumber\\
	&&+2(br)^2u^\la{\mc D}_\la\sigma_{\mu\nu}\left[H_2(br)-H_1(br)\right]+4(br)^2H_2(br)\,\omega_{(\mu|\la|}\sigma^{\la}_{\phantom{1}\nu)}\,.\label{EFmetric2}
\eea
The dependence on the boundary coordinates is left implicit; $b$ is $b(x)$ and $u_\mu$ is $u_\mu(x)$. Again, we have specialized to flat space at the boundary and  grouped second derivatives of the fluid velocity into objects transforming nicely under conformal transformations:\footnote{${\mc D}_\mu$ is the Weyl-covariant derivative, see \cite{Bhattacharyya:2008mz, Loganayagam:2008is}.}
\bea
u^\lambda{\mc D}_\lambda \sigma_{\mu\nu}&\equiv&P_\mu{}^\alpha P_\nu{}^\beta u^\lambda\partial_\lambda \sigma_{\alpha \beta}+\frac{\theta}{d-1}\sigma_{\mu\nu}\\
{\mc D}_\lambda \sigma^\lambda{}_\mu&\equiv&\left(\partial^\lambda-(d-1){\mc A}^\lambda\right) \sigma_{\mu\lambda}\\
{\mc D}_\lambda \omega^\lambda{}_\mu&\equiv&\left(\partial^\lambda-(d-3){\mc A}^\lambda\right) \omega_{\lambda\mu}\\
{\mc R}&\equiv&2(d-1)\partial_\lambda {\mc A}^\lambda-(d-2)(d-1){\mc A}^2\,.
\eea

The functions $F(br)$, $H_1(br)$, $H_2(br)$, $K_1(br)$, $K_2(br)$ and $L(br)$ are defined as in \cite{Bhattacharyya:2008mz} and compiled in Appendix~\ref{FuncsFluidsGrav}. All these functions are finite for their argument becoming $br=1$, making the metric \eqref{EFcoords} explicitly regular at the future horizon $r=1/b$. $u_\mu \ud x^\mu$ is the ingoing Eddington-Finkelstein time.

The metric \eqref{EFcoords} is dual to boundary fluid flows of the stress tensor:
\begin{equation}\label{enmomintro:eq}
\begin{split}
T_{\mu\nu} =& P\left(\eta_{\mu\nu}+d u_\mu u_\nu \right)-2\eta \sigma_{\mu\nu}-2\eta \tau_\omega \left[u^{\lambda}\mathcal{D}_{\lambda}\sigma_{\mu \nu}+\omega_{\mu}{}^{\lambda}\sigma_{\lambda \nu}+\omega_\nu{}^\lambda \sigma_{\mu\lambda} \right]\\
&+2\eta b\left[u^{\lambda}\mathcal{D}_{\lambda}\sigma_{\mu \nu}+\sigma_{\mu}{}^{\lambda}\sigma_{\lambda \nu} -\frac{\sigma_{\alpha \beta}\sigma^{\alpha \beta}}{d-1}P_{\mu \nu} \right]
\end{split}
\end{equation}
with\footnote{$H_z$ is the Harmonic number function, which is analytic in $z$ with singular points for $z$ a negative integer.}
\begin{equation}
P=\frac{1}{16\pi G_{N}b^d}\,,\quad
\eta = \frac{s}{4\pi}=\frac{1}{16\pi G_{N}b^{d-1}}\,,\quad  \tau_{\omega} =  b \int_{1}^{\infty}\frac{\xi^{d-2}-1}{\xi(\xi^{d}-1)}\ud\xi=-\frac{b}{d}H_{2/d-1} \,.
\label{AdStransp}\end{equation}

Some of the constraints amongst the Einstein equations for \eqref{EFcoords} reduce to the conservation equations for this stress tensor
\beq
\partial_\mu T^{\mu\nu}=0\,,
\label{consT}
\eeq
which have to be satisfied for \eqref{EFcoords} to be a solution. To second order in derivatives, which corresponds to neglecting the contributions beyond the shear viscosity in \eqref{enmomintro:eq}, these are the relativistic Navier-Stokes equations for the boundary fluid. Eq.~\eqref{EFcoords} then gives a gravitational dual for each solution to the relativistic Navier-Stokes equations \eqref{consT}. We have used these equations to trade derivatives of $b$ in \eqref{EFcoords} by derivatives of $u_\mu$.

Finally, when the metric settles down at late times to a uniform brane configuration, it is possible to reconstruct the event horizon of the spacetime by tracing back the known event horizon of the late time limit of the solutions \cite{Bhattacharyya:2008xc}. The uniform black brane horizon position receives second-order corrections \cite{Bhattacharyya:2008mz},
\eq
r_H=\frac1b+b\lp h_1\si_{\al\beta}\si^{\al\beta}+ h_2\omega_{\al\beta}\omega^{\al\beta}+ h_3\mc R\rp+\ldots
\label{horizonlocation}\eeq
where the coefficients $h_1$, $h_2$ and $h_3$ are given in Appendix~\ref{FuncsFluidsGrav}. We can define on this null hypersurface an area form. Hawking's area theorem guarantees that it can never decrease (see e.g.~\cite{Hawking:1973uf}). Pulling back this form to the boundary of AdS along ingoing null geodesics, this generates an entropy current for the conformal fluid \cite{Bhattacharyya:2008xc},
\eq
J^\mu_S=s\left[u^\mu+b^2u^\mu\left(A_1\si_{\al\beta}\si^{\al\beta}+A_2\omega_{\al\beta}\omega^{\al\beta}+A_3\mathcal R\right)+b^2\left(B_1\mc D_\la\sigma^{\mu\la}+B_2\mc D_\la\omega^{\mu\la}\right)+\ldots\right]
\label{entropycurrent}\eeq
with $s$ the entropy density \eqref{AdStransp}, and constants $A_1$, $A_2$, $A_3$, $B_1$ and $B_2$ given in Appendix~\ref{FuncsFluidsGrav}.
While the current provided by this construction is not unique \cite{Bhattacharyya:2008xc,Booth:2010kr,Booth:2011qy}, it satisfies the second law of thermodynamics, as the bulk gravitational dynamics ensures that the corresponding entropy production rate is always positive,\footnote{This is expressed in Weyl-covariant fashion as $\mc D_\mu J_S^\mu\geq0$, the conformal weight of the entropy current being $d$.} $\p_\mu J_S^\mu\geq0$.

\subsection{Ricci-flat metric and stress tensor \label{section:RicciHydro}}
To obtain a Ricci flat metric from \eqref{EFcoords}, we need to consider configurations with a $T^{d-p-1}$ symmetry, for which $u^\mu=(-\tilde{u}^a,\vec{0})$, where $a=0,1,\dots,p$. The minus sign is necessary for the Ricci-flat solution to be in ingoing Eddington-Finkelstein coordinates (as opposed to outgoing). The result follows straightforwardly by application of the map. A first step is the identification of the warp factor of \eqref{KKDiagonal_intro}. It is
\beq
e^{2\hat{\phi}(r,x;d)/(d-p-1)}={\mc G}_{yy}\,,
\eeq
which is the component of ${\mc G}_{\mu\nu}$ in \eqref{EFmetric2} along the homogeneous directions $\vec{y}$.

We wish to write this metric in terms of natural $(p+1)$-dimensional quantities. That is, we would like to write the derivatives of the velocities appearing in the quantities \eqref{EFmetric1} and \eqref{EFmetric2} as natural derivatives of $\tilde{u}^a$ rather than $u^\mu$. There is more to this than just the mentioned sign. For example, the shear tensor of the reduced $(p+1)$-dimensional space, which is traceless in $p+1$ dimensions is
\beq
\tilde{\sigma}_{ab}\equiv \tilde{P}_a{}^c \tilde{P}_b{}^d\partial_{(c}\tilde{u}_{d)}-\frac{\tilde{\theta}}{p}\tilde{P}_{ab}\,.
\eeq
This shear tensor is related to the $d$-dimensional one by
\beq
\tilde{\sigma}_{ab}=-\delta_a{}^\mu \delta_b{}^\nu \sigma_{\mu\nu}+\left(\frac{1}{p}+\frac{1}{n+1}\right)\theta P_{ab}\,,
\label{relationshears}\eeq
where we have applied the change $d\rightarrow -n$, as we will in the remainder of this section. Eq.~\eqref{relationshears} implies
\beq
\sigma_{\mu\nu}\sigma^{\mu\nu}=\tilde{\sigma}_{ab}\tilde{\sigma}^{ab}+\left(\frac{1}{p}+\frac{1}{n+1}\right)\,\tilde{\theta}^2\,,
\label{eqshear2}\eeq
where we used $\theta=-\tilde{\theta}$. Note, also, that $\omega_{\mu\nu}=-\delta_\mu{}^a\delta_\nu{}^b\tilde{\omega}_{ab}$, and $\omega_{\alpha\beta}\omega^{\alpha\beta}=\tilde{\omega}_{ab}\tilde{\omega}^{ab}$.

Using these identities we find that the objects involved in ${\mc G}_{yy}$ are
\beq
u^\lambda{\mc D}_\lambda \sigma_{yy}= \frac{1}{n+1}\tilde{u}^c\partial_c \tilde{\theta}-\frac{\tilde{\theta}^2}{(n+1)^2}
\eeq
and
\beq
\omega_{y\mu}=0\,,\qquad \sigma_{y\lambda}\sigma^\lambda{}_{y}=\frac{\tilde{\theta}^2}{(n+1)^2}\,.
\eeq

Collecting these results and implementing
\beq
r\rightarrow 1/r\,,\quad\quad b=r_0\,,
\eeq
${\mc G}_{yy}$ in terms of natural $(p+1)$-dimensional fluid quantities reads
\beq\begin{split}
	A={\mc G}_{yy}=&\frac{1}{r^2}\Bigg(1-2F(r_0/r)\frac{r_0\,\tilde{\theta}}{n+1}+2\frac{r_0^2\,\tilde{\theta}^2}{(n+1)^2}\left[F^2(r_0/r)-H_2(r_0/r)\right]\\
	&-2\left[H_1(r_0/r)-K_1(r_0/r)\right]\frac{r_0^2\,\sigma_{\mu\nu}\sigma^{\mu\nu}}{n+1}+2\frac{r_0^2\,\tilde{u}^c \partial_c\tilde{\theta}}{n+1}\left[H_2(r_0/r)-H_1(r_0/r)\right]\Bigg)\,,
\label{Gyy}\end{split}\end{equation}
where for reasons of space we have kept implicit the term $\sigma_{\mu\nu}\sigma^{\mu\nu}$, to be substituted according to eq.~\eqref{eqshear2}.

The Ricci-flat metric is
\beq \label{EFcoordsRF}
	\ud s^2_0=\frac{1}{{\mc G}_{yy}}\left(2\tilde{u}_a\, \ud x^a\lp -\frac{\ud r}{r^2}+{\mc V}_b \ud x^b\rp
+{\mc G}_{ab}\ud x^a \ud x^b+\ud\Omega_{n+1}^2\right)
\end{equation}
with
\bea
	{\mc V}_a&=&\frac{1}{r}{\mc A}_a-\frac1{n+2}\left[\mathcal D_\la\omega^\la_{\phantom{1}a}-\mathcal D_\la\sigma^\la_{\phantom{1}a}+\frac{{\mc R}\tilde{u}_a}{2(n+1)}\right]-\frac{2L(r_0/r)}{(r/r_0)^{n+2}}\tilde{P}_a{}^c\mathcal D_\la\sigma^\la_{\phantom{1}c}\nonumber\\
	&&+\frac{\tilde{u}_a}{2(r/r_0)^{n}}\left[\frac{1}{r^2}\left(1-(r/r_0)^{n}\right)-\frac12\omega_{\al\ba}\omega^{\al\ba}+(r_0/r)^2K_2(r_0/r)\frac{\sigma_{\al\ba}\sigma^{\al\ba}}{(n+1)}\right]\label{EFmetric1RF}
\eea
and
\bea
	{\mc G}_{ab}&=&\frac{1}{r^2}\tilde{P}_{ab}-\omega_{a\la}\omega^{\la}_{\phantom{1}b}+2\frac{r_0}{r^2}F(r_0/r)\sigma_{ab}+2(r_0/r)^2\sigma_{a\la}\sigma^{\la}_{\phantom{1}b}\left[F^2(r_0/r)-H_1(r_0/r)\right]\nonumber\\
	&&-2(r_0/r)^2\left[H_1(r_0/r)-K_1(r_0/r)\right]\frac{\sigma_{\al\ba}\sigma^{\al\ba}}{n+1}\tilde{P}_{ab}+\nonumber\\
	&&+2(r_0/r)^2u^\la{\mc D}_\la\sigma_{ab}\left[H_2(r_0/r)-H_1(r_0/r)\right]+4(r_0/r)^2H_2(r_0/r)\,\omega_{(a|\la|}\sigma^{\la}_{\phantom{1}b)}\label{EFmetric2RF}
\eea
with $d\rightarrow-n$ substituted in the functions in Appendix~\ref{FuncsFluidsGrav}.

We wish to write the tensorial objects in  ${\mc G}_{ab}$ and ${\mc V}_b$, in terms of natural $(p+1)$-dimensional quantities. The result follows straightforwardly after the substitutions
\beq
\sigma_{ab}=-\tilde{\sigma}_{ab}-\frac{n+p+1}{p(n+1)}\tilde{\theta}\tilde{P}_{ab}\,,\quad\sigma_{ay}=0\,,\quad \omega_{ab}=-\tilde{\omega}_{ab}\,,\quad {\mc A}_{a}=\tilde{u}^c\partial_c\tilde{u}_a+\frac{1}{n+1}\tilde{\theta}\,\tilde{u}_a\,.
\label{subs1}\eeq
These imply
\beq
\sigma_{a\lambda}\sigma^\lambda{}_b=\tilde{\sigma}_{ac}\tilde{\sigma}^c{}_b+2\left(\frac{1}{p}+\frac{1}{n+1}\right)\tilde{\theta}\tilde{\sigma}_{ab}+\left(\frac{1}{p}+\frac{1}{n+1}\right)^2\tilde{\theta}^2\tilde{P}_{ab}
\label{subs2}\eeq
and
\beq
\omega_{a\lambda}\omega^\lambda{}_b=\tilde{\omega}_{ac}\tilde{\omega}^c{}_{b}\,,\qquad\omega_{(a|\lambda|}\sigma^\lambda{}_{b)}= \tilde{\omega}_{(a|c|}\tilde{\sigma}^c{}_{b)}\,.
\label{subs3}\eeq
The remaining relevant objects for \eqref{EFmetric1RF} and \eqref{EFmetric2RF} are
\bea
u^\lambda{\mc D}_\lambda \sigma_{ab}&=&\tilde P_a{}^c \tilde P_b{}^d \tilde u^e\partial_e \left[\tilde{\sigma}_{cd}+\frac{n+p+1}{p(n+1)}\tilde{\theta}\tilde{P}_{cd}\right]-\frac{\tilde{\theta}}{n+1}\left[\tilde{\sigma}_{ab}+\frac{n+p+1}{p(n+1)}\tilde{\theta}\tilde{P}_{ab}\right]\\
{\mc D}_\lambda \sigma^\lambda{}_a&=&-\left(\partial^d+(n+1){\mc A}^d\right) \left[\tilde{\sigma}_{ad}+\left(\frac{1}{p}+\frac{1}{n+1}\right)\tilde{\theta}\tilde{P}_{ad}\right]\\
{\mc D}_\lambda \omega^\lambda{}_a&=&\left(\partial^c+(n+3){\mc A}^c\right) \tilde{\omega}_{ac}\\
{\mc R}&=&-2(n+1)\partial_a {\mc A}^a-(n+2)(n+1){\mc A}_b{\mc A}^b\,.
\label{subs4}\eea

To summarize, \eqref{EFcoordsRF} with \eqref{EFmetric1RF}, \eqref{EFmetric2RF} and \eqref{Gyy},  and \eqref{subs1}-\eqref{subs4} substituted and the functions as in the appendix is Ricci flat. It is the metric of a slowly fluctuating black $p$-brane with a regular future event horizon.\footnote{These are also called {\em blackfolds} \cite{Emparan:2009cs,Emparan:2009at,Camps:2012hw}.} This metric is one of the main results of this section.\footnote{The result to first order in derivatives was constructed in \cite{Camps:2010br}, that we recover upon discarding second derivatives.}

As we have discussed in sec.~\ref{section:HoloDicRicciFlat}, the stress tensor of the Ricci-flat side follows from the one in the AdS side upon the analytic continuation $d\rightarrow -n$ and $T_{ab}\rightarrow - \tilde{T}_{ab}$ and $u_a\rightarrow- \tilde{u}_a$, and it is conserved as a consequence of constraints in the Einstein equations for \eqref{EFcoordsRF}. The result is
\begin{equation}
\begin{split}
\tilde{T}_{ab} =& \tilde{P}\left(\eta_{ab}-n \tilde{u}_a \tilde{u}_b \right)+2\tilde{\eta} \sigma_{ab}+2\tilde{\eta} \tilde{\tau}_\omega \left[u^{\lambda}\mathcal{D}_{\lambda}\sigma_{ab}+2\omega_{(a|\lambda|}\sigma^\lambda{}_{b)} \right]\\
&-2\tilde{\eta} r_0\left[u^{\lambda}\mathcal{D}_{\lambda}\sigma_{ab}+\sigma_{a}{}^{\lambda}\sigma_{\lambda b} -\frac{\sigma_{\alpha \beta}\sigma^{\alpha \beta}}{d-1}\tilde{P}_{ab} \right]
\label{Tflatimp}\end{split}
\end{equation}
with\footnote{The AdS and Ricci-flat Newton constants are related by the volumes of the torus and sphere on which they are respectively compactified, $G_N=\tilde G_N/\Omega_{(n+1)}$ (see Section~\ref{section:map}).}
\begin{equation}
\tilde{P}=-\frac{\Omega_{(n+1)}r_0^{n}}{16\pi \tilde{G}_{N}},\qquad
\tilde{\eta} = \frac{\tilde{s}}{4\pi}=\frac{\Omega_{(n+1)}r_0^{n+1}}{16\pi \tilde{G}_{N}}\qquad \text{and} \qquad \tilde{\tau}_{\omega} =  \frac{r_0}{n}H_{-2/n-1} \,.
\label{pBraneTransp}\end{equation}
Note that with our conventions the pressure is negative, while the energy density and viscosities are positive.
$\tilde{\tau}_\omega$ diverges for $n=1,2$ and has to be renormalized. We will deal with this in sec.~\ref{subsection:n=1,2}.

Using the identities \eqref{subs1}-\eqref{subs4} the stress tensor \eqref{Tflatimp} can be written more explicitly as:
\begin{equation} \label{Tflat}
\begin{split}
\tilde{T}_{ab} =&\, \tilde{\varepsilon} \tilde{u}_a \tilde{u}_b+\tilde{P}\tilde{P}_{ab}-2\tilde\eta \tilde\sigma_{ab}-\tilde{\zeta}\tilde{\theta}\tilde{P}_{ab}\\
&+2\tilde{\eta} \tilde{\tau}_\omega \left[\tilde{P}_a{}^c\tilde{P}_b{}^d\tilde{u}^{e}\partial_{e}\tilde{\sigma}_{cd} - \frac{\tilde{\theta}\tilde{\sigma}_{ab}}{n+1}+2\tilde{\omega}_{(a}{}^{c}\tilde{\sigma}_{b)c }\right]+\tilde{\zeta}\tilde{\tau}_\omega\,\left[\tilde{u}^c\partial_c\tilde{\theta}-\frac{1}{n+1}\tilde{\theta}^2\right]\tilde{P}_{ab}\\
&-2\tilde{\eta} r_0\left[\tilde{P}_a{}^c\tilde{P}_b{}^d\tilde{u}^{\epsilon}\partial_{\epsilon}\tilde{\sigma}_{cd}+\left(\frac{2}{p}+\frac{1}{n+1}\right)\tilde{\theta}\tilde{\sigma}_{ab}+\tilde{\sigma}_{a}{}^{c}\tilde{\sigma}_{cb} +\frac{\tilde{\sigma}^2}{n+1}\tilde P_{ab} \right]\\
&-\tilde{\zeta} r_0\left[\tilde{u}^c\partial_c\tilde{\theta}+\left(\frac{1}{p}+\frac{1}{n+1}\right)\tilde{\theta}^2\right]\tilde{P}_{ab}\\
\end{split}
\end{equation}
with
\beq
\tilde{\varepsilon}=-(n+1)\tilde{P}\,,\qquad\tilde{\zeta}=2\tilde{\eta}\left(\frac{1}{p}+\frac{1}{n+1}\right)\,.
\label{BFprops}\eeq
The second of these relations is a direct consequence of the structure in \eqref{relationshears}. Both of them follow directly from the fact that the ADM stress tensor of black branes can be written as the analytical continuation of the stress tensor of a conformal field theory, as we have discussed. This pattern has been explained for Einstein-Dilaton systems in \cite{Kanitscheider:2009as}. Here we have generalized it to Ricci-flat black branes.

\subsection{Entropy current \label{section:EntropyCurrent}}

The AdS to Ricci-flat map that we used to derive the Ricci-flat metric \eqref{EFcoordsRF} can be extended to the location of the horizon \eqref{horizonlocation} of the AdS brane and to the entropy current \eqref{entropycurrent} of the conformal fluid. This gives us the position $r_H$ of the horizon of the black $p$-brane,
\eq
r_H=r_0\left[1-r_0^2\lp h_1\si_{\al\beta}\si^{\al\beta}+ h_2\omega_{\al\beta}\omega^{\al\beta}+ h_3\mc R\rp+\ldots\right],
\label{bfhorizonlocation}\eeq
and the expression for an entropy current for the blackfold fluid,
\eq
\tilde J^a_S=\tilde s\left[
\tu^a+r_0^2\tu^a\left(A_1\si_{\al\beta}\si^{\al\beta}+A_2\omega_{\al\beta}\omega^{\al\beta}+A_3\mathcal R\right)-\left(B_1\mc D_\la\sigma^{a\la}+B_2\mc D_\la\omega^{a\la}\right)+\ldots
\right],
\label{bfentropycurrent}\eeq
with $\tilde s=4\tilde G_{N}/\lp\Omega_{(n+1)}r_0^{n+1}\rp$ the entropy density \eqref{pBraneTransp} of the brane.
In these expressions, $\sigma^2$, $\omega^2$, $\mc R$, $\mc D_\la\sigma^{a\la}$ and $\mc D_\la\omega^{a\la}$ are given (for compactness) by equations \eqref{subs1}-\eqref{subs4} in terms of the blackfold fluid data $(r_0,\tu^a)$, and we performed the substitution $d\rightarrow-n$ into the constants of Appendix~\ref{FuncsFluidsGrav}. Notice that after applying the map on $J^\mu_S$, we still need to flip its overall sign to obtain \eqref{bfentropycurrent}, for the same reason that we needed to flip the sign of $u^\mu$. The resulting entropy current is future-directed, and implements the second law of thermodynamics for blackfolds, $\p_a\tilde J^a_S\geq0$.

\subsection{The Gregory-Laflamme instability from AdS \label{section:GL}}
The metrics \eqref{EFcoordsRF} are slowly-varying Ricci-flat black branes, and these are known to suffer from the Gregory-Laflamme instability \cite{Gregory:1993vy,Gregory:1994bj} (see \cite{Gregory:2011kh,Lehner:2011wc} for a review). In the derivative expansion, this instability is captured  by an imaginary speed of sound \cite{Emparan:2009at}, which is a direct consequence of the equation of state in \eqref{BFprops}:
\beq
\tilde{c}_s^2=\frac{d\tilde{P}}{d\tilde{\varepsilon}}=-\frac{1}{n+1}\,.
\eeq
These unstable sound modes fluctuate the pressure of the fluid, which is the thickness of the black string, $r_0$, as in the Gregory-Laflamme instability. The dispersion relation is a property of linearized fluctuations
\beq \label{HydroFluct}
r_0= \langle r_0\rangle+ \delta r_0\, e^{\Omega t+ik x}\,,\quad\quad \tilde{u}_a=\langle \tilde{u}_a\rangle+ \delta \tilde{u}_a e^{\Omega t+ik x}\,.
\eeq

The equations of motion for these linearized fluctuations  follow from the conservation of the linearization of \eqref{Tflat}, which amounts to discarding terms with products of derivatives:
\begin{equation} \label{TflatLin}
\begin{split}
\tilde{T}^\textrm{G-L}_{ab} =& \tilde{P}\left(\eta_{ab}-n\tilde{u}_a \tilde{u}_b \right)-2\tilde\eta \tilde\sigma_{ab}-\tilde{\zeta}\tilde{\theta}\tilde{P}_{ab}\\
&+2\tilde{\eta} (\tilde{\tau}_\omega - r_0) \tilde{P}_a{}^c\tilde{P}_b{}^d\tilde{u}^{e}\partial_{e}\tilde{\sigma}_{cd} +\tilde{\zeta}(\tilde{\tau}_\omega - r_0)\,\tilde{P}_{ab}\tilde{u}^c\partial_c\tilde{\theta}\\
\end{split}
\end{equation}
The dispersion relation for the sound mode that follows from conservation of \eqref{TflatLin} is
\beq
\Omega=\frac{1}{\sqrt{n+1}}k-\frac{2+n}{n(1+n)}r_0k^2+\frac{(2+n)[2+n(2\tilde{\tau}_\omega/r_0-1)]}{2n^2(1+n)^{3/2}}r_0^2k^3+O(k^4)\,,
\label{disprel}\eeq
where we have taken $\langle u_a\rangle=(-1,0,\dots,0)$ and substituted $\langle r_0 \rangle=r_0$.

Strictly speaking, the dispersion relation corresponding to the spacetimes that follow from  \eqref{EFcoordsRF} is the truncation of the above to order $k^2$. This corresponds to the Navier-Stokes approximation to the conservation equations, or keeping only the first line in \eqref{TflatLin}. Eq.~\eqref{disprel} is the dispersion relation for the spacetimes to third order in derivatives, that are one order higher in derivatives with respect to the ones we have constructed.

\FIGURE{\includegraphics[width=.7\textwidth]{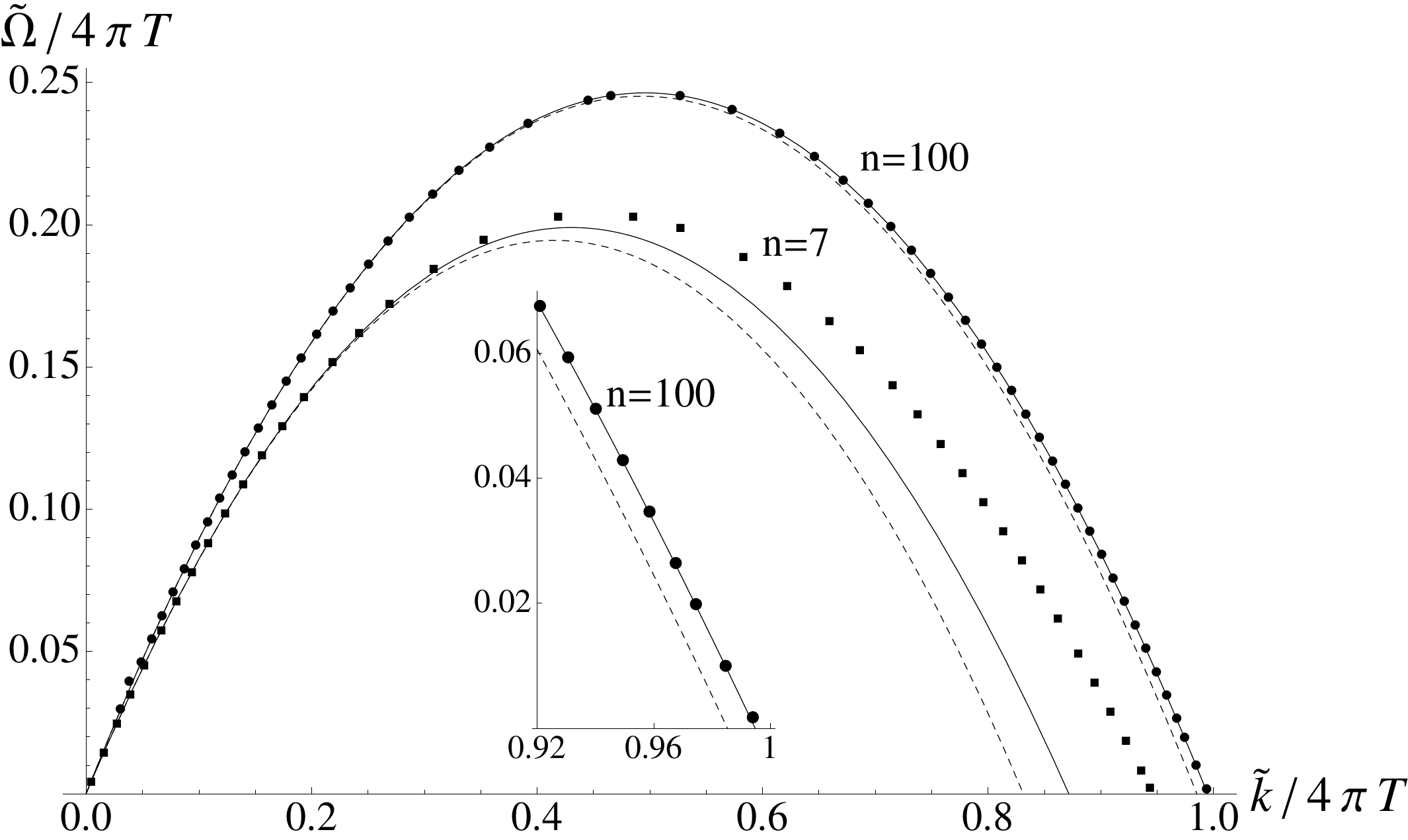}
	\label{Fig1}
	\caption{\sf{\sc Gregory-Laflamme dispersion relation, for two values of $n$: $n=7$ and $n=100$}. The small $k$ region is where the hydrodynamic expansion is valid strictly speaking, but for large $n$ the threshold mode $k_c$ is also well-described. In the insert, we show that at large $n$ the agreement between the hydrodynamic approximation and the numerics becomes excellent.}}

Figure \ref{Fig1} compares the prediction \eqref{disprel} to numerics. The analytic approximation turns out to be better in the limit of large number of transverse dimensions. To take this limit it is convenient to redefine
\be
	\tilde\Omega=n\, \Omega\,,\qquad \tilde k=\sqrt{n}\,k\,,
\ee
and to use the temperature $T=n/(4\pi\, r_0)$ as dimensionful parameter. One checks that at large $n$ \eqref{disprel} becomes
\beq
\tilde{\Omega}=\tilde{k}\left(1-\frac{\tilde{k}}{4\pi T}\right)-\frac{\tilde{k}}{n}\left(\frac{1}{2}+\frac{\tilde{k}}{4\pi T}-\frac{\tilde{k}^2}{(4\pi T)^2}
\right)+O\left(\frac{1}{n^2}\right)\,,
\label{GLdisprel}\eeq
up to corrections in powers of $\tilde{k}$ higher than 3.

Recently ref.~\cite{Emparan:2013moa} computed the Gregroy-Laflamme dispersion relation in the limit of many transverse dimensions to order $1/n^2$, without truncations in the number of derivatives. That result agrees with \eqref{GLdisprel}. This implies that the derivative expansion becomes exact in the limit of infinite dimensions for the linearized Gregory-Laflamme instability. Our metrics \eqref{EFcoordsRF} are valid in the nonlinear regime, so we expect them to fully describe the instability (linearly and nonlinearly) at large $n$.

\subsection{The $n=1,2$ cases\label{subsection:n=1,2}}

The Ricci-flat metric \eqref{EFcoordsRF} is finite, except if $d=-2$ or $d=-1$. In these cases, we find a divergence. This divergence can be regularized. After regularization, the boundary metric $\hg_{(0)}$ will no longer be the Minkowski metric but will receive a number of second-derivative corrections, and the stress-tensor (the $\hg_{(d/2)}$ term of the FG expansion) will no longer be in the Landau frame.

The reason for these divergences are obscured by the choice of Eddington-Finkelstein coordinates, so let us go back to Fefferman-Graham gauge. The details of the change of gauge are in Appendix~\ref{app:EFtoFG}. Recall that on the AdS side the asymptotic expansion at order $\rho^{(d+2 l)/2}$ is proportional to the $2l$ derivatives of the stress-tensor plus nonlinear terms. After applying the AdS/Ricci-flat map the overall radial power becomes $r^{n-2l}$.
At a given $n$, for $l>l_c=[(n+1)/2]$,\footnote{Here $[{\phantom{1}}]$ means the integer part.} the power becomes negative and this term is no longer decaying at large $r$. Moreover, logarithms will be generated when analytically continuing $r^{n-2l}$ as $l\to l_c$.

Since we are working to second order in derivatives, this is an issue only when $n=1,2$ (in which case $l_c=1$).
We will mainly focus on the $n=2$ case: the discussion for $n=1$ goes along similar lines.

\subsubsection{The $d\rightarrow-2$ case}

As mentioned previously, in FG coordinates, Asymptotically AdS metrics read as in \eqref{FGcoords}, where the terms subleading to the stress-tensor can be expressed in terms of the stress-tensor and its derivatives. Here, we reproduce the first line of \eqref{FGbeyondST}, which shows the expansion to order $\rho^d$:
\beq
\mfg_{\mu\nu}=\eta_{\mu\nu}+\frac{\rho^{d/2}}{d}T_{\mu\nu}-\frac{\rho^{d/2+1}}{2d(d+2)}\Box T_{\mu\nu}+\frac{\rho^d}{2d^2}\left(T_\mu{}^\sigma T_{\sigma\nu}-\frac{1}{4(d-1)}T_{\sigma\kappa}T^{\sigma\kappa}\eta_{\mu\nu}\right).
\eeq
Notice now that the term at order $\rho^{d/2+1}$ is expanded  as $d\rightarrow -2$  like
\beq
\rho^{d/2+1}g_{(d/2+1)}=\left[\left(\frac{1}{4(d+2)}+\frac{1}{8}(1+\log\rho)\right)+\dots\right]\Box T_{\mu\nu}.
\eeq
The order $\rho^0$ that this term produces contributes to the metric in the boundary. Because this contribution is a constant at the order in derivatives we are working, it maintains the flatness in the boundary, but it is singular. This singularity is only a coordinate singularity at the boundary that can be dealt by changing coordinates as
\beq
x^\mu\rightarrow x^\mu-\frac{1}{2}\left(\frac{1}{4(d+2)}+\frac{1}{8}\right)x^\nu \Box T^\mu{}_\nu
\eeq
which brings the boundary metric to its minkowskian form $\eta_{\mu\nu}$.

This, as any coordinate change, modifies the stress tensor
\beq
T_{\mu\nu}\rightarrow T_{\mu\nu}-\left(\frac{1}{4(d+2)}+\frac{1}{8}\right)T^\rho{}_{(\mu}\Box T_{\nu)\rho}\,.
\eeq
There is, however, yet another contribution to the stress tensor, coming from the $g_{(d+1)}$ term. So, the full stress tensor will be
\beq
\tilde{T}^{\textrm{2nd}}_{\mu\nu}=T_{\mu\nu}- \left(\frac{1}{4(d+2)}+\frac{1}{8}\right)T^\rho{}_{(\mu}\Box T_{\nu)\rho}+d\, g_{(d+1)}\,.
\eeq

The limit of this expression as $d\rightarrow -2$ is not yet finite and the stress tensor is not in Landau frame. Effectively, one changes frame by:
\beq
b^{-d}\rightarrow b^{-d}-\frac{1}{d-1} u^\mu u^\nu \tilde{T}^{\textrm{2nd}}_{\mu\nu}
\eeq
and setting to zero the viscous corrections with mixed components (that is, with one leg proportional to $u_\mu$ and the other orthogonal to it). The practical effect of this redefinition of the temperature is a contribution to the dissipative stress tensor proportional to $P_{\mu\nu}$. This comes from the redefined pressure. So the final limit that should give a nice stress tensor is:
\beq\label{RenSTd=-2}
\left.T^{\textrm{2nd}}_{\mu\nu}\right|_{d=-2}=\lim_{d\rightarrow-2}\left(T_{\mu\nu}- \left(\frac{1}{4(d+2)}+\frac{1}{8}\right)T^\rho{}_{(\mu}\Box T_{\nu)\rho}+dg_{d+1}-\frac{1}{d-1} u^\rho u^\sigma \tilde{T}^{\textrm{2nd}}_{\rho\sigma}P_{\mu\nu}\right).
\eeq
This limit is finite and we find that the second order dissipative stress tensor reads:
\beq
\begin{split}
&T^{\textrm{2nd}}_{\mu\nu}=b^4\left[\left(\frac{\theta^2}{9}-\frac{\dot{\theta}}{3}-\frac{a^2}{4}\right)P_{\mu\nu}+\frac{a_\mu a_\nu}{4}+ \frac{7}{4}u^\lambda\mathcal{D}_\lambda\sigma_{\mu\nu}+\frac{2}{3}\theta\sigma_{\mu\nu}\right.\\
&\qquad\qquad\qquad\qquad\qquad\qquad\left.+\frac{5}{4}\sigma_\mu{}^\lambda\sigma_{\lambda\nu}+\frac{1}{4}\omega_\mu{}^\lambda\omega_{\lambda\nu}+\sigma^\lambda{}_{(\mu}\omega_{\nu)\lambda}\right].
\end{split}
\eeq
Note that this has a non-zero trace
\beq
\eta^{\mu\nu}T^{\textrm{2nd}}_{\mu\nu}=\frac{b^4}{4}\left[4a^2-\frac{4}{3}\theta^2+5\sigma^2-\omega^2+4\dot{\theta}\right]=\frac{1}{16}T^{\rho\sigma}\Box T_{\rho\sigma}\,.
\eeq
Converting to tilded quantities (which includes a change of sign between $\tilde u_a$ and $\hat u_a$, as well as between $\tilde T_{ab}$ and $\hat T_{ab}$), the renormalized, reduced stress-tensor for $n=2$ is:
\bea\label{RenSTn=2}
\left.\tT_{ab}\right|_{d=-2}&=&\tilde P\left(\eta_{ab}-2\tu_a\tu_b\right)-2\tilde\eta\tilde\sigma_{ab}-\tilde\zeta\tilde\theta\tilde P_{ab}\nonumber\\	
	&&+ r_0\tilde \eta\left[\tilde \sigma _{ c (a}\tilde \omega _{b) }{}^{c}+\frac{5}{4}\tilde \sigma _{a c }\tilde \sigma _{b }{}^{c }+\frac{1}{4}\tilde\omega _{ac }\tilde \omega ^{c }{}_{b }+\frac{7}{4}\tilde u^{c }\partial_{c }\tilde \sigma _{ab }+\frac{1}{12}\tilde\theta\tilde\sigma _{ab }+\frac{1}{4}\tilde a_{a }\tilde a_{b }\right.\nn\\
	&&\left.-\frac72\tilde a^c\tilde u_{(a}\tilde \sigma_{b)c}+\tilde P_{ab}\left(\frac{\tilde \theta^2}{9}-\frac{\dot{\tilde \theta}}{3}-\frac{\tilde a^2}{4}\right)\right]\nn\\
	&&+ r_0\tilde \zeta\left[\frac{5}{4}\tilde \theta\tilde \sigma _{ab}+\tilde P_{ab}\left(\frac{7\dot{\tilde \theta}}{8}-\frac{\tilde \theta^2}{3}+\frac{5}{8p}\tilde \theta^2\right)\right],\label{RenST-2}\\
	&&\tilde P =\frac{- r_0^2}{16\pi\tilde G_N}\,,\qquad \tilde \eta =\frac{ r_0^3}{16\pi\tilde G_N}\,,\qquad \tilde \zeta =2\tilde \eta\frac{p+3}{3p}\,.
\eea
Here we obtained the renormalized expression \eqref{RenSTd=-2} by implementing the boundary diffeomorphism on the general FG metric, but we could also have done it directly on the fluids/gravity metrics. The poles as $n=2$ are removed by changing appropriately the integration bounds on the metric functions. This is done explicitly in appendix \ref{app:n=1,2} and recovers the same stress-tensor \eqref{RenSTn=2}, which serves as a nice double check of our procedure.

In app. \ref{app:FGexp-2}, we also derive the conservation equation obeyed by \eqref{RenSTn=2}, which reads:
\beq\label{RenSTn=2eom}
\partial^b \tT_{ab} + \frac{1}{8}\left(\tT^{bc}\partial_a\Box \tT_{bc}-\frac{1}{p+3}\tT^{b}\,_{b}\partial_a\Box \tT^c\,_{c}\right)
-\frac{1}{8}\tT^{bc}\partial_c\Box \tT_{ba}
+\frac{1}{8}\partial^c \tT^{ab}\Box \tT_{bc}=0\,.
\eeq
Plugging \eqref{RenSTn=2} in \eqref{RenSTn=2eom} and linearizing in amplitudes, we find the dispersion relation:
\beq\label{DispReln=2}
\Omega=\frac{1}{\sqrt{3}}k-\frac{2}{3}k^2-\frac{37}{72\sqrt{3}}k^3+O(k^4)\,.
\eeq
If we extrapolate the value of $\tau_\omega(n=2)$ from matching formula \eqref{disprel} to this we get $\tau_\omega(n=2)=-37/48$.

\subsubsection{The $d\rightarrow-1$ case}

Again, we have to deal with the contribution to the boundary metric, of order $\rho^0$. This contribution maintains the flatness of the boundary, but is diverging. This can be taken care of by performing the coordinate transformation
\eq
x^\mu\rightarrow x^\mu -\Lambda^\mu{}_\nu x^\nu
\eeq
with the matrix
\eq
\begin{split}
\Lambda_\munu=
&\frac14T^\al{}_{(\mu}\Box T_{\nu)\al}
+\frac1{32}\lp\frac12+\frac1{d+1}\rp\lp\p_\mu(T^{\al\beta})\p_\nu T_{\al\beta}-(T^{\al\beta}\Box T_{\al\beta})\eta_\munu\rp\\
&
+\frac14\lp1+\frac1{d+1}\rp\lp T^{\al\beta}\p_\al\p_{(\mu}T_{\nu)\beta}-T^{\al\beta}\p_\al\p_\beta T_\munu+(\p_\al T_\mu{}^\beta)\p_\beta T_\nu{}^\al-(\p_\al T_{(\mu}{}^\beta)\p_{\nu)} T^\al{}_\beta\rp
\\
&
-\frac1{32}\lp\frac72+\frac3{d+1}\rp T^{\al\beta}\p_\mu\p_\nu T_{\al\beta}
+\frac1{32}\lp\frac32+\frac1{d+1}\rp(\p^\kappa T^{\al\beta})(\p_\kappa T_{\al\beta})\eta_\munu
\end{split}
\eeq
before taking the $d\rightarrow-1$ limit. The procedure outlined in the previous section becomes extremely cumbersome due to the large number of terms. We have included some technical details in appendix \ref{app:FGexp-1}, where the trace of the renormalized stress-tensor is obtained in \eqref{TraceRenSd=-1} as well as the conservation equation it obeys in \eqref{eomdm1}. For simplicity, we have obtained the renormalized, reduced stress-tensor for $n=1$ using the procedure in appendix \ref{app:n=1,2}:
\bea\label{RenSTn=1}
\left.\tT_{ab}^{}\right|_{d=-1}&=&\tilde P\left(\eta_{ab}-\tu_a\tu_b\right)-2\tilde\eta\tilde\sigma_{ab}-\tilde\zeta\tilde\theta\tilde P_{ab}\nonumber\\			
&&+r_0\tilde \eta\left[\frac{13}{8}\tilde \sigma _{ c (a}\tilde \omega _{b) }{}^{c}+\frac{15}{16}\tilde \sigma _{a c }\tilde \sigma _{b }{}^{c }+\frac{9}{16}\tilde\omega _{ac }\tilde \omega ^{c }{}_{b }+\frac{7}{4}\tilde u^{c }\partial_{c }\tilde \sigma _{ab }+\frac{9}{16}\tilde\theta\tilde\sigma _{ab }\right.\nonumber\\
	&&\left.+\frac{9}{16}\tilde a_{a }\tilde a_{b }-\frac72\tilde a^c\tilde u_{(a}\tilde \sigma_{b)c}+\tilde P_{ab}\left(-\frac{5}{32}\tilde\omega^2-\frac{69}{32}\tilde\sigma^2+\frac{15}{16}\tilde\theta^2-\frac{15}{8}\dot{\tilde\theta}-\frac{3}{32}\tilde a^2\right)\right]\nn\\
	&&+r_0\tilde \zeta\left[\frac{15}{16}\tilde \theta\tilde \sigma _{ab}+\tilde P_{ab}\left(\frac78\dot{\tilde \theta}-\frac{15}{8}\tilde \theta^2+\frac{15(p+2)}{64p}\tilde \theta^2\right)\right],\label{RenST-1}\\
	&&\tilde P =\frac{-r_0}{16\pi\tilde G_N}\,,\qquad \tilde \eta =\frac{r_0^2}{16\pi\tilde G_N}\,,\qquad \tilde\zeta =\tilde\eta\frac{p+2}p\,.
\eea
Note that this encompasses the stress-tensor for the black string, which can be obtained from the previous expression by setting $p=1$. Its trace explicitly agrees with \eqref{TraceRenSd=-1}. Plugging \eqref{RenSTn=1} in its equation of motion \eqref{eomdm1} and linearizing in amplitudes, we obtain the dispersion relation for the Gregory-Laflamme instability:
\beq\label{DispReln=1}
\Omega= \frac{1}{\sqrt{2}}k-\frac{3}{2}k^2+\frac{75}{16\sqrt{2}}k^3+O(k^4)\,.
\eeq

\subsubsection{Discussion for the dispersion relation $n=1,2$}

For these values of $n$, we find that the dispersion relation no longer captures the correct physics beyond the small $k$ regime: there is no threshold mode, as the $k^3$ term in the dispersion relations \eqref{DispReln=2} and \eqref{DispReln=1} comes with the `wrong' sign. This might hint that the hydrodynamic expansion is an asymptotic expansion. Note that a different approach to this problem, which utilized the large $n$ limit without recourse to the hydrodynamic expansion \cite{Emparan:2013moa}, also found that the dispersion relation was not well-reproduced at finite $k$ for $n=1,2$. Nevertheless, both approaches agree in the regime of overlap (small $k$, large $n$). It is also interesting that the stress-tensor from which the Gregory-Laflamme dispersion relation was obtained is no longer conserved, but obeys equations of motion like \eqref{eomdm2} and \eqref{eomdm1}. On the right-hand side, a number of nonlinear contributions appear, which indeed are perfectly allowed on dimensional grounds in Ricci-flat spacetimes -- just as conformal anomalies appear on the right-hand side in the Asymptotically AdS case. This is also consistent with what we have seen happening at the level of the FG expansion: terms which were higher order in the FG coordinate for $n>2$ now mix with lower-order terms for $n=1,2$.

\section{AdS/Rindler correspondence \label{section:Rindler}}

\FIGURE{\includegraphics[width=.6\textwidth]{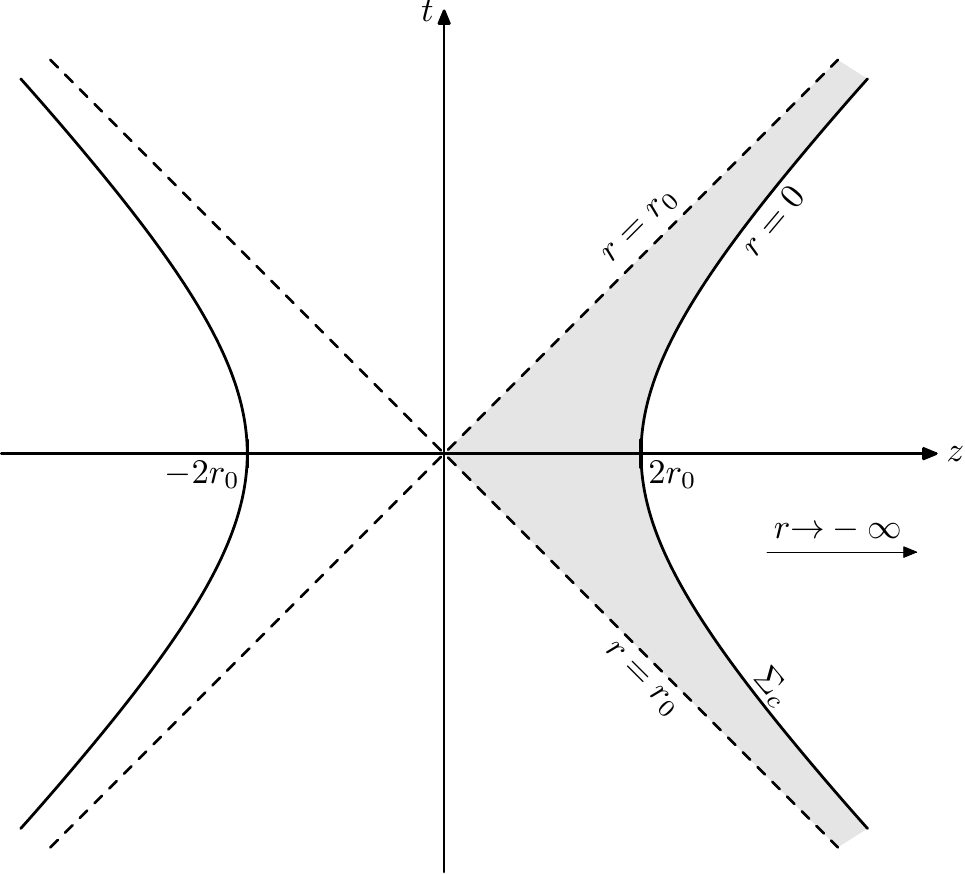}
\caption[]{\sf{\sc The Rindler wedge.} The $r<r_0$ region of the Rindler metric \eqref{RindlerEF} is mapped to the $z>|t|$ wedge of the Minkowski spacetime with metric $ds^2=-\dd t^2+\dd z^2+\dd\vec x^2$ by the change of coordinates $\tau=2r_0\log\frac{z-t}{2r_0}$, $r=r_0\lp1-\frac{z^2-t^2}{4r_0^2}\rp$, with $\tau$ defined by $\dd\tau=\tilde u_a\dd x^a$. The constant $r$ observers are uniformly accelerated and experience an acceleration horizon in $r=r_0$ (dashed lines). The cutoff surface $\Sigma_c$ (solid curve) has induced metric $\eta_{ab}$. Through the AdS/Ricci-flat map, the AdS boundary is mapped on $\Sigma_c$, and the exterior region of the AdS black brane (as well as its perturbations) is mapped to the shaded region, where $r_0\geq r>0$.}
\label{fig::Rindler}
}

When $n=-1$, the transverse sphere of the Schwarzschild $p$-brane \eqref{BlackpBrane} collapses to a point, and the metric is that of $(p+2)$-dimensional Rindler spacetime \eqref{Rindler}. Therefore, the AdS/Ricci-flat correspondence maps the AdS black branes to the Rindler wedge in this limit. Indeed, the unperturbed Schwarzschild $p$-brane metric \eqref{EFcoordsRF}, that is obtained applying the map on AdS black branes, reduces for $n\to-1$ to the flat metric
\eq
ds_0^2=\frac{r}{r_0}\tilde u_a\tilde u_b\,\dd x^a\dd x^b-2\tilde u_a\,\dd x^a\dd r+\eta_{ab}\,\dd x^a\dd x^b.
\label{RindlerEF}\eeq
This metric has an acceleration horizon in $r=r_0$ for observers on fixed $r$ orbits, and its asymptotic region lies in $r\to-\infty$. Indeed, the region $r<r_0$ -- where the velocity field $\tilde u_a$ is timelike -- covers the Rindler wedge of Minkowski spacetime with ingoing Eddington-Finkelstein-like coordinates (see fig.~\ref{fig::Rindler}). This choice of coordinates makes the metric regular on the future horizon. Finally, we will be interested in the hypersurface $\Sigma_c=\{r=0\}$, on which uniformly accelerated observers with proper acceleration $1/2r_0$ lie, and whose induced metric $\eta_{ab}$ is the Minkowski metric in one dimension less. From Eq.~\eqref{pBraneTransp}, the corresponding ADM stress tensor assumes the perfect fluid form with pressure\footnote{We set $\Omega_{0}=1$, since only one single point contributes. Also, in this section, we use the convention $16\pi\tilde G_N=1$.} $\tilde P=-1/r_0$, and reparameterizing the radial coordinate as $r=\tilde P(R-R_c)$ the Rindler metric becomes
\eq
ds_0^2=-2\tilde P\,\tilde u_a\dd x^a\dd R+\left[\eta_{ab}-\tilde P^2(R-R_c)\tilde u_a\tilde u_b\right]\dd x^a\dd x^b,
\eeq
where now the horizon is located in $R_h=R_c-1/\tilde P^2$ and $\{R=R_c\}$ is the hypersurface $\Sigma_c$ whose induced metric is $\eta_{ab}$. This is the seed metric used in \cite{Compere:2011dx,Compere:2012mt,Eling:2012ni} to study the fluid dual to Rindler spacetime. It is therefore reasonable to suppose that in the $n\rightarrow-1$ limit, the metric \eqref{EFcoordsRF} yields the Rindler metric with hydrodynamic perturbation to second order in derivatives. However, if performed naively, this procedure does not produce a finite result.

To understand the origin of the divergency, we turn to the dynamics dictated by the dual stress tensor \eqref{Tflat}. From its conservation, we obtain the fluid equations that allow to determine the expansion and the acceleration of a fluid element in terms of the collective variables $(r_0,\tilde u_a)$,\footnote{In this section we use the notation $\DD=\tilde u^a\p_a$, $\Dp_a=\tilde P_a{}^b\p_b$ and we define $\mathcal K_{ab}=\tilde P_a{}^c\tilde P_b{}^d\p_{(c}\tilde u_{d)}$. It satisfies $\p_a\tilde u_b=\mc K_{ab}+\tilde\omega_{ab}-\tilde u_a\tilde a_b\,$.}
\begin{align}
\tilde\theta&=-(n+1)\DD\log r_0+\frac{2r_0}{n}\left[{\mathcal K}^2+(n+1)(\DD\log r_0)^2\right],
\label{expansion}\\
\DD\tilde u_a&=\Dp\log r_0-\frac{2r_0}n\left[
(n+2)\DD\log r_0\,\Dp_a\log r_0+\Dp_a\DD\log r_0-\tilde P_a{}^b\p_c\mathcal K_b{}^c\right.\nn\\
&\qquad\qquad\qquad\qquad\qquad\left.-(n+1)\mathcal K_a{}^b\Dp_b\log r_0
\right].
\label{acceleration}\end{align}
Hence, when $n\rightarrow -1$, we have $\tilde\theta=0+\mathcal O(\p^2)$ and the fluid is incompressible to that order, as expected. Moreover, the equation of state \eqref{BFprops} asserts that the equilibrium energy density vanishes, $\tilde\varepsilon=0$. This property is incompatible with the Landau gauge in which the stress tensor \eqref{Tflat} is written; to obtain a finite result we need to switch gauges before performing the $n\rightarrow-1$ limit. An appropriate choice for the incompressible fluid is the isotropic gauge \cite{Compere:2011dx,Compere:2012mt}, in which the stress tensor takes the form
\eq
T_{ab}={\mc E}\,\tu_a\tu_b+\mc P\tilde P_{ab}+\tilde\Pi^\perp_{ab}\,,
\label{isotropic}\eeq
with the transverse dissipative correction $\tilde\Pi^\perp_{ab}$ satisfying $\tilde\Pi^\perp_{ab}\tu^b=0$ and containing no term proportional to $\tilde P_{ab}\,$. The pressure $\mc P$ in isotropic gauge is easily derived as the proportionality factor of $\tilde P_{ab}$ in $\tT_{ab}\,$,
\eq\begin{split}
\mc P =\tilde P\left[\vphantom{\frac2n}\right.1&-2r_0\DD\log r_0+\frac{2}{n}r_0^2\left(\vphantom{\frac21}
(n+2+H_{-\frac{2}{n}-1})(\DD\log r_0)^2\right.\\
&\qquad\qquad\qquad\qquad\qquad\left.\left.
-(n-H_{-\frac{2}{n}-1})\,\DD\DD\log r_0+\frac{n+2}{n+1}\mc K^2\right)\right],
\end{split}\eeq
where we used \eqref{expansion} and the relation $\eta=-\tilde Pr_0$ to express everything in terms of the independent variables $r_0$ and $\tu_a$. It is clear now that this corresponds to an infinite shift of the pressure $\tilde P$ because of the diverging term $\mc K^2/(n+1)$. Inverting the previous relation,
\eq\begin{split}
r_0 =\,&(-\mc P)^{\frac1n}+\frac2{n^2}(-\mc P)^{\frac2n}\DD\log\mc P+\frac2{n^4}\lp3-H_{-\frac2n-1}\rp(-\mc P)^{\frac3n}\lp\DD\log\mc P\rp^2\\
&\quad+\frac2{n^3}\lp2+n-H_{-\frac2n-1}\rp(-\mc P)^{\frac3n}\DD\DD\log\mc P-\frac2{n^2}(n+2)(-\mc P)^{\frac3n}\frac{\mc K^2}{n+1}+\mc O(\p^3)
\end{split}\label{r0P}\eeq
we can trade $r_0$ for $\mc P$, and express all quantities in terms of $\mc P$ and $\tu_a$.
The stress tensor is now in isotropic gauge, the $n\rightarrow-1$ limit is finite, and we recover the stress tensor of the fluid dual to Rindler space (in exact agreement with \cite{Compere:2012mt,Eling:2012ni}) of the form \eqref{isotropic} with energy density
\eq
\mc E=-\frac2{\mc P}\mc K^2,
\eeq
and dissipative correction
\eq\begin{split}
\tilde\Pi^\perp_{ab}=-2\mc K_{ab}
+\frac4{\mc P}\lp
\Dp_a\log\mc P\,\Dp_b\log\mc P\vphantom{\frac12}\right.
&-\tilde P_a{}^c\tilde P_b{}^d\p_c\p_d\log\mc P
-\mc K_{ab}\DD\log\mc P\\
&\left.-\frac12\mc K_a{}^c\mc K_{cb}
+\mc K_{(a}{}^c\tilde\omega_{b)c}+\tilde\omega_a{}^c\tilde\omega_{bc}
\rp.\end{split}
\eeq

It is now clear how to obtain the long wavelength corrections to the Rindler metric using the AdS/Ricci-flat correspondence. After mapping the asymptotically AdS fluid/gravity metric \eqref{EFcoords} to the Ricci-flat metric \eqref{EFcoordsRF} we need to absorb the diverging terms in the redefinition of the pressure, and then impose the gauge $g_{ra}=-\mc P\tu_a$ in order to set the corresponding Brown-York stress tensor in isotropic gauge. This is done by first rewriting the spacetime metric in terms of $(\mc P,\tu_a)$ using the fluid equations \eqref{expansion} and \eqref{acceleration}, then substituting $r_0$ with $\mc P$ using \eqref{r0P}, and finally performing a diffeomorphism to define a new radial coordinate $R$ related to $(r,x^a)$ by
\begin{align}
r(R,x^a) =  \mc P(R-R_c)&-\mc P(R-R_c)^2\left[
2\,\DD\DD\log\mc P+\lp\frac1{n+1}-\frac32+\log(\mc P(R-R_c))\rp\mc K^2
\right]\nonumber\\
&+\mc P(R-R_c)\lp n+1-2(R-R_c)\mc K^2\rp \chi(R,\mc P),
\label{Rdiffeo}\end{align}
with $\chi(R,\mc P)$ and arbitrary smooth function of its arguments.
Now the limit can be safely taken and reproduces the metric obtained in \cite{Compere:2012mt,Eling:2012ni},
\begin{align}
ds^2_0=&- 2\mc P\,\tu_a\dd x^a\dd R\nn\\
&+\left\{\vphantom{\frac{\mc P^2}2}\left[\eta_{ab}-\mc P^2 (R-R_c)\tu_a\tu_b\right]
-2\mc P(R-R_c)\tu_{(a}\lp2 \Dp_{b)}\log\mc P -\tu_{b)}\DD\log\mc P\rp\right.\nonumber\\
&
+\left[2(R-R_c)\mc K^2+\frac{\mc P^2}2(R-R_c)^2\lp\mc K^2
+2(\Dp\log\mc P)^2\rp
+\frac{\mc P^4}2(R-R_c)^3\tilde\omega^2\right]\tu_a\tu_b\nonumber\\
&+4(R-R_c)
\tu_{(a}\tilde P_{b)}{}^c\p_d\mathcal K_c{}^d+2\mc P^2(R-R_c)^2\tu_{(a}\lp\tilde P_{b)}{}^c\p_d\mathcal K_c{}^d
-(\mc K_{b)}{}^c+\tilde\omega_{b)}{}^c)\Dp_c\log\mc P\rp\nonumber\\
&+2(R-R_c)\left[
2\tilde P_a{}^c\tilde P_b{}^d\p_c\p_d\log\mc P
+2\DD\log\mc P\,\mc K_{ab}
-2\Dp_a\log\mc P\,\Dp_b\log\mc P\right.\nonumber\\
&\qquad\qquad\qquad\quad\left.\left.+\mc K_a{}^c\mc K_{cb}-2\mc K_{(a}{}^c\tilde\omega_{b)c}+2\tilde\omega_a{}^c\tilde\omega_{cb}\vphantom{\tilde P_a{}^c}\right]
-\mc P^2(R-R_c)^2\tilde\omega_a{}^c\tilde\omega_{cb}\vphantom{\frac{\mc P^2}2}\right\}\dd x^a\dd x^b.
\label{Rindlerfluidmetric}
\end{align}
Note that the resulting metric does not depend on the arbitrary function $\chi(R,\mc P)$: this function is not fixed at this order of the perturbative expansion, but would be fixed if we would extend the computation to the third order in the derivative expansion.

Interestingly, the expectation value of the dual scalar operator \eqref{non-conf_intro} in the reduced $(p+2)$-dimensional theory is given by the pressure $\mc P$ in the $n\rightarrow-1$ limit,
\eq
\hat{\mc O}_\phi=\mc P.
\eeq
Consequently, the pressure in the isotropic gauge acquires a geometrical meaning as it measures the volume variations of the torus on which we compactified the original AdS theory.

We can obtain a better understanding of the way the AdS/Ricci-flat correspondence works by comparing the boundary conditions that have been used on both sides of the duality. On the AdS side, the hydrodynamic perturbations are obtained by solving a Dirichlet problem for the metric whose value is fixed on the conformal boundary \cite{Bhattacharyya:2008mz}. When studying the hydrodynamic perturbations of Rindler spacetime, a similar Dirichlet problem is solved in the shaded region of fig.\ \ref{fig::Rindler}, for which the metric is fixed on the $\Sigma_c$ hypersurface (solid curve in fig.\ \ref{fig::Rindler}). By \eqref{Rdiffeo}, this $R=R_c$ hypersurface corresponds to $r=0$, that is in turn mapped on the asymptotic timelike boundary of AdS, where the original boundary conditions were imposed. Likewise, the regularity of the perturbations on the future event horizon of the AdS black brane is equivalent to an analogous condition on the Rindler horizon. In a nutshell, we are solving precisely the same Dirichlet problem on both sides of the correspondence.

Finally, we redefine the radial coordinate and the pressure as above, and then take the $n\rightarrow-1$ limit of the black brane horizon position $r_H$ and of its entropy current $\tilde J^a_S$, given in \eqref{bfhorizonlocation} and \eqref{bfentropycurrent} respectively. The limits are finite; however, due to the arbitrary function $\chi(R,\mc P)$ appearing in the radial reparametrization \eqref{Rdiffeo}, the coefficient of the second order term proportional to $\mc K^2$ is not fixed for the location of the perturbed Rindler horizon obtained through the map. Nevertheless, all other terms in the resulting expression,
\eq
R_H=R_c-\frac1{\mc P^2}+\frac2{\mc P^3}D\log\mc P+\frac1{\mc P^4}\lp
4DD\log\mc P-8(D\log\mc P)^2+\#\,\mc K^2-\frac{\mc\omega^2}2+(D_\perp\mc P)^2
\rp+\ldots
\label{rindlerhorizon}\eeq
are in agreement with \cite{Compere:2012mt}.
For the entropy current, the same procedure yields
\begin{align}
\mc J^a_S = & \tilde s\tu^a\lp1+\frac1{\mc P^2}\mc K^2-\frac3{2\mc P^2}\tilde\omega^2\rp\\
&+\frac{\tilde s}{\mc P^2}\lp
2D_\perp^a\log\mc PD\log\mc P-2D_\perp^aD\log\mc P-3\tilde P^{ab}\p_c\mc K^c{}_b-\lp\mc K^a{}_b+\tilde\omega^a{}_b\rp D_\perp^b\log\mc P
\rp+\ldots.\nonumber
\end{align}
This entropy current differs only by a trivial current (with parameters $t_1=-2$ and $t_2=0$ in the notation of \cite{Compere:2012mt}) from the one obtained in the literature. We therefore recover, applying the AdS/Ricci-flat correspondence, all details of the relativistic Rindler fluids as obtained in \cite{Compere:2012mt,Eling:2012ni}, and we have clarified the link with the conformal fluid of AdS/CFT, by showing how the AdS boundary conditions become the Rindler boundary conditions on the cut-off surface $\Sigma_c$.


\section{Conclusion and Outlook \label{section:ccl}}

In this work, we have presented a map relating a class of Asymptotically locally AdS spacetimes, with the bulk having a flat toroidal transverse subspace, to a class of Ricci-flat spacetimes with a transverse round sphere. Though we have focussed on the case of negative cosmological constant and round spheres, a similar map exists for positive/zero cosmological constant and different internal spaces. We also expect the map to generalize to include matter field such as gauge fields, $p$-forms etc.\ (we have already seen that the map extends to free massless fields). A more ambitious extension would be to include the moduli of the internal space. On the AdS side the moduli of the torus were included in the analysis of \cite{Gouteraux:2011qh}, while deformations of the sphere may be studied
following \cite{Skenderis:2006uy}.

A prime motivation for this work was to understand how holography works for asympotically flat spacetimes. The map implies that the Ricci-flat spacetimes that enter in the correspondence inherit holographic properties from the corresponding AdS spacetimes. While the full development of holography is left for future work, we have already seen that the holographic data are mapped in unexpected ways. In particular, the Dirichlet boundary conditions in AdS are mapped to boundary conditions at the position of a $p$-brane (which is located in the interior of spacetime) and the stress energy tensor due to this $p$-brane is the holographic stress energy tensor of the corresponding AdS spacetime.

The dimension of the $p$-brane is related to the dimension of the compactified toroidal subspace of the boundary of AdS: a $(d-p-1)$-dimensional toroidal compactification in AdS yields a Ricci-flat spacetime with asymptotically flat boundary conditions transverse to a $p$-brane. On the Ricci-flat side, different boundary conditions yield
different super-selection sectors and each of these is obtained by considering different compactifications on the AdS side. Linear fluctuations respecting the $p$-brane boundary conditions are linked with the linear fluctuations in AdS that yield holographic 2-point functions. In particular, regularity in the interior of AdS is linked with normalizability on the Ricci-flat side.

In this paper we set to zero the sources for dual operators, except in one case where we considered infinitesimal
sources. To extract the holographic data we would need to turn on arbitrary sources on the AdS side, and in particular a non-trivial boundary metric. This would allow us to study how the map acts on general boundary correlation functions. We initiated such an analysis here by studying a specific case of a 2-point function and it would be interesting to extend this to general $n$-point functions. It would also be interesting to see how the map acts on non-local observables such as Wilson loops and the entanglement entropy.

We also started analyzing the implications of symmetries. The map implies that the Ricci-flat spacetimes inherit a generalized conformal structure. We have already seen that the structure of second order hydrodynamics is controlled by this structure, and it would be interesting to extract all implications of this ``hidden conformal invariance''.

We have also applied the AdS/Ricci-flat map to black holes in Poincar\'e AdS and to second-order in viscous corrections fluids/gravity metrics. The black holes map to Schwarzschild black $p$-branes, and we derived asymptotically flat second-order hydrodynamic metrics (blackfolds). In particular, these solutions describe the nonlinear evolution of the Gregory-Laflamme instability. We showed that there is an associated entropy current with non-negative divergence and a finite, non-conformal stress tensor, whose tensorial structure is also determined by that of the parent entropy current and conformal stress tensor in AdS, respectively. From this stress tensor, we obtained the Gregory-Laflamme dispersion relation up to third order in momentum. It correctly reproduces the physics of the GL instability and captures the existence of a threshold mode. The quantitative agreement with numerical results becomes excellent for large dimensions of the transverse sphere.

A particularly interesting case to analyze further is that of the Schwarzschild black hole. This is obtained from the AdS black branes by compactifying all spatial boundary directions. As such this case has many similarities with $AdS_2$ and understanding holography is likely to require resolving many of the issues associated with $AdS_2$ holography.

Another special case is that of no sphere ($n=-1$). In this case the AdS black brane maps to Rindler spacetime, and the second order fluid/gravity metric and entropy current map to the second order hydrodynamic metric and entropy current for the Rindler fluid \cite{Compere:2011dx,Compere:2012mt,Eling:2012ni}, after a change to the isotropic gauge. In particular, the AdS boundary maps to the position of the accelerated observer, and horizons map to each other (recall that there is a scaling that changes the position of the accelerated observer \cite{Compere:2012mt} -- this is related to scaling the AdS radial direction, as discussed in the Introduction). It is likely that this case can provide the fastest route to flat space holography
since there is no transverse sphere. As such it would be very interesting to study it further by turning on sources and study correlation functions and other non-local observables.

In this paper, we mostly focused on mapping properties and observables from AdS to Ricci-flat spacetimes. It would also be very interesting to use the map the other way around. In asymptotically flat spacetimes the main observable is the S-matrix and it would be interesting to study how the AdS/Ricci-flat map acts on it. A related question is to map the structure of asymptotically flat asymptotia and the corresponding symmetry groups that act on them (such as the BMS group, etc.)\ to AdS. One may also wish to study how other asymptotically flat black holes (Kerr, Kerr-Newman, etc.)\ map to AdS.

Finally, the map was developed at the level of gravitational theories, and it would be interesting to see whether it can be extended to the full string theory. To this end, one would like to formulate the map in terms of the worldsheet of the string. If successful, this may give us a new route to understand string theory on AdS, starting from string theory on flat spacetimes.


\acknowledgments

MC and KS acknowledge support from  the John Templeton Foundation.
This publication was made possible through the support of a grant from the
John Templeton Foundation. The opinions expressed in this publication are those of the authors and do not necessarily reflect the views of the John Templeton Foundation.
JC is supported by the European Research Council grant no. ERC-2011-StG 279363-HiDGR.
B.G. gratefully acknowledges support from the European Science Foundation for the activity ``Holographic Methods for Strongly Coupled Systems'' with Short Visit Grant 5979, and wishes to thank the Newton Institute and DAMTP, Cambridge, UK for hospitality during the final stage of this work. Preliminary versions of this work were presented
in a number of workshops including the "Workshop on Holography, gauge theory and black holes" in Amsterdam, The Netherlands, the "Workshop on Quantum Aspects of Black Holes" in Seoul, Korea, the "String theory workshop" in Benasque, Spain, the GR20 conference in Warsaw, Poland, the workshop ``Fields, Strings and Holography'' in Swansea, UK, the
program ``Mathematics and Physics of the Holographic Principle'' at the Newton Institute, Cambridge, UK, the ``First Tuscan Meeting'' in Pisa, Italy and the "13th Trobada de Nadal de Fisica Teorica'' in Barcelona, Spain. We would like to thank the participants of these workshops for interesting questions and comments, as well as Roberto Emparan and Sigbj\o rn Hervik for discussions.
\vfill
\pagebreak
\appendix

\addtocontents{toc}{\protect\setcounter{tocdepth}{1}}

\section{Diagonal dimensional reduction of Einstein gravity with a cosmological constant \label{appendix:KKred}}

Our starting point is the Einstein-AdS action with a cosmological constant in $d+1$ dimensions \eqref{EinsteinAdSAction}\footnote{In this appendix, the dimensionality of each object is explicitly indicated as an index or exponent.}:
\be
    S_{d+1}=\frac1{16\pi G_N^{d+1}}\int\ud^{d+1}x\sqrt{-g_{d+1}}\left[R_{d+1}-2\Lambda\right].
\ee
The equations of motion are
\be
    G_{AB}+\Lambda g_{AB}=0\,.\label{app:EeqA1}
\ee
We wish to perform a reduction to an Einstein-Dilaton theory with the Ansatz:
\be
    \ud s^2_{d+1}=e^{2\al\phi}\ud s^2_{p+2} + R^2e^{2\ba\phi}\ud X^2_{d-p-1},
    \label{app:KKStatic1}
\ee
where $\ud X^2_{d-p-1}$ is the metric of an $(d-p-1)$-dimensional (compact) Euclidean manifold with radius $R_X$.

For a diagonal Ansatz, it is consistent to take all scalar fields along each reduced direction equal. Nonetheless, let us check that such an Ansatz is consistent by reducing Einstein's equations directly and writing out the action from which they derive.

Using the tetrad formalism, the higher-dimensional Einstein tensor $G^{d+1}_{AB}$ $(A,B,\dots=0\dots d)$ can be projected on the external $(a,b,\ldots=0\dots p+1)$ and internal $(i,j,\ldots=1\dots d-p-1)$ coordinates:
\bea
    G^{d+1}_{ab}&=&G^{p+2}_{ab}+\left[p\al^2+2(d-p-1)\al\ba-(d-p-1)\ba^2\right]\partial_a\phi\partial_b\phi \nonumber\\
    &&-\left[p\al+(d-p-1)\ba\right]\nabla_a\nabla_b\phi-\frac{g_{ab}^{p+2}}2\left\{-2\left[p\al+(d-p-1)\ba\right]\square\phi\right.\nonumber\\
&&\left.-\left[p(p-1)\al^2+ 2(d-p-1)(p-1)\al\ba+(d-p-1)(d-p)\ba^2\right]\partial\phi^2\right.\nn\\
&&\left.+X_{d-p-1}e^{2(\al-\ba)\phi}\right\}\label{app:GmunuA}\\
    G^{d+1}_{ij}&=&G^{d-p-1}_{ij}-\frac12g_{ij}^{d-p-1}e^{2(\ba-\al)\phi}\left\{R_{p+2}-2\left[(p+1)\al+(d-p-2)\ba\right]\square\phi\right.\nonumber\\
    &&\left.-\left[(p+1)p\al^2+2p(d-p-2)\al\ba+(d-p-2)(d-p-1)\ba^2\right]\partial\phi^2\right\}\label{app:GmnA}
\eea
where $G^{p+2}_{ab}$ and $G^{d-p-1}_{ij}$ are respectively the Einstein tensor associated to the $(p+2)$-dimensional metric and to the $(d-p-1)$-dimensional compact space $\mathbf X_{d-p-1}$. $X_{d-p-1}$ is its Ricci curvature and verifies
\be
X_{d-p-1}=\frac{(d-p-1)(d-p-2)}{R_X^2}\,.
\ee
Then, taking the trace of $G_{AB}^{d+1}$, one finds the Ricci scalar
\bea
    R_{d+1}e^{2\al\phi} &=& R_{p+2}+e^{2(\al-\ba)\phi}X_{d-p-1}-2((p+1)\al+(d-p-1)\ba)\Box\phi\\
    &&-\left[p(p+1)\al^2+(d-p-1)(d-p)\ba^2+2(d-p-1)p\al\ba\right]\partial\phi^2,\nn
\eea
while
\be
    \det g_{d+1}=e^{2\left[(p+2)\al+(d-p-1)\ba\right]\phi}\det g_{p+2}\,.
\ee
Setting the overall conformal factor in $\phi$ in the action to unity requires
\be
    (d-p-1)\ba=-p\al
\ee
upon which
\be
    R_{d+1}e^{2\al\phi} = R_{p+2}-2\al\Box\phi-\frac{pd}{d-p-1}\al^2\partial\phi^2+e^{2\frac{d-1}{d-p-1}\al\phi}X_{d-p-1}\,.
\ee
In order to have a canonically normalized kinetic term for the scalar, we then set
\be
    \al=-\sqrt{\frac{d-p-1}{2pd}}=-\frac\da2\quad \Leftrightarrow \quad \da=\sqrt{\frac{2(d-p-1)}{pd}}
\ee
so that the bulk action naively becomes
\bea
    S_{p+2}&=&\frac{1}{16\pi G_N^{p+2}}\int_{\mathcal M}\ud^{p+2}x\sqrt{-g_{p+2}}\left[R_{p+2}-\half\partial\phi^2-2\Lambda e^{-\da\phi}+X_{d-p-1}e^{-\frac{2\phi}{p\da}}\right]\nn\\
    && \qquad-\frac{1}{16\pi G_N^{p+2}}\int_{\partial\mathcal M}\ud^{p+1}x\sqrt{-h_{p+1}}\,\da\, n\cdot\partial\phi\,.
    \label{app:AdSMaxD+1}
\eea
where $h_{p+1}$ is the induced metric on the boundary $\partial\mathcal M$ of the manifold $\mathcal M$ and $n^a$ a unit vector normal to that boundary.

To check that this is correct, we can also replace in \eqref{app:GmunuA} and \eqref{app:GmnA}
\bea
    G^{d+1}_{ab}&=&G^{p+2}_{ab}-\frac12\partial_a\phi\partial_b\phi-\frac12g_{ab}^{p+2}\left[X_{d-p-1}e^{-\frac{2\phi}{p\da}}-\frac12\partial\phi^2\right]\label{app:GmunuA2}\\
    G^{d+1}_{ij}&=&G^{d-p-1}_{ij}-\frac12g_{ij}^{d-p-1}e^{\frac{2\phi}{p\da}}\left[R_{p+2}+\frac{2}{p\da}\square\phi-\frac12\partial\phi^2\right].\label{app:GmnA2}
\eea
and reexpress Einstein's equations \eqref{app:EeqA1}:
\bea    G^{p+2}_{ab}&=&\frac12\partial_a\phi\partial_b\phi+\frac{g_{ab}^{p+2}}2\left[X_{d-p-1}e^{-\frac{2\phi}{p\da}}-2\Lambda e^{-\da\phi}\right]\label{app:GmunuA3}\\
    G^{d-p-1}_{ij}&=&\frac{g_{ij}^{d-p-1}}2e^{\frac{2\phi}{p\da}}\left[R_{p+2}+\frac{2\square\phi}{p\da}-\frac12\partial\phi^2-2\Lambda e^{-\da\phi}\right].\label{app:GmnA3}
\eea
In \eqref{app:GmunuA3}, we recognize the lower-dimensional equation of motion for the metric, as derived from \eqref{app:AdSMaxD+1}. Taking the trace of \eqref{app:GmnA3} and replacing again in \eqref{app:GmnA3}, one finds that $\mathbf X_{d-p-1}$ must be an Einstein space, that is
\be\label{app:EinsteinSpaceCondition}
R_{d-p-1}^{ij}=\frac{X_{d-p-1}}{d-p-1}g_{d-p-1}^{ij}\,.
\ee
Note that this condition becomes very important once higher-derivative terms are included, for instance constraining the square of the Weyl tensor of $\mathbf X_{d-p-1}$, \cite{Charmousis:2012dw}.
The lower-dimensional Ricci scalar can be derived from \eqref{app:GmunuA3} or  \eqref{app:GmnA3}:
\bea
    R_{p+2}&=&\frac12\partial\phi^2+\frac{p+2}{p}2\Lambda e^{-\da\phi}-\frac{p+2}{p}X_{d-p-1}e^{-\frac{2\phi}{p\da}}\\
    R_{p+2}&=&\frac12\partial\phi^2+2\Lambda e^{-\da\phi}-\frac{d-p-3}{d-p-1}X_{d-p-1}e^{-\frac{2\phi}{p\da}}-\frac{2\square\phi}{p\da}\,.
\eea
Subtracting the two previous equations yields the dilaton equation of motion:
\be\label{app:DilEq}
    \square\phi=-2\da\Lambda e^{-\da\phi}+\frac{2}{p\da}X_{d-p-1}e^{-\frac{2\phi}{p\da}},
\ee
identical to that derived from \eqref{app:AdSMaxD+1}, while the other combination gives back the trace of Einstein's equations.

The metric Ansatz becomes
\be
    \ud s^2_{d+1}=e^{-\da\phi}\ud s^2_{(p+2)} + e^{\frac\phi\da\left(\frac{2}{p}-\da^2\right)}\ud X^2_{d-p-1}\,.
    \label{app:KKStatic}
\ee
We have also defined the lower-dimensional Newton's constant $G_N^{p+2}= G_N^{d+1}/V_{d-p-1}$, where $V_{d-p-1}$ is the volume of $\mathbf X^{d-p-1}$. The boundary term in \eqref{app:AdSMaxD+1} has no impact on the equations of motion, but would be important for the computation of the Euclidean action on-shell. It is exactly compensated by a term coming from the reduction of the Gibbons-Hawking-York boundary term.

Inspecting both the action \eqref{app:AdSMaxD+1} or the field equations \eqref{app:GmunuA3}, \eqref{app:DilEq}, it is clear that they are invariant under the exchange
\be
-2\Lambda\leftrightarrow X_{d-1}\Leftrightarrow \ell=\leftrightarrow R_X\qquad \delta\leftrightarrow \frac2p\delta
\ee
where $\ell$ is the AdS radius. Such a symmetry of the action and field equations was formerly noticed in \cite{Charmousis:2003wm}, for the special case of metrics with Weyl symmetry.

\section{Useful hydrodynamics formul\ae\ }

The velocity is denoted $u_\mu$ and is normalized $u_\mu u^\mu=-1$, while indices are raised and lowered with the Minkowski metric $\eta_{\mu\nu}$ (we restrict ourselves to the flat boundary case). As a consequence, all covariant derivatives can be replaced by ordinary partial derivatives.

We define the zeroth- and first-order derivative hydrodynamic data the usual way:
\beq
\mc A_\mu=u^\la\p_\la u_\mu-\frac{\p_\la u^\la}{d-1}u_\mu=a_\mu-\frac{\theta}{d-1}u_\mu,\qquad
P_{\mu\nu}=\eta_{\mu\nu}+u_\mu u_\nu\,.
\eeq
 The shear $\sigma_{\mu\nu}$ and vorticity $\omega_{\mu\nu}$ are the symmetric and anti-symmetric traceless pieces of the gradient of the velocity
\be
	\sigma_{\mu\nu}=P_\mu^\ka P_\nu^\la \partial_{(\ka}u_{\la)}-\frac{\partial\cdot u}{d-1}P_{\mu\nu}\,,\qquad \omega_{\mu\nu}=P_\mu^\ka P_\nu^\la \partial_{[\ka}u_{\la]}
\ee
so that the gradient of the velocity can be decomposed as
\be
	\partial_\mu u_\nu = \sigma_{\mu\nu}+\omega_{\mu\nu}+\frac{P_{\mu\nu}\theta}{d-1}-u_\mu a_\nu\,,\qquad \theta = \partial\cdot u\,,\qquad a_\mu=u\cdot\partial u_\mu
\ee
where we also defined the expansion and the acceleration.

To parameterize second-derivative data, the Weyl-invariant formalism of \cite{Bhattacharyya:2008mz} is useful, and we adopt their conventions for the Weyl invariant acceleration $\mathcal A_\mu$ and covariant derivative:
\bea
	&&\mathcal A_\mu = a_\mu -\frac{\theta u_\mu}{d-1}\,,\\
	&& P_\mu^\nu\mathcal D^\la\sigma_{\nu\la}= \partial^\la\sigma_{\mu\la}-(d-1)a^\la\sigma_{\mu\la}-\sigma^2u_\mu\,,\qquad \sigma^2=\sigma_{\mu\nu}\sigma^{\mu\nu}\,,\\
	&&P_\mu^\nu\mathcal D^\la\omega_{\nu\la}=\partial^\la\omega_{\mu\la}-(d-3)a^\la\omega_{\mu\la}+\omega^2u_\mu\,, \qquad \omega^2=\omega_{\mu\nu}\omega^{\mu\nu}\,,\\
	&&u^\la\mathcal D_\la\sigma_{\mu\nu}=\dot\sigma_{\mu\nu}+\frac{\theta\sigma_{\mu\nu}}{d-1}-2a^\la u_{\left(\mu\right.}\sigma_{\left.\nu\right)\la}\,.
\eea
Some of the second-derivative data is not independent:
\bea
	&&\dot a_\mu = u^\la\partial_\la a_\mu = \frac1{d-2}\left(P_\mu^\nu\mathcal D^\la\sigma_{\nu\la}+P_\mu^\nu\mathcal D^\la\omega_{\nu\la}\right)+a^2u_\mu\,,\\
	&&\frac{\partial_\mu\theta}{d-1} =  \frac1{d-2}\left(P_\mu^\nu\mathcal D^\la\sigma_{\nu\la}+P_\mu^\nu\mathcal D^\la\omega_{\nu\la}\right)-\frac{\dot\theta u_\mu}{d-1}+a^\la\sigma_{\mu\la}+a^\la\omega_{\mu\la}\,,\\
	&&\partial\cdot a=\sigma^2-\omega^2+\dot\theta+\frac{\theta^2}{d-1}\,.
\eea
Finally, the hydrodynamic equations allow to replace temperature derivatives:
\be
	\partial_\nu b = b\mathcal A_\nu+2b^2\left[\frac{\sigma^2u_\nu}{d(d-1)}-\frac{P_\nu^\la}d\mathcal D^\mu\sigma_{\la\mu}\right].
\ee

\section{Functions in the fluids/gravity metrics}\label{FuncsFluidsGrav}
The functions in sec.~\ref{secFluidsGravity} are \cite{Bhattacharyya:2008mz}:

\begin{equation}
F(br)\equiv \int_{br}^{\infty}\frac{y^{d-1}-1}{y(y^{d}-1)}\ud y\,,
\end{equation}
\begin{equation}
H_1(br)\equiv \int_{br}^{\infty}\frac{y^{d-2}-1}{y(y^{d}-1)}\ud y\,,
\end{equation}
\begin{equation}
\begin{split}
H_2(br)&\equiv \int_{br}^{\infty}\frac{\ud\xi}{\xi(\xi^d-1)}
\int_{1}^{\xi}y^{d-3}\ud y \left[1+(d-1)y F(y) +2 y^{2} F'(y) \right]\\
&=\frac{1}{2} F(br)^2-\int_{br}^{\infty}\frac{\ud\xi}{\xi(\xi^d-1)}
\int_{1}^{\xi}\frac{y^{d-2}-1}{y(y^{d}-1)}\ud y\,,
\end{split}
\end{equation}
\begin{equation}
K_1(br) \equiv \int_{br}^{\infty}\frac{\ud\xi}{\xi^2}\int_{\xi}^{\infty}\ud y\ y^2 F'(y)^2 \,,
\end{equation}
\begin{equation}
\begin{split}
K_2(br) \equiv \int_{br}^{\infty}\frac{\ud\xi}{\xi^2}&\left[\vphantom{\int_{\xi}^{\infty}}
1-\xi(\xi-1)F'(\xi) -2(d-1)\xi^{d-1} \right.\\
&\left. \quad +\left(2(d-1)\xi^d-(d-2)\right)\int_{\xi}^{\infty}\ud y\ y^2 F'(y)^2 \right]\,,
\end{split}
\end{equation}
\begin{equation}
L(br) \equiv \int_{br}^\infty\xi^{d-1}\ud\xi\int_{\xi}^\infty \ud y\ \frac{y-1}{y^3(y^d
-1)}\,.
\end{equation}
The constants appearing in the horizon location $r_H$ and the entropy current are respectively \cite{Bhattacharyya:2008mz},
\eq
 h_1=\frac{2(d^2+d-4)}{d^2(d-1)(d-2)}-\frac{K_{2H}}{d(d-1)}\,,\qquad
 h_2=-\frac{d+2}{2d(d-2)}\,,\qquad
 h_3=-\frac1{d(d-1)(d-2)}\,,
\eeq
and
\begin{align}
 A_1&=\frac2{d^2}(d+2)-\frac{K_{1H}d+K_{2H}}d\,,& A_2&=-\frac1{2d}\,,\\
 B_1&=-2A_3=\frac2{d(d-2)}\,,& B_2&=\frac1{d-2}\,,
\end{align}
with
\eq
\begin{split}
K_{2H}\equiv K_2(1) =\int_{1}^{\infty}\frac{\ud\xi}{\xi^2}&\left[\vphantom{\int_{\xi}^{\infty}}
1-\xi(\xi-1)F'(\xi) -2(d-1)\xi^{d-1} \right.\\
&\left. \quad +\left(2(d-1)\xi^d-(d-2)\right)\int_{\xi}^{\infty}\ud y\ y^2 F'(y)^2 \right]\,,
\end{split}
\eeq
\eq
\begin{split}
K_{1H}d+K_{2H}\equiv K_1(1)d+K_2(1)=
\int_1^\infty\frac{\ud\xi}{\xi^2}&\left[1-\xi(\xi-1)F'(\xi) -2(d-1)\xi^{d-1} \right.\\
&\left. \quad +2\left((d-1)\xi^d+1\right)\int_{\xi}^{\infty}\ud y\ y^2 F'(y)^2 \right]\,.
\end{split}
\eeq

\section{The fluids/gravity metric in FG coords\label{app:EFtoFG}}

We would like to bring the fluid/gravity metric \eqref{EFcoords}-\eqref{EFmetric2} from Eddington-Finkelstein (EF)
\be
	\ud s^2=-2u_\mu\, \ud x^\mu\lp \ud r+{\mc V}_\nu \ud x^\nu\rp
+{\mc G}_{\mu\nu}\ud x^\mu \ud x^\nu,
\ee
to Fefferman-Graham(FG) coordinates
\be
	\ud s^2 = \frac{\ud \rho^2}{4\rho^2}+\frac1\rho \mfg_{\mu\nu}(\rho,z^\la)\ud z^\mu\ud z^\nu,
\ee
so that reading off the holographic stress-tensor is straightfoward. Letting the EF coordinates be functions of the FG ones, $\left(r(\rho,z^\la),\,x^\mu(\rho,z^\la)\right)$, yields the following system of first-order differential equations:
\bea
	\frac1{4\rho^2}&=&-2u^\mu\partial_\rho x^\mu\left(\partial_\rho r+\mathcal V_\nu\partial_\rho x^\nu\right)+\mathcal G_{\mu\nu}\partial_\rho x^\mu\partial_\rho x^\nu,\label{ScalarFGEq}\\
	0&=&-2u_\nu\partial_\rho x^\nu\left(\partial_\mu r+\mathcal V_\la\partial_\mu x^\la\right)-2u_\nu\partial_\mu x^\nu\left(\partial_\rho r + \mathcal V_\la\partial_\rho x^\la\right) + 2\mathcal G_{\ka\la}\partial_\rho x^\ka\partial_\mu x^\la,\label{VectorFGEq}\\
	\frac{g_{\mu\nu}}\rho &=&\mathcal G_{\ka\la}\partial_\mu x^\ka\partial_\nu x^\la -2u_\ka\partial_\nu x^\ka\left(\partial_\mu r +\mathcal V_\la\partial_\mu x^\la\right).\label{TensorFGEq}
\eea
These can be solved order by order. More precisely, at a given order, one only needs to solve \eqref{ScalarFGEq} and \eqref{TensorFGEq} to have access to the metric in FG coordinates at that order. The remaining equation \eqref{VectorFGEq} only serves as input for the next order. Consequently, we only evaluate $x^\mu(\rho,z^\la)$ to first order in the gradient expansion, as we only wish to obtain the metric to second order in derivatives.

The change of EF coordinates $(r,x^\mu)$ to FG $(\rho,z^\mu)$ goes like:
\bea
	r&=&R_0(\rho,z^\mu)+R_1(\rho,z^\mu)+R_2(\rho,z^\mu)\,,\qquad \xi=\left(\frac{\rho}{b^2}\right)^{\frac d2}\\
	R_0(\rho,z^\mu)&=&\frac1{\sqrt{\rho}}\left(1+\frac\xi4\right)^{\frac2d}\,,\quad \partial_\rho\left(R_0\right)=\frac{-R_0\sqrt{f}}{2\rho }\,,\quad \partial_\mu\left(R_0\right)=R_0(\sqrt{f}-1)\frac{\partial_\mu b}b\\
	R_1(\rho,z^\mu)&=&\frac{R_0\,k_b\,\theta}{d-1}\,,\quad \partial_\rho\left(k_b\right)=\frac{1}{2\rho R_0\sqrt{f}}\,,\quad \partial_\mu\left(k_b\right)=\left(k_b-\frac1{R_0\sqrt{f}}\right)\frac{\partial_\mu b}b\\
	&&\partial_\rho\left(R_1\right)=\frac{\left(1-R_0k_bf\right)\theta}{2\rho\sqrt{f}(d-1)}\,,\quad \partial_\mu\left(R_1\right)=\frac{R_0k_b\partial_\mu\theta}{d-1}+\frac{(R_0k_bf-1)\theta\mathcal A_\mu}{\sqrt{f}(d-1)}\\
	\partial_\rho\left(\frac{R_2}{R_0\sqrt{f}}\right)&=&\partial_\rho\left[\left(X_{1a}+X_{1\theta}-k_b^2\right)\frac{\omega^2}4+\left(X_{1a}-X_{1\theta}+\frac{k_b^2-2X_{1a}}{\sqrt{f}}\right)\frac{a^2}2\right.\nonumber\\
	&&\left.+\left(2X_{1\theta}-k_b^2-\frac{k_b^2}{\sqrt{f}}\right)\frac{\dot\theta}{2(d-1)}+\left(1-\frac1{\sqrt{f}}\right)\left(X_{1\theta}-\frac{k_b^2}2\right)\frac{\theta^2}{(d-1)^2}\right]\nn\\
	&&+\frac{(f-1)}{4\rho f(d-1)}\left(2k_b+bK_2\right)b\sigma^2\\
	x^\mu&=&z^\mu-k_b u_\mu+X_{1a}a_\mu+X_{1\theta}\frac{\theta u_\mu}{d-1}+X_2^\mu\label{CoordChangex}\\
	\partial_\rho\left(X_{1a}\right)&=&\frac{\sqrt{f}+R_0 k_b}{2\rho R_0^2\sqrt{f}}\,,\qquad \partial_\mu\left(X_{1a}\right)=\left(2X_{1a}-\frac{\sqrt{f}+R_0 k_b}{ R_0^2\sqrt{f}}\right)\frac{\partial_\mu b}b \label{X1aEq}\\
\partial_\rho\left(X_{1\theta}\right)&=&\frac{R_0 k_b\sqrt{f}-1}{2\rho R_0^2f}\,, \qquad\partial_\mu\left(X_{1\theta}\right)=\left(2X_{1\theta}-\frac{R_0 k_b\sqrt{f}-1}{R_0^2f}\right)\frac{\partial_\mu b}b \label{X1thEq}\\
	\partial_\nu x^\mu&=& \delta_\mu^\nu-k_b\partial_\nu u^\mu-\left(k_b-\frac1{R_0\sqrt f}\right)u^\mu\frac{\partial_\nu b}{b}+X_{1a}\partial_\nu a^\mu+X_{1\theta}\left(\frac{\theta\partial_\nu u^\mu}{d-1}+\frac{\partial_\nu\theta u^\mu}{d-1}\right)\nonumber\\
	&&+a^\mu\partial_\nu\left(X_{1a}\right)+\frac{\theta u^\mu}{d-1}\partial_\nu\left(X_{1\theta}\right).
\eea

Armed with these definitions, the metric is written in FG coordinates at second order in derivatives of the velocity and temperature but {\em exactly in $\rho$} (here the index in metric elements denotes the order in derivatives):
\bea
	\frac{g_{0\mu\nu}}\rho&=&R_0^2\left[\eta_{\mu\nu}+\left(R_0b\right)^{-d}u_\mu u_\nu\right],\\
	\frac{g_{1\mu\nu}}\rho&=&\frac{2bR_0^2}{d}\log\left(1-(bR_0)^{-d}\right)\sigma_{\mu\nu}\\
	\frac{g_{2\mu\nu}}\rho&=&R_0^2\left(k_b^2+2X_{1a}-4bk_bF-2b^2H_1+2b^2F^2\right)\sigma_{\mu\la}\sigma^\la_{\phantom{1}\nu} + R_0^2\left[\left(2X_{1a}-k_b^2\right)-1\right]\omega_{\mu\la}\omega^\la_{\phantom{1}\nu}\nonumber\\
	&&+2R_0^2\left(k_b^2-2bk_b F+b^2H_2\right)\sigma_{\la\left(\mu\right.}\omega_{\left.\nu\right)}^{\phantom{1}\la}+2R_0^2\left(X_{1a}+X_{1\theta}-k_b^2\right)\frac{\theta\sigma_{\mu\nu}}{d-1}\nonumber\\
	&&+2R_0^2\left(X_{1a}-bk_b F+b^2H_2-b^2H_1\right)u^\la\mathcal D_\la\sigma_{\mu\nu}+R_0^2\left[\left(2X_{1a}-k_b^2\right)-1\right]a_\mu a_\nu\nonumber\\
	&&+R_0^2\left[2\frac{R_2}{R_0}+2\left(X_{1a}+X_{1\theta}-k_b^2\right)\frac{\theta^2}{(d-1)^2}+\frac{2X_{1a}\dot\theta}{d-1}+\frac{2b^2\sigma^2}{d-1}\left(H_1-K_1\right)\right]P_{\mu\nu}\nonumber\\
	&&+\left[2+R_0^2\left(k_b^2-2X_{1a}+2fX_{1\theta}-fk_b^2\right)\right]\left(\frac{\theta u_{\left(\mu\right.}a_{\left.\nu\right)}}{d-1}+\frac{u_{\left(\mu\right.}P_{\left.\nu\right)}^{\phantom{1}\la}\mathcal D^\ka\omega_{\la\ka}}{d-2}\right)\nonumber\\
	&&+\left[\frac{2+R_0^2\left(k_b^2-2X_{1a}+2fX_{1\theta}-fk_b^2\right)}{d-2}+\frac{4bR_0}{d}\left(R_0k_bf-1\right)\right.\nonumber\\
	&&\left.+4b^2R_0^2(1-f)L\right]u_{\left(\mu\right.}P_{\left.\nu\right)}^{\phantom{1}\la}\mathcal D^\ka\sigma_{\la\ka}+R_0^2\left[2X_{1a}-(1+f)k_b^2+2fX_{1\theta}\right] u_{\left(\mu\right.}a^\la\sigma_{\left.\nu\right)\la} \nn\\
&&+R_0^2\left[-2X_{1a}+k_b^2+f\left(2X_{1\theta}+4X_{1a}-3k_b^2\right)\right] u_{\left(\mu\right.}a^\la\omega_{\left.\nu\right)\la}+\left[d(f-1)-2f\right]R_2R_0u_\mu u_\nu\nonumber\\
	&&+(f-1)\frac{\omega^2}2u_\mu u_\nu+\left[d(f-1)R_0^2k_b^2-2(1+R_0^2fX_{1\theta})\right]\frac{\dot\theta u_\mu u_\nu}{d-1}+\left[R_0^2\left(d(f-1)-4f\right)\right.\nonumber\\
	&&\left.\times\left(X_{1\theta}-\frac{k_b^2}2\right)-2\right]\frac{\theta^2u_\mu u_\nu}{(d-1)^2}+\frac{4R_0(1-R_0k_bf)}{d(d-1)}\sigma^2u_\mu u_\nu\nonumber\\
	&&+(f-1)R_0^2(2bk_b+b^2K_2)\frac{\sigma^2u_\mu u_\nu }{(d-1)}+\left[R_0^2\left(d(f-1)-2f\right)\left(X_{1a}-\frac{k_b^2}2\right)+1\right]a^2u_\mu u_\nu\,.\nn\\
\eea

We may now expand asymptotically in $\rho\to 0$ the metric coefficients above to extract the holographic stress-tensor. With a view towards the analytic continuation for $d=-1,-2$, we also provide some of the subleading powers in $\rho$ (in the metric elements, the exponent in parentheses is the order in $\rho$ in the FG expansion, the other exponent not in parentheses the order in derivatives):
\bea
	g^{2,(0)}_{\mu\nu}&=&\eta_{\mu\nu}\label{Blaiseg0}\\
	g^{2,(0<i<d/2)}_{\mu\nu}&=&0
\eea

\be \label{Blaisegdo2}
	 b^{d-2}g^{2,(d/2)}_{\mu\nu}=\frac{2}{d}\sigma_{\mu\la}\sigma^\la_{\phantom{1}\nu}-\frac{4\tau }{db}\sigma_{\la\left(\mu\right.}\omega_{\left.\nu\right)}^{\phantom{1}\la}+\frac{2 (b-\tau )}{db}u^\la\mathcal D_\la\sigma_{\mu\nu}-\frac{2 P_{\mu\nu}\sigma^2}{(-1+d) d}
\ee

\bea
	b^{d}g^{2,(d/2+1)}_{\mu\nu}&=&-\frac{\sigma_{\mu\la}\sigma^\la_{\phantom{1}\nu}}{2+d}+\frac{\omega_{\mu\la}\omega^\la_{\phantom{1}\nu}}{2+d}+\frac{2}{2+d}\sigma_{\la\left(\mu\right.}\omega_{\left.\nu\right)}^{\phantom{1}\la}-\frac{2}{2+d}\frac{\theta\sigma_{\mu\nu}}{d-1}+\frac{a_\mu a_\nu }{2+d}\nonumber\\
	&&+\frac{P_{\mu\nu}}{2(d+2)}\left[\sigma^2-\omega^2-d\, a^2+\frac{(d-2) \dot\theta}{(d-1)}+\frac{(d-2) \theta^2}{(d-1)^2}\right]+\frac{(-2+d) \theta u_{\left(\mu\right.}a_{\left.\nu\right)}}{(2+d)(d-1)}\nonumber\\
	&&-\frac{u_{\left(\mu\right.}P_{\left.\nu\right)}^{\phantom{1}\la}\mathcal D^\ka\omega_{\la\ka} }{2+d}-\frac{u_{\left(\mu\right.}P_{\left.\nu\right)}^{\phantom{1}\la}\mathcal D^\ka\sigma_{\la\ka}}{2+d}+\frac{d  u_{\left(\mu\right.}a^\la\sigma_{\left.\nu\right)\la}}{2+d}-\frac{(4+d)  u_{\left(\mu\right.}a^\la\omega_{\left.\nu\right)\la} }{2+d}\nonumber\\
	&&\left[\frac{(d-3) \sigma^2}{2 (2+d)}+\frac{(d-2)}{2 (2+d)}\left(\dot\theta+\frac{\theta^2}{d-1}-(d+1)a^2\right)-\frac{(1+d)\omega^2}{2 (2+d)}\right]u_\mu u_\nu
\eea

\be
	 b^{2d-2}g^{2,(d)}_{\mu\nu}=\frac{4}{d^2}\sigma_{\mu\la}\sigma^\la_{\phantom{1}\nu}-\frac{4 \tau }{d^2b}\sigma_{\la\left(\mu\right.}\omega_{\left.\nu\right)}^{\phantom{1}\la}+\frac{2 (b-\tau )}{d^2b}u^\la\mathcal D_\la\sigma_{\mu\nu}-\frac{5 P_{\mu\nu}\sigma^2}{2 (-1+d) d^2}+\frac{\sigma^2 u_\mu u_\nu}{2 (-1+d) d^2}
\ee

\bea
	b^{2d}g^{2,(d+1)}_{\mu\nu}&=&-\frac{(6+d) }{8 (1+d) (2+d)}\sigma_{\mu\la}\sigma^\la_{\phantom{1}\nu}-\frac{(-2+d) }{8 (1+d) (2+d)}\omega_{\mu\la}\omega^\la_{\phantom{1}\nu}+\frac{3 }{ (1+d) (2+d)}\sigma_{\la\left(\mu\right.}\omega_{\left.\nu\right)}^{\phantom{1}\la}\nonumber\\
	&&-\frac{(4+d)  }{2 (1+d) (2+d)}\frac{\theta\sigma_{\mu\nu}}{d-1}-\frac{(-6+d)}{8 (1+d) (2+d)}u^\la\mathcal D_\la\sigma_{\mu\nu}-\frac{(-2+d) a_\mu a_\nu}{8 (1+d) (2+d)}\nonumber\\
	&&+\frac{P_{\mu\nu}}{8 (1+d) (2+d)}\left[(d-2)\,\omega^2+(d-2) (d-1)\, a^2-\frac{(d-3) (d-2)}{(d-1)}\left( \dot\theta\right.\right.\nonumber\\
	&&\left.\left.+\frac{\theta^2}{d-1}\right)-(d-6)\sigma^2\right]-\frac{(d-2) (2 d-1)\theta u_{\left(\mu\right.}a_{\left.\nu\right)} }{4 (d-1)(1+d) (2+d)}+\frac{(2 d-1)u_{\left(\mu\right.}P_{\left.\nu\right)}^{\phantom{1}\la}\mathcal D^\ka\omega_{\la\ka} }{4 (1+d) (2+d)}\nonumber\\
	&&+\frac{(-3+2 d) u_{\left(\mu\right.}P_{\left.\nu\right)}^{\phantom{1}\la}\mathcal D^\ka\sigma_{\la\ka}}{4 (1+d) (2+d)}-\frac{(-2+d) d  u_{\left(\mu\right.}a^\la\sigma_{\left.\nu\right)\la}}{2 (1+d) (2+d)}+\frac{d (4+d) u_{\left(\mu\right.}a^\la\omega_{\left.\nu\right)\la} }{2 (1+d) (2+d)}\nonumber\\
	&&+\left[\left(2+3 d+4 d^2\right)\omega^2-\left(6-13 d+4 d^2\right)\sigma^2  -(d-2)  (4 d-1)\left(\dot\theta+\frac{\theta^2}{d-1}\right)\right.\nonumber\\
	&&\left.+(d-2) \left(2+3 d+4 d^2\right) a^2\right]\frac{u_\mu u_\nu}{8 (1+d) (2+d)}
\eea

\bea
	b^{3d-2}g^{2,(3d/2)}_{\mu\nu}&=&\frac{ \left(72-6 d+d^2\right)}{24 d^3}\sigma_{\mu\la}\sigma^\la_{\phantom{1}\nu}-\frac{ \left(24-6 d+d^2\right) \tau }{12 d^3b}\sigma_{\la\left(\mu\right.}\omega_{\left.\nu\right)}^{\phantom{1}\la}+\frac{ (d-6)\sigma^2 u_\mu u_\nu}{12 d^3}\nonumber\\
	&&+\frac{ \left(24-6 d+d^2\right) (b-\tau )}{24 d^3b}u^\la\mathcal D_\la\sigma_{\mu\nu}-\frac{ \left(36-8 d+d^2\right) P_{\mu\nu}\sigma^2}{24(-1+d) d^3}
\eea

\bea
	\frac{g^{2,(3d/2+1)}_{\mu\nu}}{b^{-3d}}&=&-\frac{\left(10+d+d^2\right) }{16 (1+d) (2+d) (2+3 d)}\sigma_{\mu\la}\sigma^\la_{\phantom{1}\nu}+\frac{(-2+d) (-1+d)  }{16 (1+d) (2+d) (2+3 d)}\omega_{\mu\la}\omega^\la_{\phantom{1}\nu}\nonumber\\
	&&+\frac{\left(38-d+d^2\right)}{8 (1+d) (2+d) (2+3 d)}\sigma_{\la\left(\mu\right.}\omega_{\left.\nu\right)}^{\phantom{1}\la}-\frac{3 \left(6+3 d+d^2\right) }{8 (1+d) (2+d) (2+3 d)}\frac{\theta\sigma_{\mu\nu}}{d-1}\nonumber\\
	&&-\frac{(-14+d) }{8 (1+d) (2+d) (2+3 d)}u^\la\mathcal D_\la\sigma_{\mu\nu}+\frac{(-2+d) (-1+d) a_\mu a_\nu }{16 (1+d) (2+d) (2+3 d)}\nonumber\\
	&&+\frac{P_{\mu\nu}}{32 (1+d) (2+d) (2+3 d)}\left[-3(d-2) (d-1) \omega^2-(d-2) (d-1) (3 d-4) \, a^2\right.\nonumber\\
	&&\left.+(-2+d) (-10+3 d) \left(\dot\theta+ \frac{\theta^2}{(d-1)}\right)+\left(38-17 d+3 d^2\right)\sigma^2 \right]\nonumber\\
	&&+\frac{(-2+d) (-2+9 d)  \theta u_{\left(\mu\right.}a_{\left.\nu\right)} }{16 (1+d) (2+d) (2+3 d)}-\frac{(-1+d) (-2+9 d) u_{\left(\mu\right.}P_{\left.\nu\right)}^{\phantom{1}\la}\mathcal D^\ka\omega_{\la\ka} }{16 (1+d) (2+d) (2+3 d)}\nonumber\\
	&&-\frac{(-2+d) (-5+9 d) u_{\left(\mu\right.}P_{\left.\nu\right)}^{\phantom{1}\la}\mathcal D^\ka\sigma_{\la\ka}}{16 (1+d) (2+d) (2+3 d)} +\frac{9 (-2+d) (-1+d)  u_{\left(\mu\right.}a^\la\sigma_{\left.\nu\right)\la}}{16 (1+d) (2+d) (2+3 d)}\nonumber\\
	&&-\frac{(4+d) \left(2-3 d+9 d^2\right)u_{\left(\mu\right.}a^\la\omega_{\left.\nu\right)\la} }{16 (1+d) (2+d) (2+3 d)}-\left[\frac{3 \left(2+7 d+2 d^2+9 d^3\right)}{32 (1+d) (2+d) (2+3 d)}\omega^2\right.\nonumber\\
	&&+\frac{\left(-18+97 d-102 d^2+27 d^3\right) }{32 (1+d) (2+d) (2+3 d)}\sigma^2+\frac{(d-2) (3 d-2) (9 d-1)}{32 (1+d) (2+d) (2+3 d)}\left(\dot\theta+\frac{\theta^2}{d-1}\right)\nonumber\\
	&&\left.-\frac{3 (-2+d) \left(2+7 d+2 d^2+9 d^3\right) }{32 (1+d) (2+d) (2+3 d)}a^2\right]u_\mu u_\nu
\label{Blaiseg3do2p1}
\eea


\section{The $n=1,2$ cases\label{app:n=1,2}}

\subsection{$n=2$ in detail}

\subsubsection{Regularizing the metric}

The divergences one encounters in the metric when taking the limit $d\to-2$ are directly linked to the fact that the various metric elements depend on functions which are themselves diverging in this limit. So finiteness of the functions will ensure finiteness of the metric. As far as the $d=-2$ case is concerned, one finds that the metric functions $X_{1a},X_{1\theta},H_1,H_2,K_1,K_2,L,R_2$ need to be regularized (see Appendix~\ref{app:EFtoFG} for the definition of these functions). This should be done step by step, as some of these functions enter in the definition of others. But generically, once all parent functions have been regularized, the remaining divergence will take the guise of a pole in $d=-2$ at order $0$ in $\rho$. For instance, one can look at the expansion of $X_{1a}$ (which is a function entering in the change to FG gauge, see Appendix~\ref{app:EFtoFG}) and take the limit $d\to-2$:
\be
	X_{1a}\underset{\rho\to 0,\,d\to-2}{\sim}X_{1a0}-\frac{(1+d)b^2}{2 (2+d)}+\frac{b^4}{6 \rho }+\rho +\frac{b^2}{4}  \text{Log}\left(\frac{\rho}{b^2}\right)+O(d+2)\,.
\ee
Here, note that we have introduced an arbitrary integration constant $X_{1a0}$, which is allowed since $X_{1a}$ is determined by its radial derivative \eqref{X1aEq}, and the expansion displays the advertised pole. This pole can simply be absorbed in a change of boundary condition, setting
\be
X_{1a0}=\frac{(1+d)b^2}{2 (2+d)}
\ee
so that
\be
	X_{1a}\underset{\rho\to 0,\,d\to-2}{\sim}\frac{b^4}{6 \rho }+\rho +\frac{b^2}{4}  \log\left(\frac{\rho}{b^2}\right)+O(d+2)\,.
\ee
Doing so systematically, one arrives at a finite metric for $d=-2$ (we work with the reduced, hatted metric for compactness instead of the tilded one):\footnote{To simplify the presentation, we omit a term proportional to $\log\rho$.}
\bea
	\hat g_{ab}&=&\eta_{ab}\left(1-\frac{b^2}{2 \rho }\right)+\frac{b^2}{\rho }\hat u_a \hat u_b+\frac{b^3}{\rho }\hat \sigma_{ab}+\left[b^2-\frac{5 b^4}{8\rho}\right]\hat \omega _{ac }\hat \omega ^{c }{}_{b } +2\left[\frac{3b^2 }{2}-\frac{b^4}{\rho }\right]\hat \sigma _{c (a }\hat \omega _{b) }{}^{c }\nn\\
	&&+\left[\frac{3b^2}2 -\frac{11b^4 }{8\rho}\right]\hat \sigma _{ac }\hat \sigma _{b }{}^{c }+\left[3b^2-\frac{19b^4 }{8\rho}\right]\hat u\cdot\hat{\mathcal D}\hat \sigma _{ab }-\left[3 b^2-\frac{b^4}{2\rho}\right]\frac{\hat \theta }{3}\hat \sigma _{ab }+\left[b^2-\frac{5 b^4}{8\rho}\right]\hat a_{a }\hat a_{b }\nn\\
&&+\left[\left(\frac{7 b^2}{2}-\frac{13 b^4}{8\rho}\right)\left(\frac{\hat \theta ^2}{9}-\frac{ \dot{\hat \theta}}{3}\right)+\left(\frac{3 b^2}{4}-\frac{b^4}{2\rho}\right)\hat \omega ^2+\left(-2 b^2+\frac{5 b^4}{8\rho}\right)\hat a^2-\frac{13b^4}{24\rho}\hat \sigma ^2\right]\hat P_{ab }\nn\\
	&&+ b^2\left[\left(\frac{5}{2}-\frac{17b^2}{8\rho}\right)\hat \sigma ^2-\left(\frac{5}{4}-\frac{9 b^2}{4\rho}\right)\hat \omega ^2-\left(\frac{15}{2}-\frac{75 b^2}{8\rho}\right)\left(\frac{\hat \theta ^2}{9}-\frac{ \dot{\hat \theta}}{3}\right)+\left(3-\frac{3b^2}\rho\right)\hat a^2\right]\hat u_{a }\hat u_{c }\nn\\
	&&-\left(2 b^2-\frac{7 b^4}{4\rho}\right)\frac{\hat \theta }{3} \hat a_{(a }\hat u_{b) }+\left(4 b^2-\frac{3 b^4}\rho\right)\hat a^{c }\hat u_{(b }\hat \sigma _{a) c }+\left(2 b^2-\frac{4 b^4}\rho\right)\hat a^{c}\hat u_{(b }\hat \omega _{a) c }\nn\\
	&&+\left(\frac{b^2}2-\frac{9b^4}{16\rho}\right)\hat u_{(a }\hat P_{b) }{}^{c }\hat{\mathcal D}^d\hat \sigma _{d c }+\left(\frac{b^2}{2}-\frac{7 b^4}{16\rho}\right)\hat u_{(a }\hat P_{b )}{}^{c }\hat{\mathcal D}^d\hat \omega _{dc }+O(\rho^{-2})\,. \label{g-2}
\eea
As expected, this metric is finite, but $\hg_{(0)}\neq\eta$, and the stress-tensor (which comes at order $\rho^{-1}$) is not in the Landau frame ($\hat u^a\hg_{(-2)ab}\neq0$). Note also that the power $0$ and $-1$ terms are not the analytic continuation for $d=-2$ of the powers $0$ and $d/2$ in \eqref{Blaiseg0} and \eqref{Blaisegdo2}. Subleading powers such as $\rho^{d/2+1}$ and $\rho^{d+1}$ also contribute. One can also check that the order of limits we take ($\rho\to0$ then $d\to-2$ versus $d\to-2$ then $\rho\to0$) does not matter.

\subsubsection{Getting Minkowski on the boundary}

In order to bring the order $0$ in $\rho$ piece of the metric \eqref{g-2} to $\eta_{ab}$, we simply need to do a change of boundary coordinates, and shift
\be \label{ShifttoMink-2}
	x^a\to x^a-\frac12X^a_{2(0)}\,,
\ee
where $X^a_{20}$ is the second-derivative, zeroth-power of $\rho$ piece of \eqref{g-2}. Of course, though \eqref{ShifttoMink-2} means that $\hg_{(0)ab}=\eta_{ab}$, the change of coordinates backreacts on the order $\rho^{-1}$ piece and changes the stress-tensor. One might worry that the original change of coordinates we did, \eqref{CoordChangex}, did not yield the Minkowski metric directly. One should remember that the shift we are doing here is second-order in derivatives, and we have left $X_2^a$ undetermined. Had we calculated it, it would have contained an arbitrary integration constant in $\rho$, which we can always choose as in \eqref{ShifttoMink-2}.

\subsubsection{Going to the Landau frame}

Implementing \eqref{ShifttoMink-2}, the term in \eqref{g-2} where the stress-tensor appears becomes:
\bea
	\hg_{(-2)ab}&=&\frac{3b^2}2\hat u_a \hat u_b-\frac{b^2}2\hat P_{ab}+b^3\hat \sigma_{ab}-\frac{b^4}{8}\hat \omega _{a c }\hat \omega ^{c }{}_{b }-\frac{b^4}2\hat \sigma _{c (a }\hat \omega _{b) }{}^{c }-\frac{5b^4 }{8}\hat \sigma _{ac }\hat \sigma _{b }{}^{c }-b^4\frac{\hat \theta }{3}\hat \sigma _{ab}\nn\\
	&&-\frac{7b^4 }{8}\hat u\cdot{\hat{\mathcal D}}\hat \sigma _{ab}-\frac{ b^4}{8}\hat a_{a }\hat a_{b }+\left[\frac{b^4}{8}\left(\frac{\hat \theta ^2}{9}-\frac{ \dot{\hat \theta}}{3}\right)-\frac{b^4}{8}\hat \omega ^2-\frac{3 b^4}{8}\hat a^2-\frac{13b^4}{24}\hat \sigma ^2\right]\hat P_{ab }\nn\\
	&&+ b^2\left[\frac{13b^2}{8}\hat \sigma ^2+\frac{3b^2}8\hat \omega ^2-\frac{15 b^2}{8}\left(\frac{\hat \theta ^2}{9}-\frac{ \dot{\hat \theta}}{3}\right)+\frac{3b^2}2\hat a^2\right]\hat u_{a }\hat u_{b }+\frac{b^4\theta }{4} \hat a_{(a }\hat u_{b) }\nn\\
	&&-b^4\hat a^{c }\hat u_{(b }\hat \sigma _{a) c }-3 b^4\hat a^{c }\hat u_{(b }\hat \omega _{a) c}-\frac{5b^4}{16}\hat u_{(a }\hat P_{b) }{}^{c }\hat{\mathcal D}^d\hat \sigma _{d c }-\frac{3 b^4}{16}\hat u_{(a }\hat P_{b )}{}^{c }\hat{\mathcal D}^d\hat \omega _{d c }\,, \label{gT-2}
\eea
from which it is immediate to see that this is not in the Landau frame. To reach it,\footnote{This is for convenience, we could as well extract transport coefficients from any other frame.} two steps are needed. First, shift the velocity by a second-derivative correction which will compensate the terms in \eqref{gT-2} which have only one leg along the velocity. Then, shift the temperature to cancel out the terms with two legs along the velocity. In so doing, an extra contribution proportional to $\hat P_{ab}$ comes in, equal to $\hat u^a \hat u^b\hg_{(-2)ab}/3$, so that the final result is
\bea
	\hg_{(-2)ab}&=&-\frac{b^2}2\hat P_{ab}+\frac{3b^2}2\hat u_a \hat u_b+b^3\hat \sigma_{ab}-\frac{b^4}{8}\hat \omega _{a c }\hat \omega ^{c }{}_{b }-\frac{b^4}2\hat \sigma _{c (a }\hat \omega _{b) }{}^{c }-\frac{5b^4 }{8}\hat \sigma _{a c }\hat \sigma _{b}{}^{c }-\nn\\
	&&-\frac{7b^4 }{8}\hat u\cdot\hat{\mathcal D}\hat \sigma _{ab }-b^4\frac{\hat \theta }{3}\hat \sigma _{ab }-\frac{ b^4}{8}\hat a_{a }\hat a_{b }+\left[-\frac{b^4}{2}\left(\frac{\hat \theta ^2}{9}-\frac{ \dot{\hat \theta}}{3}\right)+\frac{ b^4}{8}\hat a^2\right]\hat P_{ab }\label{gT-2b}\,.
\eea
In the end, the renormalized, shifted metric reads
\bea
	\hat g_{ab}&=&\eta_{ab}+\frac1\rho\hg_{(-1)ab}+O(\rho^{-2})\label{g-2b}\,.
\eea
From this, one can extract the stress-tensor, see \eqref{RenST-2}.

\subsection{Renormalized stress-tensor for $n=2$}

Converting to tilded quantities (which includes a change of sign between $\tilde u_a$ and $\hat u_a$, as well as $\tilde T_{ab}$ and $\hat T_{ab}$), the renormalized, reduced stress-tensor for $d=-2$ is:
\bea\label{app:RenST-2}
\left.\tT_{ab}\right|_{d=-2}&=&\tilde P\left(\eta_{ab}-2\tu_a\tu_b\right)-2\tilde\eta\tilde\sigma_{ab}-\tilde\zeta\tilde\theta\tilde P_{ab}\nonumber\\	
	&&+r_0\tilde \eta\left[\tilde \sigma _{ c (a}\tilde \omega _{b) }{}^{c}+\frac{5}{4}\tilde \sigma _{a c }\tilde \sigma _{b }{}^{c }+\frac{1}{4}\tilde\omega _{ac }\tilde \omega ^{c }{}_{b }+\frac{7}{4}\tilde u^{c }\partial_{c }\tilde \sigma _{ab }+\frac{1}{12}\tilde\theta\tilde\sigma _{ab }+\frac{1}{4}\tilde a_{a }\tilde a_{b }\right.\nn\\
	&&\left.-\frac72\tilde a^c\tilde u_{(a}\tilde \sigma_{b)c}+\tilde P_{ab}\left(\frac{\tilde \theta^2}{9}-\frac{\dot{\tilde \theta}}{3}-\frac{\tilde a^2}{4}\right)\right]\nn\\
	&&+r_0\tilde \zeta\left[\frac{5}{4}\tilde \theta\tilde \sigma _{ab}+\tilde P_{ab}\left(\frac{7\dot{\tilde \theta}}{8}-\frac{\tilde \theta^2}{3}+\frac{5}{8p}\tilde \theta^2\right)\right],\\
	&&\tilde P =\frac{-r_0^2}{16\pi\tilde G_N}\,,\qquad \tilde \eta =\frac{r_0^3}{16\pi\tilde G_N}\,,\qquad \tilde \zeta =2\tilde \eta\frac{p+3}{3p}\,.
\eea

\subsection{Renormalized stress-tensor for $n=1$}

Using the procedure outlined in the previous section, the renormalized, reduced stress-tensor for $d=-1$ is:
\bea\label{app:RenST-1}
\left.\tT_{ab}^{}\right|_{d=-1}&=&\tilde P\left(\eta_{ab}-\tu_a\tu_b\right)-2\tilde\eta\tilde\sigma_{ab}-\tilde\zeta\tilde\theta\tilde P_{ab}\nonumber\\			
&&+r_0\tilde \eta\left[\frac{13}{8}\tilde \sigma _{ c (a}\tilde \omega _{b) }{}^{c}+\frac{15}{16}\tilde \sigma _{a c }\tilde \sigma _{b }{}^{c }+\frac{9}{16}\tilde\omega _{ac }\tilde \omega ^{c }{}_{b }+\frac{7}{4}\tilde u^{c }\partial_{c }\tilde \sigma _{ab }+\frac{9}{16}\tilde\theta\tilde\sigma _{ab }\right.\nonumber\\
	&&\left.+\frac{9}{16}\tilde a_{a }\tilde a_{b }-\frac72\tilde a^c\tilde u_{(a}\tilde \sigma_{b)c}+\tilde P_{ab}\left(-\frac{5}{32}\tilde\omega^2-\frac{69}{32}\tilde\sigma^2+\frac{15}{16}\tilde\theta^2-\frac{15}{8}\dot{\tilde\theta}-\frac{3}{32}\tilde a^2\right)\right]\nn\\
	&&+r_0\tilde \zeta\left[\frac{15}{16}\tilde \theta\tilde \sigma _{ab}+\tilde P_{ab}\left(\frac78\dot{\tilde \theta}-\frac{15}{8}\tilde \theta^2+\frac{15(p+2)}{64p}\tilde \theta^2\right)\right],\\
	&&\tilde P =\frac{-r_0}{16\pi\tilde G_N}\,,\qquad \tilde \eta =\frac{r_0^2}{16\pi\tilde G_N}\,,\qquad \tilde\zeta =\tilde\eta\frac{p+2}p\,.
\eea
Note that this encompasses the stress-tensor for the black string, which can be obtained from the previous expression by setting $p=1$.

\section{Fefferman-Graham expansions beyond the stress tensor \label{app:SubleadingFG}}
In this section units are such that $16\pi G_N=1$ and $\ell=1$.

Let us consider AAdS space times with flat boundary metrics and in an expansion to second derivatives. The FG expansion reads
\beq
\begin{split}
&\ud s^2=\frac{\ud\rho^2}{4\rho^2}+\frac{1}{\rho}\mfg_{\mu\nu} \ud x^\mu \ud x^\nu\\
&\mfg=g_{(0)}+\rho^{d/2}g_{(d/2)}+\rho^{d/2+1}g_{(d/2+1)}+\rho^d g_{(d)}+\rho^{d+1}g_{(d+1)}+\cdots
\end{split}
\label{FGcoords}\eeq
with
\beq
\begin{split}
\mfg_{\mu\nu}&=\eta_{\mu\nu}+\frac{\rho^{d/2}}{d}T_{\mu\nu}-\frac{\rho^{d/2+1}}{2d(d+2)}\Box T_{\mu\nu}+\frac{\rho^d}{2d^2}\left(T_\mu{}^\sigma T_{\sigma\nu}-\frac{1}{4(d-1)}T_{\sigma\kappa}T^{\sigma\kappa}\eta_{\mu\nu}\right)\\&+\frac{\rho^{d+1}}{2d^2(d+1)(d+2)}\Bigg[\frac{d-2}{4(d-1)}T^{\rho\sigma}\partial_\mu\partial_\nu T_{\rho\sigma}- T^{\rho\sigma}\partial_\sigma\partial_{(\mu}T_{\nu)\rho}-\frac{d}{4(d-1)}(\partial_\mu T_{\rho\sigma})\partial_\nu T^{\rho\sigma}\\&-
(\partial^\rho T^{\sigma}{}_\mu)\partial_\sigma T_{\rho\nu}+T^{\rho\sigma}\partial_\rho\partial_\sigma T_{\mu\nu}+(\partial^\rho T^\sigma{}_{(\mu})\partial_{\nu)}T_{\rho\sigma}-(d+1)T^\rho{}_{(\mu}\Box T_{\nu)\rho}\\&
+\frac{\eta_{\mu\nu}}{4(d-1)}\left(dT_{\rho\sigma}\Box T^{\rho\sigma}+(\partial_\kappa T_{\rho\sigma})\partial^\kappa T^{\rho\sigma}\right)\Bigg]
\\&+\rho^{3 d/2}T^3+\rho^{3 d/2+1}T^2\partial\partial T+\cdots\,.
\label{FGbeyondST}\end{split}
\eeq
$T_{\mu\nu}$ is the boundary stress tensor and it is traceless and conserved. Let us prove the result above.

The equations determining $\mfg_{\mu\nu}$ in \eqref{FGcoords} are:
\begin{eqnarray}
\rho \,[2 \mfg^{\prime\prime} - 2 \mfg^\prime \mfg^{-1} \mfg^\prime + \textrm{Tr}\,
(\mfg^{-1} \mfg^\prime)\, \mfg^\prime] -\textrm{Ric} (\mfg) - (d - 2)\,
\mfg^\prime - \textrm{Tr} \,(\mfg^{-1} \mfg^\prime)\, \mfg & = & 0\,, \label{eqg}\\
\nabla_\mu\, \textrm{Tr} \,(\mfg^{-1} \mfg^\prime) - \nabla^\nu \mfg_{\mu\nu}^\prime  & = & 0\,, \label{mixedFG}\\
\textrm{Tr} \,(\mfg^{-1} \mfg^{\prime\prime}) - \frac{1}{2} \textrm{Tr} \,(\mfg^{-1} \mfg^\prime
\mfg^{-1}
\mfg^\prime) & = & 0\,. \label{trace}
\end{eqnarray}

Eq.~\eqref{trace} implies that the $d/2+1$ term is traceless. Eq.~\eqref{eqg} is
\beq
g_{(d/2+1)}=\frac{1}{d+2}\left(\textrm{Ric}(g_{(0)}+\rho^{d/2} g_{(d/2)})\right)_{\rho^{d/2}}=-\frac{1}{2d(d+2)}\Box T_{\mu\nu}\,.
\eeq

Now we compute the trace of the $d$ term:
\beq
\begin{split}
\Tr\left(\eta^{-1}g_{(d)}\right)=&\frac{1}{d(d-1)}\left[\frac{d^2}{8}\Tr\left(\eta^{-1}g_{(d/2)}\eta^{-1}g_{(d/2)}\right)+\frac{d}{2}\left(\frac{d}{2}-1\right)\Tr(\eta^{-1}g_{(d/2)}\eta^{-1}g_{(d/2)})\right]\\
=&\frac{3d-4}{8d^2(d-1)}T_{\rho\sigma}T^{\rho\sigma}\,.
\end{split}\eeq
Eq.~\eqref{eqg} gives
\beq
\begin{split}
g_{(d)}=&\frac{1}{d^2}\left(\frac{d^2}{2}g_{(d/2)}\eta^{-1}g_{(d/2)}+d\Tr\left(\eta^{-1}g_{(d)}\right)\eta-\frac{d}{2}\Tr\left(\eta^{-1}g_{(d/2)}\eta^{-1}g_{(d/2)}\right)\eta\right)\\
=&\frac{1}{2d^2}\left(T_{\mu}{}^\rho T_{\rho\nu}-\frac{1}{4(d-1)}T_{\rho\sigma}T^{\rho\sigma}\eta_{\mu\nu}\right).
\end{split}\eeq

Now we look at the $(d+1)$ term. From  \eqref{trace} we determine its trace by collecting the $\rho^{d+1}$ terms:
\beq
\begin{split}
\textrm{Tr}(\eta^{-1}g_{(d+1)})=&\frac{1}{2(d+1)}\Bigg[\left(\frac{d}{2}+1\right)\Tr\left(g_{\left(\frac d2\right)}\eta^{-1}g_{\left(\frac d2+1\right)}\eta^{-1}\right)\\
&+\left(\frac{d}{2}-1\right)\Tr\left(g_{\left(\frac d2+1\right)}\eta^{-1}g_{\left(\frac d2\right)}\eta^{-1}\right)+\left(\frac{d}{2}+1\right)\Tr\left(\eta^{-1}g_{\left(\frac d2\right)}\eta^{-1}g_{\left(\frac d2+1\right)}\right)\Bigg]\\
=&-\frac{3d+2}{8d^2(d+1)(d+2)}T^{\rho\sigma}\Box T_{\rho\sigma}\,.
\end{split}\eeq
Now we examine \eqref{eqg} at the order $\rho^{d}$. Relevant contributions are as follows:
\beq
\begin{split}
-2\left(\mfg^\prime \mfg^{-1} \mfg^{\prime}\right)_{\mu\nu}=&\cdots-\rho^{d-1}d\left(\frac{d}{2}+1\right)\left[g_{\left(\frac d2+1\right)\mu\sigma}\eta^{\sigma\rho} g_{\left(\frac d2\right)\rho\nu}+g_{\left(\frac d2\right)\mu\sigma}\eta^{\sigma\rho} g_{\left(\frac d2+1\right)\rho\nu}\right]+\cdots\\=&\cdots+\frac{\rho^{d-1}}{2d}T_{\rho(\mu}\Box {T_{\nu)}}^\rho+\cdots
\end{split}\eeq
and
\beq\begin{split}
-\textrm{Tr}(\mfg^{-1}\mfg^\prime)\mfg_{\mu\nu}=&\dots+\rho^{d}(d+1)\left[\Tr\left(\eta^{-1}g_{\left(\frac d2\right)}\eta^{-1}g_{\left(\frac d2+1\right)}\right)\eta_{\mu\nu}-\Tr(\eta^{-1}g_{(d+1)})\eta_{\mu\nu}\right]+\cdots\\=&\cdots-\frac{\rho^{d}}{8d^2}T_{\rho\sigma}\Box T^{\rho\sigma}\eta_{\mu\nu}+\cdots
\end{split}\end{equation}
The equation for $g_{d+1}$ becomes:
\beq
g_{(d+1)}=\frac{1}{(d+1)(d+2)}\left(\frac{1}{\rho}\textrm{Ric}(\mfg)+\frac{1}{\rho}\Tr(\mfg^{-1}\mfg)\mfg+2\mfg^\prime \mfg^{-1} \mfg^\prime\right)_{\rho^{d-1}}\,,
\label{gdp1}\eeq
where we have to collect the $\rho^{d-1}$ contribution of the rhs.

There remains to compute the Ricci tensor at the relevant order. There are two terms contributing to it at order $\rho^d$, one coming from $g_{d/2}^2$ and one coming from $g_{d}$. Both of them can be treated as perturbations of flat space (second and first order respectively). Formul\ae\ 4.4.51 and 4.4.4 of Wald \cite{Wald:1984rg} are relevant with $\gamma= g_{(d/2)}$ and $\gamma=g_{(d)}$ respectively.

Thus,
\beq
\begin{split}
\textrm{Ric}(\mfg)_{\mu\nu}=\cdots&+\rho^d\Bigg[\frac{1}{2}g_{(d/2)}^{\rho\sigma}\partial_\mu\partial_\nu g_{(d/2)\rho\sigma}-g_{(d/2)}^{\rho\sigma}\partial_\rho\partial_{(\mu}g_{(d/2)\nu)\sigma}\\
&+\frac{1}{4}(\partial_\mu g_{(d/2)\rho\sigma})\partial_\nu g_{(d/2)}^{\rho\sigma}+
(\partial^\rho g_{(d/2)}^\sigma{}_{\nu})\partial_{[\rho}g_{(d/2)\sigma]\mu}\\
&+\frac{1}{2}g_{(d/2)}^{\rho\sigma}\partial_\rho\partial_\sigma g_{(d/2)\mu\nu}\\
&+\partial^\rho\partial_{(\nu}g_{(d)\mu)\rho}-\frac{1}{2}\Box g_{(d)\mu\nu}-\frac{1}{2}\partial_\mu\partial_\nu\Tr(\eta^{-1}g_{(d)})\Bigg]+\cdots\,,
\end{split}\eeq
where indices are raised with $\eta^{\mu\nu}$ and we used that $\partial^\mu g_{(d/2)\mu\nu}=0$ and $\eta^{\mu\nu}g_{(d/2)\mu\nu}=0$. We get
\beq
\begin{split}
\textrm{Ric}(\mfg)_{\mu\nu}=\cdots&+\frac{\rho^d}{2d^2}\Bigg[\frac{d-2}{4(d-1)}T^{\rho\sigma}\partial_\mu\partial_\nu T_{\rho\sigma}-
T^{\rho\sigma}\partial_\sigma\partial_{(\mu}T_{\nu)\rho}\\
&-\frac{d}{4(d-1)}(\partial_\mu T_{\rho\sigma})\partial_\nu T^{\rho\sigma}-(\partial^\rho T^{\sigma}{}_\mu)\partial_\sigma T_{\rho\nu}\\
&+T^{\rho\sigma}\partial_\rho\partial_\sigma T_{\mu\nu}+
(\partial^\rho T^{\sigma}{}_{(\mu})\partial_{\nu)}T_{\rho\sigma}-T^{\rho}{}_{(\mu}\Box T_{\nu)\rho}\\
&+\frac{\eta_{\mu\nu}}{4(d-1)}\left(T_{\rho\sigma}\Box T^{\rho\sigma}+
(\partial_\kappa T_{\rho\sigma})\partial^{\kappa}T^{\rho\sigma}\right)\Bigg]+\cdots\,.
\end{split}\eeq

The result of adding everything up in \eqref{gdp1} gives the result \eqref{FGbeyondST}.

\section{FG expansion for $d=-2$\label{app:FGexp-2}}
In this section units are such that $16\pi G_N=1$ and $\ell=1$.

We record here the equivalent expansion for the $d=-2$ ($n=2$) case. Logs will appear. We will make the assumption that we are in a weak field region, namely that the metric reads $\eta_{AB}+\textrm{corrections}$. In particular this will require that the logs have a certain size, and the range of $\rho$ for which this will hold is not arbitrarily large. Let us discuss this in some detail.

We sit at a coordinate distance $\rho$ from the object sourcing the gravitational field, which is characterized by an energy density $r_0^2$. The object is not homogeneous on the scale $\lambda$. We are in the long-wavelength expansion, meaning that
\beq
r_0\ll \lambda\,.
\eeq
Consider the range of $\rho$ satisfying
\beq
\frac{r_0^2}{\lambda^2}\log\rho\ll \frac{r_0^2}{\rho}\ll 1\,.
\label{rhodm2}\eeq
In this regime we can expand the metric \`{a} la FG
\beq
\ud s^2=\frac{\ud\rho^2}{4\rho^2}+\frac{1}{\rho}\left[\eta_{\mu\nu}+\frac{1}{\rho}\, T_{\mu\nu}+\log\rho\,\Box T_{\mu\nu}+\frac{1}{\rho^2} \left(T^2\right)_{\mu\nu}+\frac{\log\rho}{\rho}\,\left(T\partial\partial T\right)_{\mu\nu}+\dots\right]\ud x^\mu\ud x^\nu
\eeq
were the regime of $\rho$ \eqref{rhodm2} is such that each term in this expansion is subdominant with respect to the previous one. When the separation between $r_0$ and $\lambda$ is parametric as in a derivative expansion, \eqref{rhodm2} is satisfied over a parametrically large range of $\rho$ (but not for $\rho\rightarrow\infty$).

Solving the equations \eqref{eqg}-\eqref{trace} for $d=-2$ order by order in $\rho$ (and assuming \eqref{rhodm2}) we find:
\beq
\begin{split}
\ud s^2=\frac{\ud\rho^2}{4\rho^2}+\frac{1}{\rho}\Bigg[&\eta_{\mu\nu}-\frac{1}{2\rho}\, T_{\mu\nu}+\frac{\log\rho}{8}\,\Box T_{\mu\nu}+\frac{1}{8\rho^2} \left(T_\mu{}^\sigma T_{\sigma\nu}+\frac{1}{12}T_{\sigma\kappa}T^{\sigma\kappa}\eta_{\mu\nu}\right)\\&+\ell(\rho)\,\left(T\partial\partial T\right)_{\mu\nu}+\dots\Bigg]\ud x^\mu\ud x^\nu
\end{split}\eeq
with an `unreduced' $T_{\mu\nu}$:
\be \label{UnredT-2}
T_{2\mu\nu}=\sigma _{\kappa(\mu }\omega _{\nu) }{}^{\kappa }+\frac{5}{4}\sigma _{\mu \kappa }\sigma _{\nu }{}^{\kappa }+\frac{1}{4}\omega _{\mu \kappa }\omega ^{\kappa }{}_{\nu }+\frac{7}{4}u^{\lambda }D_{\lambda }\sigma _{\mu \nu }+\frac{2}{3}\theta \sigma _{\mu \nu }+\frac{1}{4}a_{\mu }a_{\nu }+P_{\mu \nu }\left(\frac{\theta ^2}{9}-\frac{\dot\theta}{3}-\frac{a^2}{4}\right)
\ee

Let us call the term multiplying $\ell(\rho)$, which is the analogous term to $g_{d+1}$ before, $L_{\mu\nu}$. Eq.~\eqref{trace} gives the equation for its trace
\beq
\ell^{\prime\prime}(\rho)\Tr (\eta^{-1}L)=-\frac{1}{8}\frac{\log\rho-1}{\rho^3}T^{\sigma\kappa}\Box T_{\sigma\kappa}.
\eeq
Solving for $\ell$ we find
\beq
\ell=\frac{\log\rho+1/2}{\rho}\,,
\eeq
which actually means that $\ell(\rho) L_{\mu\nu}$ has two contributions:
\beq
\ell(\rho)L_{\mu\nu}=\frac{\log\rho}{\rho}L^{0}_{\mu\nu}+\frac{1}{2\rho}L^1_{\mu\nu}
\eeq
and they satisfy
\beq
\eta^{\mu\nu} L^{0}_{\mu\nu}=\eta^{\mu\nu}L^{1}_{\mu\nu}=-\frac{1}{16}T^{\mu\nu}\Box T_{\mu\nu}\,.
\label{TraceL}\eeq

Eq.~\eqref{eqg} fixes $L^0$ completely but leaves $L^{1}$ free (as this of the same type than the stress tensor). The result in eq.~\eqref{FGbeyondST} is essentially valid:
\beq
\begin{split}
L^0_{\mu\nu}=-\frac{1}{16}\Bigg[&\frac{1}{3}T^{\rho\sigma}\partial_\mu\partial_\nu T_{\rho\sigma}- T^{\rho\sigma}\partial_\sigma\partial_{(\mu}T_{\nu)\rho}-\frac{1}{6}(\partial_\mu T_{\rho\sigma})\partial_\nu T^{\rho\sigma}\\&-
(\partial^\rho T^{\sigma}{}_\mu)\partial_\sigma T_{\rho\nu}+T^{\rho\sigma}\partial_\rho\partial_\sigma T_{\mu\nu}+(\partial^\rho T^\sigma{}_{(\mu})\partial_{\nu)}T_{\rho\sigma}+T^\rho{}_{(\mu}\Box T_{\nu)\rho}\\&
-\frac{\eta_{\mu\nu}}{12}\left(-2T_{\rho\sigma}\Box T^{\rho\sigma}+(\partial_\kappa T_{\rho\sigma})\partial^\kappa T^{\rho\sigma}\right)\Bigg]\,.
\label{Ldm2}\end{split}
\eeq

If we now decide to call the stress tensor all contributions at the order $1/\rho$, we find that \eqref{TraceL} changes the tracelessness condition to:
\beq
T^{\mu}{}_\mu=\frac{1}{16}T^{\rho\sigma}\Box T_{\rho\sigma}
\label{anomalydm2}\eeq
which has the looks of some sort of conformal anomaly.

At $d=-2$ this becomes
\beq
T^{\mu\nu}\Box T_{\mu\nu}=\frac{b^4}{4}\left[4a^2-\frac{4}{3}\theta^2+5\sigma^2-\omega^2+4\dot{\theta}\right]\,.
\eeq

This new $T_{\mu\nu}$ receiving this traceful second derivative correction obeys a set of effective equations of motion, that can be derived by collecting the $\rho^{-2}$ term of eq.~\eqref{mixedFG}. They are, schematically, of the form
\beq
\partial\cdot T=\partial(T\partial\partial T)\,.
\eeq

At order $1/\rho^2$, we get:
\beq
\Tr\left(\mfg^{-1}\mfg^\prime\right)=\dots+\frac{1}{32\rho^2}T^{\kappa\sigma}\Box T_{\kappa\sigma}+\dots\,,
\eeq
and
\beq
\nabla^{\nu}\mfg^\prime_{\mu\nu}=\dots+\frac{1}{2\rho^2}\partial^\nu T_{\mu\nu}+\frac{1}{\rho^2}\partial^\nu L^{0}_{\mu\nu}+
\frac{1}{16\rho^2}T^{\rho\nu}\partial_\rho\Box T_{\mu\nu}+\frac{1}{32\rho^2}(\Box T^{\rho\sigma})\partial_{\mu}T_{\rho\sigma}+ \dots\,.
\eeq
Collecting in eq.~\eqref{mixedFG}:
\beq
\partial^\nu T_{\mu\nu}+2\partial^\nu L^{0}_{\mu\nu}+
\frac{1}{8}T^{\rho\nu}\partial_\rho\Box T_{\mu\nu}-\frac{1}{16}T^{\kappa\sigma}\partial_\mu\Box T_{\kappa\sigma}=0\,,
\eeq
which simplifies to
\beq
\partial^\nu T_{\mu\nu} - \frac{1}{8}T^{\nu\rho}\partial_\mu\Box T_{\nu\rho}
+\frac{1}{8}T^{\nu\rho}\partial_\rho\Box T_{\nu\mu}
-\frac{1}{8}\partial^\rho T^{\mu\nu}\Box T_{\nu\rho}=0\,.
\label{eomdm2}\eeq

In terms of the $(p+1)$-dimensional stress tensor they read
\beq
\partial^b \tT_{ab} + \frac{1}{8}\left(\tT^{bc}\partial_a\Box \tT_{bc}-\frac{1}{p+3}\tT^{b}\,_{b}\partial_a\Box \tT^c\,_{c}\right)
-\frac{1}{8}\tT^{bc}\partial_c\Box \tT_{ba}
+\frac{1}{8}\partial^c \tT^{ab}\Box \tT_{bc}=0\,,
\eeq
where we took into account the sign flip in the stress tensor induced by the map and that in the term in parenthesis the compact directions in the KK reduction play a role.

\section{FG expansion for $d=-1$\label{app:FGexp-1}}
In this section units are such that $16\pi G_N=1$ and $\ell=1$.

Pretty much as in the previous section, we record here the special case $d=-1$. Same introductory considerations about the range of $\rho$ apply.
\beq
\begin{split}
\mfg_{\mu\nu}&=\eta_{\mu\nu}-\rho^{-1/2}T_{\mu\nu}+\frac{\rho^{1/2}}{2}\Box T_{\mu\nu}+\frac{\rho^{-1}}{2}\left(T_\mu{}^\sigma T_{\sigma\nu}+\frac{1}{8}T_{\sigma\kappa}T^{\sigma\kappa}\eta_{\mu\nu}\right)\\
&+\frac{\log{\rho}}{2}\Bigg[\frac{3}{8}T^{\rho\sigma}\partial_\mu\partial_\nu T_{\rho\sigma}- T^{\rho\sigma}\partial_\sigma\partial_{(\mu}T_{\nu)\rho}-\frac{1}{8}(\partial_\mu T_{\rho\sigma})\partial_\nu T^{\rho\sigma} -
(\partial^\rho T^{\sigma}{}_\mu)\partial_\sigma T_{\rho\nu}\\
&+T^{\rho\sigma}\partial_\rho\partial_\sigma T_{\mu\nu}+(\partial^\rho T^\sigma{}_{(\mu})\partial_{\nu)}T_{\rho\sigma}-\frac{\eta_{\mu\nu}}{8}\left(-T_{\rho\sigma}\Box T^{\rho\sigma}+(\partial_\kappa T_{\rho\sigma})\partial^\kappa T^{\rho\sigma}\right)\Bigg] \\&
-\frac{\rho^{-3/2}}{6}\lp T_\mu{}^\rho T_\rho{}^\si T_{\si\nu}+\frac{7}{16}T^{\rho\si}T_{\rho\si}T_{\mu\nu}\rp+\ell(\rho)T^2\partial\partial T\dots
\label{FGbeyondSTdm1}\end{split}
\eeq
The equation for the trace of the $T^2\partial\partial T$ term becomes:
\beq
\begin{split}
\ell^{\prime\prime}(\rho)\Tr(T^2\partial\partial T)=&\frac{3}{32}\left(-1+\frac{1}{2}\log\rho\right)\rho^{-5/2}\Big[ T^{\mu\nu}\partial_\mu T^{\rho\sigma}\partial_\nu T_{\rho\sigma}-8\,T^{\mu\nu}\partial_\mu T^{\rho\sigma}\partial_\rho T_{\nu\sigma}\\&+8\,T_{\mu}{}^{\nu}\partial^\rho T^{\mu\sigma}\partial_\sigma T_{\nu\rho}
+8\,T^{\mu\rho}T^{\sigma\nu}\partial_\rho\partial_\sigma T_{\mu\nu}-11\,T^{\mu\nu}T^{\rho\sigma}\partial_\rho\partial_\sigma T_{\mu\nu}\Big]
\\&-\frac{3}{16}\rho^{-5/2}\,T^{\mu\rho}T_{\rho}{}^{\nu}\Box T_{\mu\nu}\,,
\end{split}\eeq
and its solution is
\beq
\begin{split}
\ell(\rho)\Tr(T^2\partial\partial T)=&\frac{1}{8}\left(\frac{1}{3}+\frac{1}{2}\log\rho\right)\rho^{-1/2}\Big[ T^{\mu\nu}\partial_\mu T^{\rho\sigma}\partial_\nu T_{\rho\sigma}-8\,T^{\mu\nu}\partial_\mu T^{\rho\sigma}\partial_\rho T_{\nu\sigma}\\&+8\,T_{\mu}{}^{\nu}\partial^\rho T^{\mu\sigma}\partial_\sigma T_{\nu\rho}
+8\,T^{\mu\rho}T^{\sigma\nu}\partial_\rho\partial_\sigma T_{\mu\nu}-11\,T^{\mu\nu}T^{\rho\sigma}\partial_\rho\partial_\sigma T_{\mu\nu}\Big]
\\&-\frac{1}{4}\rho^{-1/2}\,T^{\mu\rho}T_{\rho}{}^{\nu}\Box T_{\mu\nu}\,.
\end{split}\eeq

The terms proportional to $\rho^{-1/2}$ are a contribution to the trace of the stress tensor:
\beq\label{TraceRenSd=-1}
\begin{split}
\eta^{\mu\nu}T_{\mu\nu}=&-\frac{1}{24}\Big[ T^{\mu\nu}\partial_\mu T^{\rho\sigma}\partial_\nu T_{\rho\sigma}-8\,T^{\mu\nu}\partial_\mu T^{\rho\sigma}\partial_\rho T_{\nu\sigma}+8\,T_{\mu}{}^{\nu}\partial^\rho T^{\mu\sigma}\partial_\sigma T_{\nu\rho}
\\&+8\,T^{\mu\rho}T^{\sigma\nu}\partial_\rho\partial_\sigma T_{\mu\nu}-11\,T^{\mu\nu}T^{\rho\sigma}\partial_\rho\partial_\sigma T_{\mu\nu}\Big]
+\frac{1}{4}\,T_{\mu}{}^{\nu}\partial^\rho T^{\mu\sigma}\partial_\sigma T_{\nu\rho}\\=&
\frac{b^3}{8}\left(6\,a^2-15\,\theta^2+30\,\dot{\theta}+42\,\sigma^2-2\,\omega^2\right)\,.
\end{split}\eeq
This agrees with the trace of the stress tensor\eqref{app:RenST-1}
\bea \label{UnredT-1}
T_{2\mu\nu}&=&\frac{13}{16}\left(\sigma _{\mu \kappa }\omega _{\nu }{}^{\kappa }+\sigma _{\nu \kappa }\omega _{\mu }{}^{\kappa }\right)+\frac{15}{16}\sigma _{\mu \kappa }\sigma _{\nu }{}^{\kappa }+\frac{9}{16}\omega _{\mu \kappa }\omega ^{\kappa }{}_{\nu }+\frac{7}{4}u^{\lambda }D_{\lambda }\sigma _{\mu \nu }+\frac{23}{16}\theta \sigma _{\mu \nu }\nonumber\\
	&&+\frac{9}{16}a_{\mu }a_{\nu }+P_{\mu \nu }\left(-\frac{5}{32}\omega ^2-\frac{69}{32}\sigma ^2+\frac{15}{16}\theta ^2-\frac{15}{8}\dot\theta-\frac{3}{32}a^2\right).
\label{dstresstensordm1}\eea

The effective equation of motion the stress tensor satisfies is
\beq\begin{split}
& \partial_\mu T^{\mu}{}_\nu
+ \frac{1}{4}\, {T}^{\mu \rho} {T}_{\mu}\,^{\sigma} {\partial}_{\rho}\,\Box{{T}_{\nu \sigma}}\,
+ \frac{1}{2}\, {T}^{\mu \rho} {T}^{\sigma \kappa} {\partial}_{\mu \rho \sigma}{{T}_{\nu \kappa}}\,
- \frac{37}{48}\, {T}^{\mu \rho} {T}^{\sigma \kappa} {\partial}_{\nu \mu \rho}{{T}_{\sigma \kappa}}\,
 \\
&
+ \frac{1}{3}\,{T}^{\mu \rho}\,  {T}^{\sigma \kappa} {\partial}_{\nu \mu \sigma}{{T}_{\rho \kappa}}\, - \frac{1}{2}\, {T}^{\mu \rho} {\partial}_{\mu}{{T}_{\nu}\,^{\sigma}}\, \Box{{T}_{\rho \sigma}}\,
+ \frac{1}{4}\, {T}^{\mu \rho} {\partial}_{\mu}{{T}_{\rho}\,^{\sigma}}\,  \Box{{T}_{\nu \sigma}}\,
+ \frac{11}{24}\, {T}^{\mu \rho} {\partial}_{\mu}{{T}^{\sigma \kappa}}\,  {\partial}_{\nu \rho}{{T}_{\sigma \kappa}}\, \\
&
- \frac{4}{3}\, {T}^{\mu \rho} {\partial}_{\mu}{{T}^{\sigma \kappa}}\,  {\partial}_{\nu \sigma}{{T}_{\rho \kappa}}\,
+ {T}^{\mu \rho} {\partial}_{\mu}{{T}^{\sigma \kappa}}\,  {\partial}_{\sigma \kappa}{{T}_{\nu \rho}}\,
+ \frac{1}{4}\, {T}^{\mu \rho} {\partial}_{\nu}{{T}_{\mu}\,^{\sigma}}\,  \Box{{T}_{\rho \sigma}}\,
 - \frac{13}{48}\, {T}^{\mu \rho} {\partial}_{\nu}{{T}^{\sigma \kappa}}\,  {\partial}_{\mu \rho}{{T}_{\sigma \kappa}}\, \\
&
 + \frac{7}{6}\, {T}^{\mu \rho} {\partial}_{\nu}{{T}^{\sigma \kappa}}\,  {\partial}_{\mu \sigma}{{T}_{\rho \kappa}}\,
 - \frac{37}{48}\, {T}^{\mu \rho} {\partial}_{\nu}{{T}^{\sigma \kappa}}\,  {\partial}_{\sigma \kappa}{{T}_{\mu \rho}}\,
 + \frac{7}{6}\, {T}^{\mu \rho} {\partial}^{\sigma}{{T}_{\mu}\,^{\kappa}}\,  {\partial}_{\nu \kappa}{{T}_{\rho \sigma}}\,
 \\
& - \frac{1}{3}\, {T}^{\mu \rho} {\partial}^{\sigma}{{T}_{\mu}\,^{\kappa}}\,  {\partial}_{\nu \rho}{{T}_{\sigma \kappa}}\,
 - {T}^{\mu \rho} {\partial}^{\sigma}{{T}_{\mu}\,^{\kappa}}\,  {\partial}_{\rho \kappa}{{T}_{\nu \sigma}}\,
 + \frac{1}{32}\, {T}^{\mu \rho} {\partial}^{\sigma}{{T}_{\mu \rho}}\, \Box{{T}_{\nu \sigma}}\,
 - \frac{1}{2}\, {T}^{\mu \rho} {\partial}^{\sigma}{{T}_{\nu}\,^{\kappa}}\,  {\partial}_{\mu \kappa}{{T}_{\rho \sigma}}\,
\\
& - \frac{1}{2}\, {T}^{\mu \rho} {\partial}^{\sigma}{{T}_{\nu}\,^{\kappa}}\,  {\partial}_{\mu \rho}{{T}_{\sigma \kappa}}\,  %
 + \frac{1}{2}\, {T}^{\mu \rho} {\partial}^{\sigma}{{T}_{\nu}\,^{\kappa}}\,  {\partial}_{\mu \sigma}{{T}_{\rho \kappa}}\,
 - \frac{3}{8}\, {T}^{\mu \rho} {\partial}^{\sigma}{{T}_{\nu}\,^{\kappa}}\,  {\partial}_{\sigma \kappa}{{T}_{\mu \rho}}\,
 + \frac{1}{2}\, {T}_{\nu}\,^{\mu} {\partial}_{\mu}{{T}^{\rho \sigma}}\,  \Box{{T}_{\rho \sigma}}\,
\\
&
 - \frac{1}{2}\, {T}_{\nu}\,^{\mu} {\partial}^{\rho}{{T}_{\mu}\,^{\sigma}}\, \Box{{T}_{\rho \sigma}}\,
 - \frac{1}{4}\, {T}_{\nu}\,^{\mu} {\partial}^{\rho}{{T}^{\sigma \kappa}}\,  {\partial}_{\mu \rho}{{T}_{\sigma \kappa}}\,
  + \frac{1}{8}\,{\partial}^{\mu}{{T}_{\nu}\,^{\rho}}\,  {\partial}_{\mu}{{T}^{\sigma \kappa}}\,  {\partial}_{\rho}{{T}_{\sigma \kappa}}\,
   \\
&
 - \frac{1}{2}\,{\partial}^{\mu}{{T}_{\nu}\,^{\rho}}\,  {\partial}_{\mu}{{T}^{\sigma \kappa}}\,  {\partial}_{\sigma}{{T}_{\rho \kappa}}\,
    - \frac{1}{2}\,{\partial}^{\mu}{{T}_{\nu}\,^{\rho}}\,   {\partial}_{\rho}{{T}^{\sigma \kappa}}\,  {\partial}_{\sigma}{{T}_{\mu \kappa}}\,
     + {\partial}^{\mu}{{T}_{\nu}\,^{\rho}}\,  {\partial}^{\sigma}{{T}_{\mu}\,^{\kappa}}\,  {\partial}_{\kappa}{{T}_{\rho \sigma}}\,\\
&
      - \frac{1}{48}\, {\partial}^{\mu}{{T}^{\rho \sigma}}\,  {\partial}^{\kappa}{{T}_{\rho \sigma}}\,  {\partial}_{\nu}{{T}_{\mu \kappa}}\,
       + \frac{1}{6}\, {\partial}^{\mu}{{T}^{\rho \sigma}}\,  {\partial}_{\nu}{{T}_{\mu}\,^{\kappa}}\,  {\partial}_{\rho}{{T}_{\kappa \sigma}}\,
        - \frac{5}{12}\, {\partial}_{\nu}{{T}^{\mu \rho}}\,  {\partial}^{\sigma}{{T}_{\mu}\,^{\kappa}}\,  {\partial}_{\kappa}{{T}_{\rho \sigma}}=0\,,
\end{split}\label{eomdm1}\eeq
and the relevant terms for the GL dispersion relation are
\beq
\begin{split}
&0=\partial_\mu T^{\mu}{}_\nu + \frac{1}{4}\, {T}^{\mu \rho} {T}_{\mu}\,^{\sigma} {\partial}_{\rho}\Box{{T}_{\nu \sigma}}\,\\
&\qquad\qquad
+ \frac{1}{2}\, {T}^{\mu \rho} {T}^{\sigma \kappa} {\partial}_{\mu \rho \sigma}{{T}_{\nu \kappa}}\,
- \frac{37}{48}\, {T}^{\mu \rho} {T}^{\sigma \kappa} {\partial}_{\nu \mu \rho}{{T}_{\sigma \kappa}}\,
+ \frac{1}{3}\,{T}^{\mu \rho}  {T}^{\sigma \kappa} {\partial}_{\nu \mu \sigma}{{T}_{\rho \kappa}}\,,
\end{split}
\eeq
which in terms of the reduced stress tensor read:
\beq\begin{split}
\partial_a \tT^{a}{}_b &
+ \frac{1}{4}\, {\tT}^{ac} {\tT}_{a}\,^{d} {\partial}_{c}\Box{{\tT}_{bd}}\,
+ \frac{1}{2}\, {\tT}^{ac} {\tT}^{de} {\partial}_{acd}{{\tT}_{be}}\,  \\&
- \frac{37}{48}\, {\tT}^{ac}\left( {\tT}^{de} {\partial}_{b a c}{{\tT}_{d e}}
-\frac{1}{p+2}\,\tT^d\,_d {\partial}_{b a c} \tT^e\,_e\right)\,
+ \frac{1}{3}\, {\tT}^{a c} {\tT}^{d e} {\partial}_{b a d}{{\tT}_{c e}}=0\,,
\end{split}\eeq
where we needed to use the fact that for the term in brackets the compact directions also play a role.



\begin{thebibliography}{99}

\bibitem{Maldacena:1997re}
  J.~M.~Maldacena,
  ``The Large N limit of superconformal field theories and supergravity,''
  Adv.\ Theor.\ Math.\ Phys.\  {\bf 2}, 231 (1998)
  \hre{hep-th}{9711200}.



\bibitem{Itzhaki:1998dd}
  N.~Itzhaki, J.~M.~Maldacena, J.~Sonnenschein and S.~Yankielowicz,
  ``Supergravity and the large N limit of theories with sixteen supercharges,''
Phys.\ Rev.\ D {\bf 58}, 046004 (1998)   \hre{hep-th}{9802042}.

\bibitem{Wiseman:2008qa}
  T.~Wiseman and B.~Withers,
  ``Holographic renormalization for coincident Dp-branes,''
 JHEP {\bf 0810}, 037 (2008), \hri{0807.0755}{[hep-th]}.

\bibitem{Kanitscheider:2008kd}
  I.~Kanitscheider, K.~Skenderis and M.~Taylor,
  ``Precision holography for non-conformal branes,''
JHEP {\bf 0809}, 094 (2008)  \hri{0807.3324}{[hep-th]}.


\bibitem{Kanitscheider:2009as}
  I.~Kanitscheider and K.~Skenderis,
  ``Universal hydrodynamics of non-conformal branes,''
  JHEP {\bf 0904} (2009) 062
  \hri{0901.1487}{[hep-th]}.


\bibitem{Caldarelli:2012hy}
  M.~M.~Caldarelli, J.~Camps, B.~Gout\'eraux and K.~Skenderis,
  ``AdS/Ricci-flat correspondence and the Gregory-Laflamme instability,''
  Phys.\ Rev.\ D {\bf 87}, 061502 (2013)
  \hri{1211.2815}{[hep-th]}.

\bibitem{Hervik:2007zz}
  S.~Hervik,
  ``Ricci Nilsoliton Black Holes,''
  J.\ Geom.\ Phys.\  {\bf 58} (2008) 1253
  \hri{0707.2755}{[hep-th]}.

\bibitem{Lauret}
  J.~Lauret,
  ``Einstein solvmanifolds and nilsolitons,''
  \hri{0806.0035}{[math.DG]}

\bibitem{BrittoPacumio:1999sn}
  R.~Britto-Pacumio, A.~Strominger and A.~Volovich,
  ``Holography for coset spaces,''
  JHEP {\bf 9911} (1999) 013
  \hre{hep-th}{9905211}.

\bibitem{Taylor:2000xf}
  M.~Taylor,
  ``Holography for degenerate boundaries,''
  \hre{hep-th}{0001177}.

\bibitem{Gouteraux:2011qh}
  B.~Gout\'eraux, J.~Smolic, M.~Smolic, K.~Skenderis and M.~Taylor,
  ``Holography for Einstein-Maxwell-dilaton theories from generalized dimensional reduction,''
  JHEP {\bf 1201} (2012) 089
  \hri{1110.2320}{[hep-th]}.

\bibitem{Smolic:2013gx}
  M.~Smolic,
  ``Holography and hydrodynamics for EMD theory with two Maxwell fields,''
  JHEP {\bf 1303}, 124 (2013)
  \hri{1301.6020}{[hep-th]}.


\bibitem{Gouteraux:2011ce}
  B.~Gout\'eraux and E.~Kiritsis,
  ``Generalized Holographic Quantum Criticality at Finite Density,''
  JHEP {\bf 1112} (2011) 036
  \hri{1107.2116}{[hep-th]}.

\bibitem{Gouteraux:2012yr}
  B.~Gout\'eraux and E.~Kiritsis,
  ``Quantum critical lines in holographic phases with (un)broken symmetry,''
  JHEP {\bf 1304} (2013) 053
  \hri{1212.2625}{[hep-th]}.

\bibitem{de Haro:2000xn}
  S.~de Haro, S.~N.~Solodukhin and K.~Skenderis,
 ``Holographic reconstruction of space-time and renormalization in the AdS / CFT correspondence,''
  Commun.\ Math.\ Phys.\  {\bf 217} (2001) 595
  \hre{hep-th}{0002230}.

\bibitem{Henningson:1998gx}
  M.~Henningson and K.~Skenderis,
  ``The Holographic Weyl anomaly,''
  JHEP {\bf 9807} (1998) 023
  \hre{hep-th}{9806087}.

\bibitem{Compere:2011dx}
  G.~Compere, P.~McFadden, K.~Skenderis and M.~Taylor,
  ``The Holographic fluid dual to vacuum Einstein gravity,''
  JHEP {\bf 1107} (2011) 050
  \hri{1103.3022}{[hep-th]}.


\bibitem{Emparan:2009at}
  R.~Emparan, T.~Harmark, V.~Niarchos and N.~A.~Obers,
  ``Essentials of Blackfold Dynamics,''
  JHEP {\bf 1003} (2010) 063
  \hri{0910.1601}{[hep-th]}.

\bibitem{Gregory:1993vy}
  R.~Gregory and R.~Laflamme,
  ``Black strings and p-branes are unstable,''
  Phys.\ Rev.\ Lett.\  {\bf 70} (1993) 2837
  \hre{hep-th}{9301052}.

\bibitem{Gregory:1994bj}
  R.~Gregory and R.~Laflamme,
  ``The Instability of charged black strings and p-branes,''
  Nucl.\ Phys.\ B {\bf 428} (1994) 399
  \hre{hep-th}{9404071}.

\bibitem{Bhattacharyya:2008jc}
  S.~Bhattacharyya, V.~E.~Hubeny, S.~Minwalla and M.~Rangamani,
 ``Nonlinear Fluid Dynamics from Gravity,''
  JHEP {\bf 0802} (2008) 045
  \hri{0712.2456}{[hep-th]}.

\bibitem{Bhattacharyya:2008mz}
  S.~Bhattacharyya, R.~Loganayagam, I.~Mandal, S.~Minwalla and A.~Sharma,
  ``Conformal Nonlinear Fluid Dynamics from Gravity in Arbitrary Dimensions,''
  JHEP {\bf 0812} (2008) 116
  \hri{0809.4272}{[hep-th]}.

\bibitem{Camps:2010br}
  J.~Camps, R.~Emparan and N.~Haddad,
  ``Black Brane Viscosity and the Gregory-Laflamme Instability,''
  JHEP {\bf 1005} (2010) 042
  \hri{1003.3636}{[hep-th]}.

\bibitem{Figueras}
P. Figueras, private communication.


\bibitem{Bredberg:2011jq}
  I.~Bredberg, C.~Keeler, V.~Lysov and A.~Strominger,
  ``From Navier-Stokes To Einstein,''
  JHEP {\bf 1207} (2012) 146
  \hri{1101.2451}{[hep-th]}.


\bibitem{Compere:2012mt}
  G.~Compere, P.~McFadden, K.~Skenderis and M.~Taylor,
  ``The relativistic fluid dual to vacuum Einstein gravity,''
  JHEP {\bf 1203} (2012) 076
  \hri{1201.2678}{[hep-th]}.


\bibitem{Eling:2012ni}
  C.~Eling, A.~Meyer and Y.~Oz,
  ``The Relativistic Rindler Hydrodynamics,''
  JHEP {\bf 1205} (2012) 116
  \hri{1201.2705}{[hep-th]}.




\bibitem{Baier:2007ix}
  R.~Baier, P.~Romatschke, D.~T.~Son, A.~O.~Starinets and M.~A.~Stephanov,
  ``Relativistic viscous hydrodynamics, conformal invariance, and holography,''
  JHEP {\bf 0804} (2008) 100
  \hri{0712.2451}{[hep-th]}.

\bibitem{VanAcoleyen:2011mj}
  K.~Van Acoleyen and J.~Van Doorsselaere,
  ``Galileons from Lovelock actions,''
  Phys.\ Rev.\ D {\bf 83} (2011) 084025
  \hri{1102.0487}{[gr-qc]}.


\bibitem{Charmousis:2012dw}
  C.~Charmousis, B.~Gout\'eraux and E.~Kiritsis,
  ``Higher-derivative scalar-vector-tensor theories: black holes, Galileons, singularity cloaking and holography,''
JHEP {\bf 1209} (2012) 011
  \hri{1206.1499}{[hep-th]}.


\bibitem{Skenderis:1999nb}
  K.~Skenderis and S.~N.~Solodukhin,
  ``Quantum effective action from the AdS / CFT correspondence,''
  Phys.\ Lett.\ B {\bf 472}, 316 (2000)
  \hre{hep-th}{9910023}.



\bibitem{Rangamani:2009xk}
  M.~Rangamani,
  ``Gravity and Hydrodynamics: Lectures on the fluid-gravity correspondence,''
  Class.\ Quant.\ Grav.\  {\bf 26} (2009) 224003
  \hri{0905.4352}{[hep-th]}.

\bibitem{Hubeny:2011hd}
  V.~E.~Hubeny, S.~Minwalla and M.~Rangamani,
  ``The fluid/gravity correspondence,''
in ``Black Holes in Higher Dimensions,''  Cambridge University Press (2012, editor: G. Horowitz)
  \hri{1107.5780}{[hep-th]}.


\bibitem{Loganayagam:2008is}
  R.~Loganayagam,
  ``Entropy Current in Conformal Hydrodynamics,''
  JHEP {\bf 0805} (2008) 087
  \hri{0801.3701}{[hep-th]}.


\bibitem{Bhattacharyya:2008xc}
  S.~Bhattacharyya, V.~E.~Hubeny, R.~Loganayagam, G.~Mandal, S.~Minwalla, T.~Morita, M.~Rangamani and H.~S.~Reall,
  ``Local Fluid Dynamical Entropy from Gravity,''
  JHEP {\bf 0806} (2008) 055
  \hri{0803.2526 [hep-th]}.



\bibitem{Hawking:1973uf}
  S.~W.~Hawking and G.~F.~R.~Ellis,
  ``The Large scale structure of space-time,''
  Cambridge University Press, Cambridge, 1973


\bibitem{Booth:2010kr}
  I.~Booth, M.~P.~Heller and M.~Spalinski,
  ``Black Brane Entropy and Hydrodynamics,''
  Phys.\ Rev.\ D {\bf 83} (2011) 061901
  \hri{1010.6301}{[hep-th]}.


\bibitem{Booth:2011qy}
  I.~Booth, M.~P.~Heller, G.~Plewa and M.~Spalinski,
  ``On the apparent horizon in fluid-gravity duality,''
  Phys.\ Rev.\ D {\bf 83} (2011) 106005
  \hri{1102.2885}{[hep-th]}.



\bibitem{Emparan:2009cs}
  R.~Emparan, T.~Harmark, V.~Niarchos and N.~A.~Obers,
  ``World-Volume Effective Theory for Higher-Dimensional Black Holes,''
  Phys.\ Rev.\ Lett.\  {\bf 102} (2009) 191301
  \hri{0902.0427}{[hep-th]}.

\bibitem{Camps:2012hw}
  J.~Camps and R.~Emparan,
  ``Derivation of the blackfold effective theory,''
  JHEP {\bf 1203} (2012) 038
   [Erratum-ibid.\  {\bf 1206} (2012) 155]
  \hri{1201.3506}{[hep-th]}

\bibitem{Gregory:2011kh}
  R.~Gregory,
  ``The Gregory-Laflamme instability,''
in ``Black Holes in Higher Dimensions,''  Cambridge University Press (2012, editor: G. Horowitz)
  \hri{1107.5821}{[gr-qc]}.

\bibitem{Lehner:2011wc}
  L.~Lehner and F.~Pretorius,
  ``Final State of Gregory-Laflamme Instability,''
in ``Black Holes in Higher Dimensions,''  Cambridge University Press (2012, editor: G. Horowitz)
  \hri{1106.5184}{[gr-qc]}.


\bibitem{Emparan:2013moa}
  R.~Emparan, R.~Suzuki and K.~Tanabe,
  ``The large D limit of General Relativity,''
JHEP {\bf 1306} (2013) 009
  \hri{1302.6382}{[hep-th]}.






\bibitem{Skenderis:2006uy}
  K.~Skenderis and M.~Taylor,
  ``Kaluza-Klein holography,''
  JHEP {\bf 0605}, 057 (2006)
  \hre{hep-th}{0603016}.


\bibitem{Charmousis:2003wm}
  C.~Charmousis and R.~Gregory,
  ``Axisymmetric metrics in arbitrary dimensions,''
  Class.\ Quant.\ Grav.\  {\bf 21} (2004) 527
  \hre{gr-qc}{0306069}.




\bibitem{Wald:1984rg}
  R.~M.~Wald,
  {\em``General Relativity,''}
  Chicago, Usa: Univ. Pr. ( 1984) 491p







\end{thebibliography}
\end{document}